\documentclass[12pt]{article}

\usepackage{amsfonts}
\usepackage{amssymb}

\usepackage[utf8]{inputenc}
\usepackage[T1]{fontenc}

\usepackage{graphicx}
\usepackage{caption}
\usepackage{xcolor}

\newtheorem{proposition}{Proposition}[section]
\newtheorem{remark}[proposition]{Remark}
\newtheorem{conjecture}[proposition]{Conjecture}
\newtheorem{theorem}[proposition]{Theorem}
\newtheorem{lemma}[proposition]{Lemma}
\newtheorem{corollary}[proposition]{Corollary}

\newtheorem{definition}[proposition]{Definition}

\newtheorem{axiom[proposition]}{Axiome}

\input{tcilatex}

\renewcommand{\theequation}{\arabic{equation}}

\renewcommand{\theequation}{\arabic{section}.\arabic{equation}}

\begin{document}

\font\fifteen=cmbx10 at 15pt
\font\twelve=cmbx10 at 12pt

\begin{titlepage}

\begin{center}

\vspace{2 cm}
%\textbf{A model with coexistence of two kinds of Bose condensations\\
%Quantum Interpretation of Thermodynamic Behaviour of the Bogoliubov Weakly Imperfect Bose-Gas\\
%Exact Solution of the Bogoliubov Hamiltonian\\ for Weakly Imperfect Bose-Gas}
\textbf{Non-conventional Dynamical Bose Condensation}

\vspace{1cm}

\hfill \textit{To the memory of Viatcheslav Borisovich Priezzhev}

\vspace{1cm}

\textbf{Valentin A. Zagrebnov}

\bigskip

Aix-Marseille Universit\'{e}, CNRS, Centrale Marseille, I2M \\
Institut de Math\'{e}matiques de Marseille (UMR 7373) - AMU,\\
Centre de Math\'{e}matiques et Informatique - Technop\^{o}le Ch\^{a}teau-Gombert\\
39, rue F. Joliot Curie, 13453 Marseille Cedex 13, France \\
{\textit{Valentin.Zagrebnov(at)univ-amu.fr}}

\setcounter{footnote}{0}

\vspace{1cm}

{\bf Abstract}

\end{center}

The paper presents a review of results concerning the non-conventional dynamical condensation
versus conventional Bose-Einstein condensation, including the case of generalised
van den Berg-Lewis-Pul\'{e} condensate.
The review is based on detailed discussion of two models: a simple toy model and the
Bogoliubov Weakly Imperfect Bose-Gas model, which was invented for explanation of superfluity
of liquid ${\rm{^4He}}$, but which is also instructive for analysis of non-conventional
condensation regarding some recent interpretations of experimental data.

\end{titlepage}

%%%%%%%%%%%%%%%%%%%%%%%%%%%%%%%%%%%%%%%%%%%%%%%%%%%%%%%%%%%%%%%%%%%%%%%%%%%%%%%%%%%%%%%%%%%%%%%%
%\tableofcontents
%%%%%%%%%%%%%%%%%%%%%%%%%%%%%%%%%%%%%%%%%%%%%%%%%%%%%%%%%%%%%%%%%%%%%%%%%%%%%%%%%%%%%%%%%%%%%%%%
%%%%%%%%%%%%%%%%%%%%%%%%%%%%%%%%%%%%%%%%%%%%%%%%%%%%%%%%%%%%%%%%%%%%%%%%%%%%%%%%%%%%%%%%%%%%%%%%
%\setcounter{footnote}{0}
%%%%%%%%%%%%%%%%%%%%%%%%%%%%%%%%%%%%%%%%%%%%%%%%%%%%%%%%%%%%%%%%%%%%%%%%%%%%%%%%%%%%%%%%%%%%%%%%
%\numberwithin{equation}{section}
%\renewcommand{\theequation}{\arabic{section}.\arabic{equation}}
%%%%%%%%%%%%%%%%%%%%%%%%%%%%%%%%%%%%%%%%%%%%%%%%%%%%%%%%%%%%%%%%%%%%%%%%%%%%%%%%%%%%%%%%%%%%%%%%
%\renewcommand{\thefootnote}{\arabic{footnote}}
%\setcounter{footnote}{0} \renewcommand{\thefootnote}{\arabic{footnote}}
%\numberwithin{equation}{section}
%\renewcommand{\theequation}{\arabic{section}.\arabic{equation}}
%%%%%%%%%%%%%%%%%%%%%%%%%%%%%%%%%%%%%%%%%%%%%%%%%%%%%%%%%%%%%%%%%%%%%%%%%%%%%%%%%%%%%%%%%%%%%%%%
\section{Introduction}\label{sec:0}
%%%%%%%%%%%%%%%%%%%%%%%%%%%%%%%%%%%%%%%%%%%%%%%%%%%%%%%%%%%%%%%%%%%%%%%%%%%%%%%%%%%%%%%%%%%%%%%%
\textbf{1.1} If one would summarise shortly the last half-century mathematical results concerning of
what is called the Bose-Einstein condensation (BEC), then it is compulsory to distinguish two
different domains of research in this field. This is a relatively recent activity related to
artificial boson systems in magneto-optical \textit{traps} \cite{LSSY05} and another domain, which
is a traditional study of \textit{homogeneous} boson systems \cite{Ver11}.

\hspace{0.4cm}
The latter comes back to Einstein's prediction in 1925 of condensate in the perfect Bose-Gas (PBG)
\cite{Ein25}, then supported after criticism in 1927 \cite{Uhl27} by F.London \cite{Lon38} after
discovery of the superfluidity \cite{Kap38}, \cite{AM38},
and, finally, was seriously bolstered by experimental observation \cite{KAZPP74,AZKPP75,DKPP78} of
condensate in the superfluid phase of the liquid Helium $^4\rm{He}$.
%\cite{BKKP92}.
The most striking was a quite accurate coincidence of the critical temperature of condensation
$T_c$ and the temperature $T_{\lambda}$ of the superfluidity $\lambda$-point, see
\cite{DKPP78,BKKP90} and \cite{BKKP92,BBKKKPPSY94}.
Even thought these data strongly support the Bogoliubov-Landau theory of superfluidity
of the liquid Helium $^4\rm{He}$, which is based on the hypothesis of BEC, the mathematical
theory of this phenomenon is still far from being complete.

\hspace{0.4cm}
Although the BEC, or \textit{generalised} BEC (gBEC) \cite{vdBLP86} in PBG, are studied in
great details, analysis of condensate in the interacting Bose-gas is a more delicate problem.
Recall that effective quantum attraction between bosons, which is behind of the BEC in the PBG,
makes this system unstable with respect to any direct \textit{attractive} interaction between
particles. So, efforts around the question: "Why do interacting bosons condense?", were
essentially concentrated around \textit{repulsive} interaction between particles. The studying
of stability of the conventional BEC (or gBEC) in the imperfect Bose-Gas (IBG) with a direct
fast-decreasing two-body repulsive interaction is still in progress \cite{SU09,BU10,Su15,WW17}.
Whereas, if one counterbalances direct attractive interaction by a repulsion stabilising
the boson system, this attraction may be the origin of a new mechanism of condensation called the
\textit{non-conventional} condensation. Implicitly this type of condensation was
introduced for the first time in \cite{vdBLP88} on the basis of their rigorous analysis of
Bose condensation in the Huang-Yang-Luttinger (HYL) model \cite{HuangYangLuttinger}.

\hspace{0.4cm}
We note that it was Thouless \cite{Thouless}, who presented an instructive
"back-of-the-envelope" calculations, which argue that a new kind of Bose
condensation may occur in the HYL model of the hard-sphere Bose-gas \cite{HuangYangLuttinger}.
Ten years after \cite{BZ-PL98}, the non-conventional condensation was discovered also in the
Bogoliubov Weakly Imperfect Bose Gas (WIBG), see \cite{BZ98JP,BZ00} and review \cite{ZagBru01}.

\hspace{0.4cm}
The difference between \textit{conventional} and \textit{non-conventional}
condensations reflects the difference in the mechanism of their formation.
The {conventional} condensation is a consequence of the balance
between \textit{entropy} and \textit{kinetic energy}, whereas the {non-conventional}
condensation results from the balance between \textit{entropy} and \textit{interaction energy}.
This difference has an important consequence: the conventional condensation would occur if it
occurs in the PBG, whilst non-conventional condensation occurs \textit{due} to interaction.
The latter motivated its another name: the \textit{dynamical} condensation
\cite{BZ-PL98,BZ98JP}.
As a consequence, the dynamical condensation may occur in \textit{low-dimensional}
boson systems, as well as to exhibit the \textit{first-order} phase transition. The both HYL
and WIBG models manifest these properties.

\smallskip

\noindent
\textbf{1.2} The aim of the paper is to give an introduction into the non-conventional dynamical
condensation for homogeneous boson systems.

\hspace{0.4cm}
To this end we first introduce in the next Section \ref{sec:1} a simple \textit{toy} model that
manifests the outlined above peculiarities of this kind of condensate. Properties of this model
and description of condensates of different types are presented in Sections \ref{sec:2} and
\ref{sec:3}. Section \ref{sec:4} is reserved for comments and concluding remarks.

\hspace{0.4cm}
Section \ref{section 0} is devoted to quantum mechanical origin of the \textit{effective off-diagonal}
interaction in the Hamiltonian of the Bogoliubov WIBG. This is important step to understanding
the origin of the non-conventional condensation in this model. Further details are presented in
Sections \ref{section 1}-\ref{section 4}. Few concluding remarks are in Section \ref{section 5}.
%\vspace{2cm}
%%%%%%%%%%%%%%%%%%%%%%%%%%%%%%%%%%%%%%%%%%%%%%%%%%%%%%%%%%%%%%%%%%%%%%%%%%%%%%%%%%%%%%%%%%%%%%%%
\section{Toy model}\label{sec:1}
%%%%%%%%%%%%%%%%%%%%%%%%%%%%%%%%%%%%%%%%%%%%%%%%%%%%%%%%%%%%%%%%%%%%%%%%%%%%%%%%%%%%%%%%%%%%%%%%
\textbf{2.1} Recall that since the first description by Einstein \cite{Ein25} in 1925, it is
known that \textit{conventional} Bose-Einstein condensation with macroscopic occupation
of a single level is a very subtle matter. For example, its magnitude
strongly depends on the shape of container or on the way to take the
thermodynamic limit, see e.g. \cite{vdBLP86,ZUK} and Section \ref{classification des
condensations}. It was Casimir \cite{casimir} who showed that in a long
prism it is possible for condensation in the Perfect Bose Gas (PBG) to occur
in a \textit{narrow band} rather than in a single level. This was an example of
\textit{generalised} BEC (gBEC), a concept introduced earlier by Girardeau \cite{Girardeau}.
The first rigorous treatment of this observation for the PBG is due to van den Berg, Lewis
and Pul\`{e} in series of papers: \cite{vdBL82}--\cite{Pule83} and \cite{vdBLP86}. They proposed
a classification of \textit{types} of gBEC. Then condensate in a single level is the gBEC of type I,
see Section \ref{classification des condensations}.

\hspace{0.4cm}
The feature of the conventional BEC (generalised or not) is that it appears in non-interacting
system of bosons as soon as total particle density
gets larger that some \textit{critical} value. Therefore, behind of conventional BEC
there is a \textit{saturation mechanism} related to the Bose
statistics of particles. In \cite{vdBLdeS} it was demonstrated that the very same
mechanism is responsible for BEC in a system of
bosons with mean-field repulsive interaction commonly called the Mean-Field imperfect Bose-gas.
Moreover, in \cite{MiVer} it was shown that, instead of geometry of
container, a judicious choice of repulsive interaction may split initial
single level condensation (type I) into \textit{non-extensive} (type III)
condensation, when \textit{no} levels are macroscopically occupied.
Therefore, the concept of conventional gBEC caused by the mechanism of saturation fits
well for bosons with repulsive interaction.

\hspace{0.4cm}
Since bosons are very sensitive to attraction, there exists non-conventional dynamical condensation
induced by this interaction \cite{BZ98JP,BZ99Ph,BZ00JP,BZ00}. Again, this kind of condensation shows
when total particle density (or chemical potential) becomes larger some
\textit{critical} value, but it is attractive interaction (and not simply Bose
statistics) that defines the value of dynamical condensate and its behaviour.
To escape the collapse, the  attractive interaction in a boson system should
be stabilised by a repulsion. Therefore, the conventional and non-conventional condensations
may \textit{coexist}.

\hspace{0.4cm}
Our toy model manifests these two kinds of condensations.
The non-conventional one is due to an \textit{attraction term} in Hamiltonian of the model.
This condensation starts at the single lowest level for moderate densities
(negative chemical potentials) and saturates after some \textit{critical} density.
It is after this threshold that the {conventional} BEC shows up to absorb the
increasing total particle density (the saturation mechanism).
At the threshold one has coexistence of these two kinds of condensations.
Moreover the repulsive interaction in our model is such that BEC splits up into
\textit{non-extensive} one, i.e., into the gBEC of type III.

\hspace{0.4cm}
Since known Bose-systems  manifesting condensation (e.g., superfluid $^4\rm{He}$) are
far from to be perfect, we hope that our toy model would give more insight into
possible scenarios for condensations in real systems. For example, in
condensate of sodium atoms in trap the interaction seems to predominate compare with
kinetic energy \cite{gazboson1}. Therefore, condensation in trapped alkali
dilute-gases \cite{gazboson2,gazboson3}, may be a combination of \textit{non-conventional} and
\textit{conventional} BECs.

\smallskip

\noindent
\textbf{2.2} To fix notations and definitions we recall first the Mean-Field (MF) imperfect Bose-gas
model introduced by Huang \cite{Hu63} Ch.5.2.6. It is a system of identical bosons of
mass $m$ enclosed in a cube $\Lambda \subset \mathbb{R}^{d}$ of volume $%
V=\left| \Lambda \right| $ centered at the origin defined by the
Hamiltonian:
\begin{equation}
H_{\Lambda }^{MF}= \ \stackunder{k\in \Lambda ^{*}}{\sum }\varepsilon
_{k}a_{k}^{*}a_{k}+\frac{\lambda }{2 V}N_{\Lambda }^{2}\, , \ \text{ }\varepsilon
_{k}:= \hbar^{2}k^{2}/2m,\text{ }\lambda >0,  \label{newdiagmodel-1}
\end{equation}
where $N_{\Lambda }= \ \stackunder{k\in \Lambda ^{*}}{\sum }a_{k}^{*}a_{k}%
\equiv \stackunder{k\in \Lambda ^{*}}{\sum }N_{k}$ is the particle-number
operator and $\varepsilon _{k}$correspond to the one-particle
kinetic-energy. Here $\{ a_{k}^{\#}\} _{k\in \Lambda ^{*}}$ are
the boson creation/annihilation operators in the boson Fock space $\mathcal{F%
}_{\Lambda }$ over $L^{2}\left( \Lambda \right) $ corresponding to the
second quantisation in the box $\Lambda = \ \stackunder{\alpha =1}{\stackrel{d}{\times }}L$
with periodic boundary conditions, i.e. to the dual
\[
\Lambda ^{*}=\{ k\in \mathbb{R}^{d}:\text{ }k_{\alpha }=\frac{2\pi
n_{\alpha }}{L}\text{, }n_{\alpha }=0,\pm 1,\pm 2,...\text{; }\alpha
=1,2,...,d \, \} .
\]
Then for $d>2$, given temperature $\theta :=\beta ^{-1}$ and total particle
density $\rho >\rho _{c}^{P}\left( \theta \right) $ (here $\rho
_{c}^{P}\left( \theta \right) := \rho ^{P}\left( \theta ^{-1},\mu
=0\right) $ where $\rho ^{P}\left( \beta ,\mu \right) $ is the particle
density of the PBG in the grand-canonical ensemble) the
MF model manifests a \textit{conventional} BEC of \textit{type I }\cite{Dav72,FaVerb80,vdBLdeS},
i.e. a macroscopic occupation only of the single-particle ground-state level $k=0$.
See Section \ref{classification des condensations} for classification
of {conventional} BEC.

\hspace{0.4cm}
However, in \cite{MiVer} it was shown that the MF model (\ref{newdiagmodel-1}) perturbed by
the \textit{repulsive} diagonal interaction
\begin{equation}
\widetilde{U}_{\Lambda }=\frac{\lambda }{2V}\stackunder{k\in \Lambda ^{*}}{%
\sum }N_{k}^{2},\text{ }\lambda >0,  \label{newdiagmodel-2}
\end{equation}
demonstrates the BEC which occurs again for densities
$\rho >\rho _{c}^{P}\left( \theta \right) $ (or $\mu > \lambda \rho
_{c}^{P}\left( \theta \right) =: \mu _{c}^{MF}\left( \theta \right) $),
but now it splits up into BEC of \textit{\ type III. } This is a \textit{non-extensive}
condensation, when \textit{no} single-particle levels are macroscopically occupied
(cf. Section \ref{classification des condensations}). This model for $\lambda >0$ was introduced
in \cite{Schr} and  we call it the \ Michoel-Schr\"{o}der-Verbeure (MSV) model:
\begin{equation}
H_{\Lambda }^{MSV}\equiv H_{\Lambda }^{MF}+\widetilde{U}_{\Lambda }.
\label{newdiagmodel-3}
\end{equation}
Then the conventional BEC of type III means that
\[
\stackunder{\Lambda }{\lim }\frac{\left\langle N_{k}\right\rangle
_{H_{\Lambda }^{MSV}}}{V}=0,\text{ }k\in \Lambda ^{*},
\]
for any $\rho $, whereas the double limit
\[
\stackunder{\delta \rightarrow +0}{\lim }\stackunder{\Lambda }{\lim }%
\frac{1}{V}\stackunder{\left\{ k\in \Lambda ^{*},0\leq \left\| k\right\|
\leq \delta \right\} }{\sum }\left\langle N_{k}\right\rangle _{H_{\Lambda
}^{MSV}}=\rho -\rho _{c}^{P}\left( \theta \right) >0,
\]
for $\mu >\mu _{c}^{MF}\left( \theta \right) $. Here we denote by $%
\left\langle -\right\rangle _{H_{\Lambda }^{MSV}}\left( \beta ,\mu \right) $,
$\beta \geq 0$, $\mu \in \Bbb{R}^{1}$, the grand-canonical Gibbs state for
the Hamiltonian $H_{\Lambda }^{MSV}$.

\hspace{0.4cm}
Note that Hamiltonian $H_{\Lambda}^{MF}-\widetilde{U}_{\Lambda }/2$ coincides for
$\lambda = 2 a$ with Hamiltonian $H_{\Lambda}^{HYL}$ for the HYL model rigorously
studied in \cite{vdBLP88}. There it was shown that HYL model manifests non-conventional
condesation of the type I that occurs only at zero-mode $k=0$.

\hspace{0.4cm}
The fact that a gentle repulsive interaction may produce a
generalised {non-extensive} BEC without any change of corresponding pressure has been
shown in \cite{BZ99Ph}. This was done in context of the model:
\begin{equation}
H_{\Lambda }^{0}= \ \stackunder{k\in \Lambda ^{*}\backslash \left\{ 0\right\} }{%
\sum }\varepsilon _{k}a_{k}^{*}a_{k}+\varepsilon _{0}a_{0}^{*}a_{0}+\frac{%
g_{0}}{2V}a_{0}^{*}a_{0}^{*}a_{0}a_{0},  \label{newdiagmodel1}
\end{equation}
with $\varepsilon _{0}\left( \neq \varepsilon _{k=0}\right) \in \Bbb{R}^{1}$
and $g_{0}>0,$ perturbed by the interaction
\begin{equation}
U_{\Lambda }=\frac{1}{V}\stackunder{k\in \Lambda ^{*}\backslash \left\{
0\right\} }{\sum }g_{k}\left( V\right) a_{k}^{*}a_{k}^{*}a_{k}a_{k},\text{ }%
0<g_{-}\leq g_{k}\left( V\right) \leq \gamma _{k}V^{\alpha _{k}},
\label{diagmodel2}
\end{equation}
with $\alpha _{k}\leq \alpha _{+}<1$ and $0<\gamma _{k}\leq \gamma _{+}$.
The perturbation $U_{\Lambda }$ (similar to the interaction $\widetilde{U}%
_{\Lambda }$ when $g_{k}=\lambda $) leads to Hamiltonian:
\begin{equation}
H_{\Lambda }^{BZ}:= H_{\Lambda }^{0}+U_{\Lambda }.  \label{diagmodel1}
\end{equation}
In contrast to the MSV model, the grand-canonical pressure for (\ref{diagmodel1}):
\begin{equation}
p_{\Lambda }^{BZ}\left( \beta ,\mu \right) = p_{\Lambda }\left[
H_{\Lambda }^{BZ}\right] := \frac{1}{\beta V}\ln {\rm{Tr}}_{\mathcal{F}%
_{\Lambda }}e^{-\beta \left( H_{\Lambda }^{BZ}-\mu N_{\Lambda }\right) }
\label{diagmodel3}
\end{equation}
is defined in the thermodynamic limit only in domain $Q=\left\{ \mu \leq
0\right\} \times \left\{ \theta \geq 0\right\} $ and it is equal to
\begin{equation}
p^{BZ}\left( \beta ,\mu \right) := \text{ }\stackunder{\Lambda }{\lim }%
p_{\Lambda }^{BZ}\left( \beta ,\mu \right) =p^{P}\left( \beta ,\mu
\right) -\stackunder{\rho _{0}\geq 0}{\inf }\left[ \left( \varepsilon
_{0}-\mu \right) \rho _{0}+\frac{g_{0}\rho _{0}^{2}}{2}\right] ,
\label{diagmodel4}
\end{equation}
see \cite{BZ99Ph}. Here $p^{P}\left( \beta ,\mu \right) $ is the pressure of the
PBG in thermodynamic limit. Note that the pressure (\ref{diagmodel4}) is
independent of parameters $\left\{ g_{k}\left( V\right) \right\} _{k\in
\Lambda ^{*}\backslash \left\{ 0\right\} }$, i.e. of the interaction (\ref
{diagmodel2}).
%%%%%%%%%%%%%%%%%%%%%%%%%%%%%%%%%%%%%%%%%%%%%%%%%%%%%%%%%%%%%%%%%%%%%%%%%%%%%%%%%%%%%%%%%%%%
\begin{remark} \label{newdiagmodelth0}
Let domain $D_{\varepsilon _{0}}$ be defined by
\begin{equation}
D_{\varepsilon _{0}} := \left\{ \left( \theta ,\mu \right) \in
Q:p^{P}\left( \beta ,\mu \right) <p^{BZ}\left( \beta ,\mu \right) \right\} .
\label{diagmodel18}
\end{equation}
Then the thermodynamic limit (\ref{diagmodel4}) says that to insure $%
D_{\varepsilon _{0}}\neq \left\{ \emptyset \right\} $ the parameter $%
\varepsilon _{0}$ must be negative, i.e.
\begin{equation}
D_{\varepsilon _{0}}=\left\{ \left( \theta ,\mu \right) \in Q:\varepsilon
_{0}<\mu \leq 0\right\} .  \label{diagmodel19}
\end{equation}
Below we consider only the case $\varepsilon _{0}<0$ and $d>2$.
\end{remark}
%%%%%%%%%%%%%%%%%%%%%%%%%%%%%%%%%%%%%%%%%%%%%%%%%%%%%%%%%%%%%%%%%%%%%%%%%%%%%%%%%%%%%%%%%%%%%%

\hspace{0.4cm}
We denote by $\rho _{\Lambda }^{BZ}\left( \beta ,\mu \right) $ the total
particle density in the grand-canonical ensemble for the model $H_{\Lambda
}^{BZ}$:
\begin{equation}
\rho _{\Lambda }^{BZ}\left( \beta ,\mu \right) := \left\langle \frac{%
N_{\Lambda }}{V}\right\rangle _{H_{\Lambda }^{BZ}}\left( \beta ,\mu \right) .
\label{diagmodel42}
\end{equation}
Then $\rho ^{BZ}\left( \beta ,\mu \right) := $ $\stackunder{\Lambda }{%
\lim }\rho _{\Lambda }^{BZ}\left( \beta ,\mu \right) $ is the corresponding
thermodynamic limit which, according to \cite{vdBLP88}, is equal to:
\begin{equation}
\rho ^{BZ}\left( \beta ,\mu \right) =\rho ^{P}\left( \beta ,\mu \right) ,
\label{diagmodel42.1}
\end{equation}
for $\left( \theta ,\mu \leq \varepsilon _{0}\right) $, and to
\begin{equation}
\rho ^{BZ}\left( \beta ,\mu \right) =\rho ^{P}\left( \beta ,\mu \right) +%
\frac{\mu -\varepsilon _{0}}{g_{0}},  \label{diagmodel42.2}
\end{equation}
for $\left( \theta ,\varepsilon _{0}<\mu <0\right) $. We remark that for $d>2$
there is a finite critical density
\begin{equation}
\rho _{c}^{BZ}\left( \theta \right) := \text{ }\stackunder{\mu \leq 0}{%
\sup }\rho ^{BZ}\left( \theta ^{-1},\mu \right) =\rho ^{BZ}\left( \theta
^{-1},\mu =0\right) =\rho _{c}^{P}\left( \theta \right) -\frac{\varepsilon
_{0}}{g_{0}}<+\infty ,  \label{diagmodel45.1}
\end{equation}
in this model.
%%%%%%%%%%%%%%%%%%%%%%%%%%%%%%%%%%%%%%%%%%%%%%%%%%%%%%%%%%%%%%%%%%%%%%%%%%%%%%%%%%%%%%%%%%%%%%%%%%%%
\begin{proposition} \label{newdiagmodel-th5}\emph{\cite{vdBLP88}}
Let $\rho >\rho _{c}^{BZ}\left( \theta \right) $ ($d>2$) and $0<g_{-}\leq $
$g_{k}\left( V\right) \leq \gamma_{k}V^{\alpha _{k}}$ for
$k\in \Lambda ^{*}\backslash \left\{ 0\right\} ,$
with $\alpha _{k}\leq \alpha _{+}<1$ and $0<\gamma _{k}\leq \gamma _{+}$.
Then for any $\varepsilon _{0}<0$ we have:\newline
$\left( i\right) $ a condensation in the mode $k=0$ (even if $d<3$), i.e.
\begin{equation}
\rho _{0}^{BZ}\left( \theta ,\mu \right) := \text{ }\stackunder{\Lambda
}{\lim }\left\langle \frac{a_{0}^{*}a_{0}}{V}\right\rangle _{H_{\Lambda
}^{BZ}}=\left\{
\begin{array}{c}
0,\text{ for }\left( \theta ,\mu \right) \in Q\backslash D_{\varepsilon _{0}}
\\
\left( \frac{\mu -\varepsilon _{0}}{g_{0}}\right) ,\text{ for }\left( \theta
,\mu \right) \in D_{\varepsilon _{0}}
\end{array}
\right\} ;  \label{diagmodel28}
\end{equation}
$\left( ii\right) $ for any $\varepsilon _{0}\in \Bbb{R}^{1}$%
\begin{equation}
\stackunder{\Lambda }{\lim }\left\langle \frac{a_{k}^{*}a_{k}}{V}%
\right\rangle _{H_{\Lambda }^{BZ}}=0\text{, }k\in \Lambda ^{*}\backslash
\left\{ 0\right\} ,  \label{diagmodel66}
\end{equation}
i.e. there is no macroscopic occupation of modes $k\neq 0$ but we have a non-extensive BEC:
\begin{equation}
\stackunder{\delta \rightarrow 0^{+}}{\lim }\stackunder{\Lambda }{\lim }%
\frac{1}{V}\stackunder{\left\{ k\in \Lambda ^{*}:\text{ }0<\left\| k\right\|
<\delta \right\} }{\sum }\left\langle N_{k}\right\rangle _{H_{\Lambda
}^{BZ}}=\rho -\rho _{c}^{BZ}\left( \theta \right) >0.  \label{diagmodel67}
\end{equation}
For $\varepsilon _{0}<0$ this type III BEC coexists with non-conventional dynamical BEC of type I
in the mode $k=0$ if $\left( \theta ,\mu \right) \in D_{\varepsilon _{0}}$.
\end{proposition}
%%%%%%%%%%%%%%%%%%%%%%%%%%%%%%%%%%%%%%%%%%%%%%%%%%%%%%%%%%%%%%%%%%%%%%%%%%%%%%%%%%%%%%%%%%%%%%%%%%
Therefore, Proposition \ref{newdiagmodel-th5} demonstrates for densities
$\rho >\rho _{c}^{BZ}\left( \theta \right) $ \textit{coexistence} of \textit{two} kinds
of condensates in the model (\ref{diagmodel1}) :\\
- a non-conventional BEC in the single mode $k=0$ due to the term $\left(
\varepsilon _{0}a_{0}^{*}a_{0}\right) $, which mimics for $\varepsilon _{0}<0$
\textit{attraction} by an external potential \cite{PaZag}\textit{\ }giving
rise to a \textit{non-conventional} condensation of {type I} (cf. Section \ref
{classification des condensations});\\
- a \textit{conventional} BEC due to \textit{%
saturation} of the total particle density, where (similar to the MSV model)
the type III of this condensation is due to the elastic repulsive
interaction $U_{\Lambda }$ (\ref{diagmodel2}) of bosons in modes $k\neq 0$.
Therefore, the interaction (\ref{diagmodel2}) is decisive for formation of
the non-extensive BEC in the model (\ref{diagmodel1}), whereas it has no impact on the
value of pressure (\ref{diagmodel4}).

\hspace{0.4cm}
Below we study a \textit{toy model}, which is a modification of the model (\ref{diagmodel1}).
It is, similar to (\ref{newdiagmodel-3}), stabilised by the MF-interaction (\ref{newdiagmodel-1}):
\begin{equation}
H_{\Lambda }= \ \stackunder{k\in \Lambda ^{*}\backslash \left\{ 0\right\} }{%
\sum }\varepsilon _{k}a_{k}^{*}a_{k}+\varepsilon _{0}a_{0}^{*}a_{0}+\frac{%
g_{0}}{2V}N_{0}^{2}+\frac{\lambda }{2 V}N_{\Lambda }^{2}+\frac{g}{2V}%
\stackunder{k\in \Lambda ^{*}\backslash \left\{ 0\right\} }{\sum }N_{k}^{2}.
\label{newdiagmodel2}
\end{equation}
Here $\lambda >0,$ $g_{0}>0,$ $g>0$ but $\varepsilon _{0}<0$. Note that
the toy model $H_{\Lambda }$ for $\varepsilon _{0}=0$ and for $\lambda =g=g_{0}$
coincides with the MSV model (\ref{newdiagmodel-3}), whilst for $\varepsilon _{0}= 0$ and
$g_{0}= g = - a, \lambda = 2 a > 0$ one gets:
\begin{equation}
H_{\Lambda }^{HYL}= \ \stackunder{k\in \Lambda ^{*}}{%
\sum }\varepsilon _{k}a_{k}^{*}a_{k}+ \frac{a}{2 V}N_{\Lambda }^{2}+
\frac{a}{2V}\{N_{\Lambda }^{2} -
\stackunder{k\in \Lambda ^{*}}{\sum }N_{k}^{2}\}.
\label{HYL}
\end{equation}
%%%%%%%%%%%%%%%%%%%%%%%%%%%%%%%%%%%%%%%%%%%%%%%%%%%%%%%%%%%%%%%%%%%%%%%%%%%%%%%%%%%%%%%%%%%%
\begin{remark} \label{HYL-bis}
Note that in the toy model \emph{(\ref{newdiagmodel2})} the effect favouring non-conventional
condensation of bosons at zero-mode is due to the kinetic-energy term, which is \textit{enhanced}
by "interaction"-energy term for $\varepsilon _{0}<0$, see Section \ref{sec:3}.
On the other hand, the Huang-Yang-Luttinger model \emph{(\ref{HYL})} is the MF Bose-gas
\emph{(\ref{newdiagmodel-1})} (that manifests a conventional zero-mode BEC) perturbed by
interaction energy, see the last term in \emph{(\ref{HYL})}. Then the effect favouring particle
accumulation at zero-mode by the kinetic energy term is \textit{enhanced} by this last
interaction-energy term since it has the smallest value when all bosons occupy the same energy-level.
Therefore, it produces a non-conventional dynamical zero-mode BEC, see \emph{\cite{vdBLP88}}.
\end{remark}
%%%%%%%%%%%%%%%%%%%%%%%%%%%%%%%%%%%%%%%%%%%%%%%%%%%%%%%%%%%%%%%%%%%%%%%%%%%%%%%%%%%%%%%%%%%%
%\renewcommand{\thefootnote}{\arabic{footnote}}
%\setcounter{footnote}{0} \renewcommand{\thefootnote}{\arabic{footnote}}
%\numberwithin{equation}{section}
%\renewcommand{\theequation}{\arabic{section}.\arabic{equation}}
\setcounter{equation}{0}
\section{Free-energy density and pressure}\label{sec:2}
%%%%%%%%%%%%%%%%%%%%%%%%%%%%%%%%%%%%%%%%%%%%%%%%%%%%%%%%%%%%%%%%%%%%%%%%%%%%%%%%%%%%%%%%%%%%
\textbf{3.1} First we consider the toy model (\ref{newdiagmodel2}) in
\textit{canonical} ensemble $\left( \beta ,\rho \right) $. This simplifies
essentially the thermodynamic study of the model. Let
$f_{\Lambda }\left(\beta ,\rho = {n}/{V}\right) $ be the corresponding free-energy density:
\begin{equation}
f_{\Lambda }\left( \beta ,\rho \right) := -\frac{1}{\beta V}\ln {\rm{Tr}}_{\,
\mathcal{H}_{\Lambda ,S}^{n}}\left( e^{-\beta H_{\Lambda }}\right) ,
\label{newdiagmodel7}
\end{equation}
where $\mathcal{H}_{\Lambda ,S}^{n}:= S\left( \stackunder{i=1}{\stackrel{%
n}{\otimes }}L^{2}\left( \Lambda \right) \right) $ is symmetrised $n$%-particle Hilbert space.
%%%%%%%%%%%%%%%%%%%%%%%%%%%%%%%%%%%%%%%%%%%%%%%%%%%%%%%%%%%%%%%%%%%%%%%%%%%%%%%%%%%%%%%%%%%%%%%%%
\begin{theorem}
\label{newdiagmodelth1}Let $\lambda >0$ , $g>0$, $g_{0}>0$ and $\varepsilon_{0}<0$. Then
\begin{equation}
f\left( \beta ,\rho \right) := \text{ }\stackunder{\Lambda }{\lim }%
f_{\Lambda }\left( \beta ,\rho \right) =\frac{\lambda}{2} \rho ^{2}+\stackunder{\rho
_{0}\in \left[ 0,\rho \right] }{\inf }\left\{ \varepsilon _{0}\rho _{0}+%
\frac{g_{0}}{2}\rho _{0}^{2}+f^{P}\left( \beta ,\rho -\rho _{0}\right)
\right\} ,  \label{newdiagmodel8}
\end{equation}
is independent of $g$ provided $g>0$. Here $f^{P}\left( \beta,\rho \right) $ is thermodynamic limit
of the PBG free-energy
\begin{equation}
f^{P}\left( \beta ,\rho \right) := \text{ }\stackunder{\Lambda }{\lim }%
f_{\Lambda }^{P}\left( \beta ,\rho \right) \, , \label{newdiagmodel9}
\end{equation}
where
\begin{equation}
f_{\Lambda }^{P}\left( \beta ,\rho \right) := -\frac{1}{\beta V}\ln
\stackunder{\left\{ n_{k}=0,1,2,...\right\} _{k\in \Lambda ^{*}}}{\sum }%
e^{-\beta ( \stackunder{k\in \Lambda ^{*}}{\sum }\varepsilon
_{k}n_{k}) }\delta _{\stackunder{k\in \Lambda ^{*}}{\tsum }%
n_{k}=\left[ \rho V\right] } \, ,  \label{newdiagmodel9bis}
\end{equation}
and $\left[ x\right] $ denotes the integer part of $x\geq 0$.
\end{theorem}
%%%%%%%%%%%%%%%%%%%%%%%%%%%%%%%%%%%%%%%%%%%%%%%%%%%%%%%%%%%%%%%%%%%%%%%%%%%%%%%%%%%%%%%%%%%%%%%%%%%%
\textit{Proof.} By (\ref{newdiagmodel2}) and (\ref{newdiagmodel7}) we get
\begin{equation}
f_{\Lambda }\left( \beta ,\rho \right) =-\frac{1}{\beta V}\ln \{
\stackunder{n_{0}=0}{\stackrel{\left[ \rho V\right] }{\sum }}e^{-\beta
Vh\left( \rho ,\frac{n_{0}}{V}\right) }\} +\frac{\lambda}{2} \rho ^{2},
\label{newdiagmodel3bis}
\end{equation}
where
\begin{equation}
h_{\Lambda }\left( \rho ,\rho _{0}\right) := \varepsilon _{0}\rho _{0}+%
\frac{g_{0}}{2}\rho _{0}^{2}-\frac{1}{\beta V}\ln \stackunder{\left\{
n_{k}=0,1,2,...\right\} _{k\in \Lambda ^{*}\backslash \left\{ 0\right\} }}{%
\sum }e^{-\beta \left( \stackunder{k\in \Lambda ^{*}\backslash \left\{
0\right\} }{\sum }\left[ \varepsilon _{k}n_{k}+\frac{g}{2V}n_{k}^{2}\right]
\right) }\delta _{\stackunder{k\neq 0}{\tsum }n_{k}=\left[ \rho V\right]
-\left[ \rho _{0}V\right] }.  \label{newdiagmodel3}
\end{equation}
By (\ref{newdiagmodel3bis}) one gets the estimate
\[
\frac{\lambda}{2} \rho ^{2}+\stackunder{\rho _{0}\in \left[ 0,\rho \right] }{\inf }%
h_{\Lambda }\left( \rho ,\rho _{0}\right) -\frac{1}{\beta V}\ln \left(
\left[ \rho V\right] +1\right) \leq f_{\Lambda }\left( \beta ,\rho \right)
\leq \frac{\lambda}{2} \rho ^{2}+\stackunder{\rho _{0}\in \left[ 0,\rho \right] }{\inf
}h_{\Lambda }\left( \rho ,\rho _{0}\right) ,
\]
which gives in thermodynamic limit:
\begin{equation}
f\left( \beta ,\rho \right) := \text{ }\stackunder{\Lambda }{\lim }%
f_{\Lambda }\left( \beta ,\rho \right) =\frac{\lambda}{2} \rho ^{2}+\text{ }%
\stackunder{\Lambda }{\lim }\stackunder{\rho _{0}\in \left[ 0,\rho \right] }{%
\inf }h_{\Lambda }\left( \rho ,\rho _{0}\right) .  \label{newdiagmodel5}
\end{equation}
Notice that (\ref{newdiagmodel3}) can be rewritten as
\begin{equation}
h_{\Lambda }\left( \rho ,\rho _{0}\right) =\varepsilon _{0}\rho _{0}+\frac{%
g_{0}}{2}\rho _{0}^{2}-\frac{1}{\beta V}\ln \langle e^{-\frac{\beta g}{%
2V}\stackunder{k\in \Lambda ^{*}\backslash \left\{ 0\right\} }{\sum }%
n_{k}^{2}}\rangle _{\widetilde{H}_{\Lambda }^{P}}\left( \beta ,\rho
-\rho _{0}\right) ,  \label{newdiagmodel5bis}
\end{equation}
where $\left\langle -\right\rangle _{\widetilde{H}_{\Lambda }^{P}}\left(
\beta ,\rho -\rho _{0}\right) $ is the canonical Gibbs state for the PBG
with \textit{excluded }mode $k=0$ for density $\rho -\rho _{0}$, with the
corresponding free-energy density $\widetilde{f}_{\Lambda }^{P}\left( \beta
,\rho \right) $ defined by (\ref{newdiagmodel9bis}) for $k\in \Lambda
^{*}\backslash \left\{ 0\right\} $. Since
\[
\stackunder{\Lambda }{\lim }\widetilde{f}_{\Lambda }^{P}\left( \beta ,\rho
\right) =\text{ }\stackunder{\Lambda }{\lim }f_{\Lambda }^{P}\left( \beta
,\rho \right) ,
\]
the Jensen inequality
\[
\langle e^{-\frac{\beta g}{2V}\stackunder{k\in \Lambda ^{*}\backslash
\left\{ 0\right\} }{\sum }n_{k}^{2}}\rangle_{\widetilde{H}_{\Lambda
}^{P}}\geq e^{-\frac{\beta g}{2V}\langle \stackunder{k\in \Lambda
^{*}\backslash \left\{ 0\right\} }{\sum }n_{k}^{2}\rangle_{\widetilde{%
H}_{\Lambda }^{P}}}
\]
and (\ref{newdiagmodel5bis}) imply the estimate
\begin{equation}
\stackunder{\Lambda }{\lim }h_{\Lambda }\left( \rho ,\rho _{0}\right) \leq
\varepsilon _{0}\rho _{0}+\frac{g_{0}}{2}\rho _{0}^{2}+f^{P}\left( \beta
,\rho -\rho _{0}\right) .  \label{newdiagmodel6}
\end{equation}
Moreover, since
\[
e^{-\frac{\beta g}{2V}n_{k}^{2}}\leq 1,
\]
by (\ref{newdiagmodel3}) we have
\[
h_{\Lambda }\left( \rho ,\rho _{0}\right) \geq \varepsilon _{0}\rho _{0}+%
\frac{g_{0}}{2}\rho _{0}^{2}+f_{\Lambda }^{P}\left( \beta ,\rho -\rho
_{0}\right) ,
\]
which together with (\ref{newdiagmodel6}) gives (\ref{newdiagmodel8}). \hfill $\square$
%%%%%%%%%%%%%%%%%%%%%%%%%%%%%%%%%%%%%%%%%%%%%%%%%%%%%%%%%%%%%%%%%%%%%%%%%%%%%%%%%%%%%%%%%%%%%%%%%%%%
\begin{remark}
\label{newdiagmodelth3}Let denote by $f_{\Lambda }^{BZ}\left( \beta ,\rho
\right) $ the free-energy density corresponding to $H_{\Lambda }^{BZ}$ (\ref
{diagmodel1}) with $g_{k}\left( V\right) =g/2$, i.e.
\[
f_{\Lambda }^{BZ}\left( \beta ,\rho \right) := -\frac{1}{\beta V}\ln {\rm{Tr}}_{%
\mathcal{H}_{\Lambda ,S}^{n}}\left( e^{-\beta H_{\Lambda }^{BZ}}\right) .
\]
Then (\ref{diagmodel1}) (\ref{newdiagmodel2}) and (\ref{newdiagmodel7})
imply that
\[
f_{\Lambda }\left( \beta ,\rho \right) =\frac{\lambda}{2} \rho ^{2}+f_{\Lambda
}^{BZ}\left( \beta ,\rho \right) ,
\]
from which by Theorem \ref{newdiagmodelth1} we deduce
\begin{equation}
f^{BZ}\left( \beta ,\rho \right) := \text{ }\stackunder{\Lambda }{\lim }%
f_{\Lambda }^{BZ}\left( \beta ,\rho \right) = \ \stackunder{\rho _{0}\in \left[
0,\rho \right] }{\inf }\left\{ \varepsilon _{0}\rho _{0}+\frac{g_{0}}{2}\rho
_{0}^{2}+f^{P}\left( \beta ,\rho -\rho _{0}\right) \right\} ,
\label{newdiagmodel13}
\end{equation}
and
\begin{equation}
f\left( \beta ,\rho \right) =\frac{\lambda}{2} \rho ^{2}+f^{BZ}\left( \beta ,\rho
\right) .  \label{newdiagmodel13bis}
\end{equation}
By explicit calculation one checks convexity of $f^{BZ}\left( \beta ,\rho
\right) $ as a function of $\rho $. Therefore, the same is true for $f\left(
\beta ,\rho \right) ,$ see (\ref{newdiagmodel8}) and (\ref{newdiagmodel13bis}).
\end{remark}
%%%%%%%%%%%%%%%%%%%%%%%%%%%%%%%%%%%%%%%%%%%%%%%%%%%%%%%%%%%%%%%%%%%%%%%%%%%%%%%%%%%%%%%%%%%%%%%%%%
\begin{remark}
\label{newdiagmodelth4}Since the pressure $p^{BZ}\left( \beta ,\mu \right) $
is a Legendre transform of the corresponding free-energy density $%
f^{BZ}\left( \beta ,\rho \right) $, we get from (\ref{newdiagmodel13}) that
\begin{eqnarray*}
p^{BZ}\left( \beta ,\mu \right) &=&\text{ }\stackunder{\rho \geq 0}{\sup }%
\left\{ \mu \rho -f^{BZ}\left( \beta ,\rho \right) \right\} \\
&=&\stackunder{\rho _{0}\geq 0}{\sup }\left\{ \stackunder{\rho \geq \rho _{0}%
}{\sup }\left\{ \mu \rho _{0}-\varepsilon _{0}\rho _{0}-\frac{g_{0}}{2}\rho
_{0}^{2}+\mu \left( \rho -\rho _{0}\right) -f^{P}\left( \beta ,\rho -\rho
_{0}\right) \right\} \right\} \\
&=&\stackunder{\rho _{0}\geq 0}{\sup }\left\{ p^{P}\left( \beta ,\mu \right)
-\left( \varepsilon _{0}-\mu \right) \rho _{0}-\frac{g_{0}}{2}\rho
_{0}^{2}\right\}
\end{eqnarray*}
which coincides with (\ref{diagmodel4}) found in {\rm{\cite{BZ99Ph}}}.
\end{remark}
%%%%%%%%%%%%%%%%%%%%%%%%%%%%%%%%%%%%%%%%%%%%%%%%%%%%%%%%%%%%%%%%%%%%%%%%%%%%%%%%%%%%%%%%%%%%%%%%%%%%
\smallskip

\noindent
\textbf{3.2} Now we consider our model (\ref{newdiagmodel2}) in the \textit{%
grand-canonical ensemble} $\left( \beta ,\mu \right) $. Let
\[
p_{\Lambda }\left( \beta ,\mu \right) := \frac{1}{\beta V}\ln {\rm{Tr}}_{%
\mathcal{F}_{\Lambda }}e^{-\beta \left( H_{\Lambda }-\mu N_{\Lambda }\right)}.
\]
be the grand-canonical pressure corresponding (\ref{newdiagmodel2}).
%%%%%%%%%%%%%%%%%%%%%%%%%%%%%%%%%%%%%%%%%%%%%%%%%%%%%%%%%%%%%%%%%%%%%%%%%%%%%%%%%%%%%%%%%%%%%%%%%%%%
\begin{theorem}
\label{newdiagmodelth2}Let $\lambda >0$ , $g_{0}>0$, $g>0$,and $\varepsilon
_{0}<0$, then:\newline
$\left( i\right) $ the domain of stability of $H_{\Lambda }$, i.e.
\begin{equation}
\widetilde{Q}:= \left\{ \left( \theta \geq 0,\mu \in \Bbb{R}^{1}\right) :%
\text{ }\stackunder{\Lambda }{\lim }p_{\Lambda }\left( \beta ,\mu \right)
<+\infty \right\} ,  \label{newdiagmodel11bisbis}
\end{equation}
is equal to $\widetilde{Q}=\left\{ \theta \geq 0\right\} \times \left\{ \mu
\in \Bbb{R}^{1}\right\} $;\newline
$\left( ii\right) $ in the thermodynamic limit one gets
\begin{equation}
p\left( \beta ,\mu \right) := \text{ }\stackunder{\Lambda }{\lim }%
p_{\Lambda }\left( \beta ,\mu \right) =\text{ }\stackunder{\alpha \leq 0}{%
\inf }\left\{ p^{BZ}\left( \beta ,\alpha \right) +\frac{\left( \mu -\alpha
\right) ^{2}}{2\lambda }\right\}  \label{newdiagmodel11}
\end{equation}
for $\left( \theta ,\mu \right) \in \widetilde{Q}$, where $p^{BZ}\left(
\beta ,\mu \right) $ is the pressure defined by (\ref{diagmodel4}).
Therefore the pressure (\ref{newdiagmodel11}) is independent of the
parameter $g$ provided it is positive.
\end{theorem}
%%%%%%%%%%%%%%%%%%%%%%%%%%%%%%%%%%%%%%%%%%%%%%%%%%%%%%%%%%%%%%%%%%%%%%%%%%%%%%%%%%%%%%%%%%%%%%%%%%%%%
\textit{Proof.} $\left( i\right) $ Notice that the Hamiltonian $H_{\Lambda }$
(\ref{newdiagmodel2}) is superstable, i.e. there are $B=-\varepsilon _{0}$
and $C=\lambda/2 $ such that
\begin{equation}
H_{\Lambda }\geq -N_{\Lambda }B+\frac{C}{V}N_{\Lambda }^{2}
\label{newdiagmodel11bis}
\end{equation}
for any box $\Lambda $. Therefore by (\ref{newdiagmodel11bis}) we obtain
that the infinite volume limit (\ref{newdiagmodel11}) exists for any $\mu
\in \Bbb{R}^{1}$.

$\left( ii\right) $ Since the pressure $p\left( \beta ,\mu \right) $ is in
fact a Legendre transform of the corresponding free-energy density $f\left(
\beta ,\rho \right) $ (\ref{newdiagmodel8}) or (\ref{newdiagmodel13bis}), by
Theorem \ref{newdiagmodelth1} we get
\begin{equation}
p\left( \beta ,\mu \right) =\text{ }\stackunder{\rho \geq 0}{\sup }\left\{
\mu \rho -f\left( \beta ,\rho \right) \right\} =\text{ }\stackunder{\rho
\geq 0}{\sup }\left\{ \mu \rho -\frac{\lambda}{2} \rho ^{2}-f^{BZ}\left( \beta ,\rho
\right) \right\} ,  \label{newdiagmodel12}
\end{equation}
with $f^{BZ}\left( \beta ,\rho \right) $ defined by (\ref{newdiagmodel13}).
Straightforward calculations give that
\[
\stackunder{\alpha \leq 0}{\inf }\left\{ \alpha \rho +\frac{\left( \mu
-\alpha \right) ^{2}}{2\lambda }-f^{BZ}\left( \beta ,\rho \right) \right\}
=\mu \rho -\frac{\lambda}{2} \rho ^{2}-f^{BZ}\left( \beta ,\rho \right)
\]
and thus (\ref{newdiagmodel12}) takes the form:
\begin{equation}
p\left( \beta ,\mu \right) =\text{ }\stackunder{\rho \geq 0}{\sup }\left\{
\stackunder{\alpha \leq 0}{\inf }\left\{ \alpha \rho +\frac{\left( \mu
-\alpha \right) ^{2}}{2\lambda }-f^{BZ}\left( \beta ,\rho \right) \right\}
\right\}  \label{newdiagmodel14}
\end{equation}
Notice that the $\stackunder{\rho \geq 0}{\sup }$ and $\stackunder{\alpha
\leq 0}{\inf }$ do not commute in general. However, convexity of the
free-energy density $f^{BZ}\left( \beta ,\rho \right) $ (cf. Remark \ref
{newdiagmodelth3}) implies that
\begin{equation}
F\left( \rho ,\alpha \right) := \alpha \rho +\frac{\left( \mu -\alpha
\right) ^{2}}{2\lambda }-f^{BZ}\left( \beta ,\rho \right)
\label{newdiagmodel15bis}
\end{equation}
is a strictly concave function of $\rho $ and a strictly convex function of $%
\alpha $. This ensures the uniqueness of the stationary point $%
\left( \widetilde{\rho },\widetilde{\alpha }\right) $ corresponding to
\begin{eqnarray*}
\partial _{\alpha }F\left( \widetilde{\rho },\widetilde{\alpha }\right) &=&0,
\\
\partial _{\rho }F\left( \widetilde{\rho },\widetilde{\alpha }\right) &=&0.
\end{eqnarray*}
Therefore
\begin{equation}
F\left( \widetilde{\rho },\widetilde{\alpha }\right) =\text{ }\stackunder{%
\rho \geq 0}{\sup }\left\{ \stackunder{\alpha \leq 0}{\inf }\left\{ F\left(
\rho ,\alpha \right) \right\} \right\} =\text{ }\stackunder{\alpha \leq 0}{%
\inf }\left\{ \stackunder{\rho \geq 0}{\sup }\left\{ F\left( \rho ,\alpha
\right) \right\} \right\} .  \label{newdiagmodel15}
\end{equation}
Since
\[
\stackunder{\rho \geq 0}{\sup }F\left( \rho ,\alpha \right) =\left\{ \frac{%
\left( \mu -\alpha \right) ^{2}}{2\lambda }+p^{BZ}\left( \beta ,\alpha
\right) \right\} ,
\]
(\ref{newdiagmodel14})-(\ref{newdiagmodel15}) imply (\ref{newdiagmodel11}). \hfill $\square$
%%%%%%%%%%%%%%%%%%%%%%%%%%%%%%%%%%%%%%%%%%%%%%%%%%%%%%%%%%%%%%%%%%%%%%%%%%%%%%%%%%%%%%%%%%%%%%%%%%%
\setcounter{equation}{0}
%%%%%%%%%%%%%%%%%%%%%%%%%%%%%%%%%%%%%%%%%%%%%%%%%%%%%%%%%%%%%%%%%%%%%%%%%%%%%%%%%%%%%%%%%%%%%%%%%%%
\section{Bose condensations}\label{sec:3}
%%%%%%%%%%%%%%%%%%%%%%%%%%%%%%%%%%%%%%%%%%%%%%%%%%%%%%%%%%%%%%%%%%%%%%%%%%%%%%%%%%%%%%%%%%%%%%%%%%%
\textbf{4.1} Let $\rho _{\Lambda }\left( \beta ,\mu \right) $ denote the
grand-canonical total particle density corresponding to the model (\ref{newdiagmodel2}), i.e.
\begin{equation}
\rho _{\Lambda }\left( \beta ,\mu \right) := \left\langle \frac{%
N_{\Lambda }}{V}\right\rangle _{H_{\Lambda }}=\partial _{\mu }p_{\Lambda
}\left( \beta ,\mu \right) ,  \label{newdiagmodel16}
\end{equation}
where $\left\langle -\right\rangle _{H_{\Lambda }}\left( \beta ,\mu \right) $
represents the grand-canonical Gibbs state for the Hamiltonian $H_{\Lambda }$
(\ref{newdiagmodel2}).
%%%%%%%%%%%%%%%%%%%%%%%%%%%%%%%%%%%%%%%%%%%%%%%%%%%%%%%%%%%%%%%%%%%%%%%%%%%%%%%%%%%%%%%%%%%%%%%%%%%
\begin{theorem}
\label{newdiagmodelth6}For $\left( \theta ,\mu \right) \in \widetilde{Q}$ (%
\ref{newdiagmodel11bisbis}) we have
\begin{equation}
\rho \left( \beta ,\mu \right) := \text{ }\stackunder{\Lambda }{\lim }%
\rho _{\Lambda }\left( \beta ,\mu \right) =\rho ^{BZ}\left( \beta ,%
\widetilde{\alpha }\left( \beta ,\mu \right) \right) .
\label{newdiagmodel21}
\end{equation}
Here $\widehat{\alpha }\left( \beta ,\mu \right) \leq 0$ is a unique
solution of equation
\begin{equation}
\rho ^{BZ}\left( \beta ,\alpha \right) +\frac{\left( \alpha -\mu \right) }{%
\lambda }=0,  \label{newdiagmodel23}
\end{equation}
when $\mu \leq \mu _{c}^{BZ}\left( \theta \right) := \lambda \rho
_{c}^{BZ}\left( \theta \right) $, whereas for $\mu >\mu _{c}^{BZ}\left(
\theta \right) $ one gets
\begin{equation}
\rho \left( \beta ,\mu \right) := \text{ }\stackunder{\Lambda }{\lim }%
\rho _{\Lambda }\left( \beta ,\mu \right) =\frac{\mu }{\lambda }\ .
\label{newdiagmodel21bis}
\end{equation}
Here $\rho _{c}^{BZ}\left( \theta \right) $ is defined above by (\ref
{diagmodel45.1}).
\end{theorem}
%%%%%%%%%%%%%%%%%%%%%%%%%%%%%%%%%%%%%%%%%%%%%%%%%%%%%%%%%%%%%%%%%%%%%%%%%%%%%%%%%%%%%%%%%%%%%%%%%%%
\textit{\textit{Proof.}} Let $\widetilde{\alpha }\left( \beta ,\mu \right)
\leq 0$ be defined by (\ref{newdiagmodel11}), i.e.
\begin{equation}
p\left( \beta ,\mu \right) =\text{ }\stackunder{\alpha \leq 0}{\inf }\left\{
p^{BZ}\left( \beta ,\alpha \right) +\frac{\left( \mu -\alpha \right) ^{2}}{%
2\lambda }\right\} =p^{BZ}\left( \beta ,\widetilde{\alpha }\left( \beta ,\mu
\right) \right) +\frac{\left( \mu -\widetilde{\alpha }\left( \beta ,\mu
\right) \right) ^{2}}{2\lambda }.  \label{newdiagmodel19}
\end{equation}
Since
\begin{equation}
\partial _{\alpha }\left[ p^{BZ}\left( \beta ,\alpha \right) +\frac{\left(
\mu -\alpha \right) ^{2}}{2\lambda }\right] =\rho ^{BZ}\left( \beta ,\alpha
\right) +\frac{\left( \alpha -\mu \right) }{\lambda },
\label{newdiagmodel20}
\end{equation}
then for $\mu \leq \mu _{c}^{BZ}\left( \theta \right) =\lambda \rho
_{c}^{BZ}\left( \theta \right) $ (cf. (\ref{diagmodel45.1})) there exists a
unique solution $\widehat{\alpha }\left( \beta ,\mu \right) \leq 0$ of (\ref
{newdiagmodel23}) which coincides with $\widetilde{\alpha }\left( \beta ,\mu
\right) $ in (\ref{newdiagmodel11}). Since $\left\{ p_{\Lambda }\left( \beta
,\mu \right) \right\} _{\Lambda }$ are convex functions of $\mu \in \Bbb{R}%
^{1}$ then combining (\ref{newdiagmodel16}) and (\ref{newdiagmodel19}) with
the Griffiths Lemma ({\cite{Grif}}, Section \ref{Griffiths lemma}) we obtain
the thermodynamic limit for the total particle density
\[
\rho \left( \beta ,\mu \right) =\partial _{\mu }p\left( \beta ,\mu \right) =%
\frac{\left( \mu -\widetilde{\alpha }\left( \beta ,\mu \right) \right) }{%
\lambda }.
\]
This together with (\ref{newdiagmodel23}) gives (\ref{newdiagmodel21}).%
\newline
Now let $\mu >\mu _{c}^{BZ}\left( \theta \right) $. Then by definitions of $%
\mu _{c}^{BZ}\left( \theta \right) $ and $\rho _{c}^{BZ}\left( \theta
\right) $ (see (\ref{diagmodel45.1})) one gets
\[
\partial _{\alpha }\left[ p^{BZ}\left( \beta ,\alpha \right) +\frac{\left(
\mu -\alpha \right) ^{2}}{2\lambda }\right] =\rho ^{BZ}\left( \beta ,\alpha
\right) +\frac{\left( \alpha -\mu \right) }{\lambda }\leq 0.
\]
This implies that
\begin{equation}
p\left( \beta ,\mu \right) =\text{ }\stackunder{\alpha \leq 0}{\inf }\left\{
p^{BZ}\left( \beta ,\alpha \right) +\frac{\left( \mu -\alpha \right) ^{2}}{%
2\lambda }\right\} =p^{BZ}\left( \beta ,0\right) +\frac{\mu ^{2}}{2\lambda },
\label{newdiagmodel22}
\end{equation}
i.e. $\widetilde{\alpha }\left( \beta ,\mu \right) =0$. Therefore, by the
Griffiths lemma and (\ref{newdiagmodel16}), (\ref{newdiagmodel22}) we get (%
\ref{newdiagmodel21bis}). \hfill $\square$
%%%%%%%%%%%%%%%%%%%%%%%%%%%%%%%%%%%%%%%%%%%%%%%%%%%%%%%%%%%%%%%%%%%%%%%%%%%%%%%%%%%%%%%%%%%%%%%%%%%%
\begin{theorem}\label{newdiagmodelth7}
Let $\varepsilon _{0}<0$. Then we have:
\begin{equation}
\rho _{0}\left( \theta ,\mu \right) := \text{ }\stackunder{\Lambda }{%
\lim }\left\langle \frac{a_{0}^{*}a_{0}}{V}\right\rangle _{H_{\Lambda
}}\left( \beta ,\mu \right) =\left\{
\begin{array}{c}
0,\text{ for }\left( \theta ,\mu \right) \in \widetilde{Q}\backslash
\widetilde{D}_{\varepsilon _{0}} \\
\left( \frac{\widetilde{\alpha }\left( \beta ,\mu \right) -\varepsilon _{0}}{%
g_{0}}\right) ,\text{ for }\left( \theta ,\mu \right) \in \widetilde{D}%
_{\varepsilon _{0}}
\end{array}
\right\} ,  \label{newdiagmodel17}
\end{equation}
with $\widetilde{\alpha }\left( \beta ,\mu \right) $ defined by equation (%
\ref{newdiagmodel19}). Here domain $\widetilde{D}_{\varepsilon
_{0}}$ is defined by:
\begin{equation}
\widetilde{D}_{\varepsilon _{0}}=\left\{ \left( \theta ,\mu \right) \in
\widetilde{Q}:\varepsilon _{0}<\widetilde{\alpha }\left( \beta ,\mu \right)
\right\} =\left\{ \left( \theta ,\mu \right) \in \widetilde{Q}:\widetilde{%
\mu }_{0}\left( \theta \right) <\mu \right\} ,  \label{newdiagmodel25}
\end{equation}
where we denote by $\widetilde{\mu }_{0}\left( \theta
\right) $ a unique solution of the equation
\begin{equation}
\widetilde{\alpha }\left( \beta ,\mu \right) = \varepsilon_{0}.
\label{newdiagmodel25bis}
\end{equation}
\end{theorem}
%%%%%%%%%%%%%%%%%%%%%%%%%%%%%%%%%%%%%%%%%%%%%%%%%%%%%%%%%%%%%%%%%%%%%%%%%%%%%%%%%%%%%%%%%%%%%%%%%%%%
\textit{\textit{Proof.} }Since $\left\{ p_{\Lambda }\left( \beta ,\mu
\right) \right\} _{\Lambda }$ are convex functions of $\varepsilon _{0}\in
\Bbb{R}^{1}$, then by
\begin{equation}
\left\langle \frac{a_{0}^{*}a_{0}}{V}\right\rangle _{H_{\Lambda }}\left(
\beta ,\mu \right) =-\partial _{\varepsilon _{0}}p_{\Lambda }\left( \beta
,\mu \right) ,  \label{newdiagmodel18}
\end{equation}
and by the Griffiths lemma we obtain that
\begin{equation}
\stackunder{\Lambda }{\lim }\left\langle \frac{a_{0}^{*}a_{0}}{V}%
\right\rangle _{H_{\Lambda }}\left( \beta ,\mu \right) =-\partial
_{\varepsilon _{0}}p\left( \beta ,\mu \right) .  \label{newdiagmodel24}
\end{equation}
For $\mu \leq \mu _{c}^{BZ}\left( \theta \right) =\lambda \rho
_{c}^{BZ}\left( \theta \right) $ there is a unique $\widetilde{\alpha }%
\left( \beta ,\mu \right) \leq 0$ defined by (\ref{newdiagmodel19}) which
verifies (\ref{newdiagmodel23}), whereas for $\mu >\mu _{c}^{BZ}\left(
\theta \right) $ according to (\ref{newdiagmodel22}) we obtain $\widetilde{%
\alpha }\left( \beta ,\mu \right) =0$. Notice that by (\ref{diagmodel4}) for
$\mu \leq \varepsilon _{0}$ we have
\[
p^{BZ}\left( \beta ,\mu \right) =p^{P}\left( \beta ,\mu \right) .
\]
Therefore, by (\ref{newdiagmodel25}), (\ref{newdiagmodel24}) one gets from (%
\ref{newdiagmodel19}) and (\ref{newdiagmodel22}) that
\[
\stackunder{\Lambda }{\lim }\left\langle \frac{a_{0}^{*}a_{0}}{V}%
\right\rangle _{H_{\Lambda }}\left( \beta ,\mu \right) =\left\{
\begin{array}{c}
0,\text{ for }\widetilde{\alpha }\left( \beta ,\mu \right) \leq \varepsilon
_{0}<0 \\
\left( \frac{\widetilde{\alpha }\left( \beta ,\mu \right) -\varepsilon _{0}}{%
g_{0}}\right) ,\text{ for }\varepsilon _{0}\leq \widetilde{\alpha }\left(
\beta ,\mu \right)
\end{array}
\right\} ,
\]
i.e. (\ref{newdiagmodel17}). \hfill $\square$
%%%%%%%%%%%%%%%%%%%%%%%%%%%%%%%%%%%%%%%%%%%%%%%%%%%%%%%%%%%%%%%%%%%%%%%%%%%%%%%%%%%%%%%%%%%%%%%%%%%%

\hspace{0.4cm}
Hence by Theorem \ref{newdiagmodelth7}, the domain $\widetilde{D}%
_{\varepsilon _{0}}$ (\ref{newdiagmodel25}) can be described as
\begin{equation}
\widetilde{D}_{\varepsilon _{0}}=\left\{ \left( \theta ,\mu \right) \in
\widetilde{Q}:\text{ }\rho _{0}\left( \theta ,\mu \right) := \text{ }%
\stackunder{\Lambda }{\lim }\left\langle \frac{a_{0}^{*}a_{0}}{V}%
\right\rangle _{H_{\Lambda }}>0\right\} .  \label{newdiagmodel26}
\end{equation}
Notice that in contrast to $D_{\varepsilon _{0}}$, see (\ref{diagmodel18}), (%
\ref{diagmodel19}), the domain $\widetilde{D}_{\varepsilon _{0}}$ has a
temperature dependent boundary and extends to positive $\mu $.
This macroscopic occupation of the mode $k=0$ (\ref{newdiagmodel17}) is a
\textit{non-conventional} Bose condensation which occurs in the model (\ref
{newdiagmodel2}) due to the \textit{attraction} term $\varepsilon_{0}a_{0}^{*}a_{0}$, for
$\varepsilon _{0}<0$ (cf. Section \ref{classification des condensations}). It is
\textit{similar} to the \textit{first} stage of
condensation manifested by the model $H_{\Lambda }^{BZ}$ (\ref{diagmodel1})
with $g_{k}\left( V\right) =g/2$) although in the latter case it is possible
only for $\mu \leq 0$, see \cite{BZ99Ph}. In particular, we have again a
saturation of the condensate density in the mode $k=0$:
\begin{equation}
\stackunder{\mu \in \Bbb{R}^{1}}{\sup }\rho _{0}\left( \theta ,\mu \right)
=\rho _{0}\left( \theta ,\mu \geq \mu _{c}^{BZ}\left( \theta \right) \right)
=-\frac{\varepsilon _{0}}{g_{0}},  \label{newdiagmodel28}
\end{equation}
cf. (\ref{diagmodel28}). Notice that for any $\mu $%
\[
\stackunder{\beta \rightarrow 0^{+}}{\lim }\widetilde{\alpha }\left( \beta
,\mu \right) =-\infty .
\]
Thus, in contrast to the model (\ref{diagmodel1}) (with $g_{k}\left(
V\right) =g/2$), the non-conventional condensation in the model (\ref
{newdiagmodel2}) depends on the temperature. There is $\widetilde{\theta }%
_{0}\left( \mu \right) $ (solution of the equation $\widetilde{\alpha }%
\left( \theta ^{-1},\mu \right) =\varepsilon _{0}$, (\ref{newdiagmodel25bis}%
)) such that
\begin{equation}
\rho _{0}\left( \theta ,\mu \right) =\frac{\widetilde{\alpha }\left( \beta
,\mu \right) -\varepsilon _{0}}{g_{0}}>0,  \label{newdiagmodel27}
\end{equation}
for $\theta \leq \widetilde{\theta }_{0}\left( \mu \right) $ and
\begin{equation}
\rho _{0}\left( \theta ,\mu \right) =0,  \label{newdiagmodel27bis}
\end{equation}
for $\theta >\widetilde{\theta }_{0}\left( \mu \right) $. This is another
way to describe the phase diagram of the model (\ref{newdiagmodel2}): $%
\widetilde{\theta }_{0}\left( \mu \right) $ is simply the inverse function
to $\widetilde{\mu }_{0}\left( \theta \right) $.
%%%%%%%%%%%%%%%%%%%%%%%%%%%%%%%%%%%%%%%%%%%%%%%%%%%%%%%%%%%%%%%%%%%%%%%%%%%%%%%%%%%%%%%%%%%%%%%%%%%%
\smallskip

\noindent
\textbf{4.2} Similar to (\ref{diagmodel1}), in the model (\ref{newdiagmodel2}) for $d>2$
we encounter for large total particle densities \textit{another kind} of
condensation: a conventional \textit{non-extensive} (i.e. type III) BEC
in the vicinity of $k=0$ (see Section \ref{classification des condensations}).
In order to control this condensation we introduce an auxiliary Hamiltonian
\begin{equation}
H_{\Lambda ,\gamma }:= H_{\Lambda }-\gamma \stackunder{\left\{ k\in
\Lambda ^{*}:\left\| k\right\| \geq \delta \right\} }{\sum }a_{k}^{*}a_{k},
\label{newdiagmodel53}
\end{equation}
for a fixed $\delta >0,$ and we set
\begin{equation}
p_{\Lambda }\left( \beta ,\mu ,\gamma \right) := \frac{1}{\beta V}\ln
{\rm{Tr}}_{\mathcal{F}_{\Lambda }}e^{-\beta H_{\Lambda ,\gamma }\left( \mu \right)}.
\label{newdiagmodel54}
\end{equation}
%%%%%%%%%%%%%%%%%%%%%%%%%%%%%%%%%%%%%%%%%%%%%%%%%%%%%%%%%%%%%%%%%%%%%%%%%%%%%%%%%%%%%%%%%%%%%%%%%%%
\begin{remark}
\label{newdiagmodelth10}Let $\gamma <\varepsilon _{\delta }:=
\varepsilon _{\left\| k\right\| =\delta }$. Then the system with Hamiltonian
$H_{\Lambda ,\gamma }$ has the same properties as the model $H_{\Lambda }$
modulo the free-particle spectrum transformation:
\begin{equation}
\varepsilon _{k}\rightarrow \varepsilon _{k,\gamma }:= \varepsilon
_{k}-\gamma .\chi _{\left[ \delta ,+\infty \right) }\left( \left\| k\right\|
\right)  \label{newdiagmodel54bis}
\end{equation}
where $\chi _{A}\left( x\right) $ is the characteristic function of domain $%
A $. In particular, the results of Theorems \ref{newdiagmodelth1} and \ref
{newdiagmodelth2} remain unchanged. For $\left( \theta ,\mu \right) \in
\widetilde{Q}$ and $\gamma <\varepsilon _{\delta }$ we have
\begin{equation}
p\left( \beta ,\mu ,\gamma \right) := \text{ }\stackunder{\Lambda }{\lim
}p_{\Lambda }\left( \beta ,\mu ,\gamma \right) =\text{ }\stackunder{\alpha
\leq 0}{\inf }\left\{ p^{BZ}\left( \beta ,\alpha ,\gamma \right) +\frac{%
\left( \mu -\alpha \right) ^{2}}{2\lambda }\right\} ,  \label{newdiagmodel55}
\end{equation}
where $p^{BZ}\left( \beta ,\mu ,\gamma \right) $ is the pressure (\ref
{diagmodel4}) but with the free-particle spectrum (\ref{newdiagmodel54bis}):
\begin{eqnarray}
p^{BZ}\left( \beta ,\mu ,\gamma \right) &=&p^{P}\left( \beta ,\mu ,\gamma
\right) -\stackunder{\rho _{0}\geq 0}{\inf }\left[ \left( \varepsilon
_{0}-\mu \right) \rho _{0}+\frac{g_{0}\rho _{0}^{2}}{2}\right]  \nonumber \\
&=&\frac{1}{\beta \left( 2\pi \right) ^{d}}\stackunder{k\in \Bbb{R}^{d}}{%
\int }\ln \left[ \left( 1-e^{-\beta \left( \varepsilon _{k,\gamma }-\mu
\right) }\right) ^{-1}\right] d^{d}k  \nonumber \\
&&-\stackunder{\rho _{0}\geq 0}{\inf }\left[ \left( \varepsilon _{0}-\mu
\right) \rho _{0}+\frac{g_{0}\rho _{0}^{2}}{2}\right] .
\label{newdiagmodel56}
\end{eqnarray}
\end{remark}
%%%%%%%%%%%%%%%%%%%%%%%%%%%%%%%%%%%%%%%%%%%%%%%%%%%%%%%%%%%%%%%%%%%%%%%%%%%%%%%%%%%%%%%%%%%%%%%%%%%%
\begin{theorem}\label{newdiagmodelth8}
For any $\left( \theta ,\mu \right) \in \widetilde{Q}$ we have
\begin{equation}
\stackunder{\Lambda }{\lim }\left\langle \frac{a_{k}^{*}a_{k}}{V}%
\right\rangle _{H_{\Lambda }}=0\text{, }k\in \Lambda ^{*}\backslash \left\{
0\right\} ,  \label{newdiagmodel39}
\end{equation}
i.e., there is no macroscopic occupation of modes $k\neq 0,$ whereas for $%
\mu >\mu _{c}^{BZ}\left( \theta \right) =\lambda \rho _{c}^{BZ}\left(
\theta \right) $ the model $H_{\Lambda }\ $(\ref{newdiagmodel2}) manifests a
generalised (non-extensive) BEC for those modes:
\begin{equation}
\stackunder{\delta \rightarrow 0^{+}}{\lim }\stackunder{\Lambda }{\lim }%
\frac{1}{V}\stackunder{\left\{ k\in \Lambda ^{*}:\text{ }0<\left\| k\right\|
<\delta \right\} }{\sum }\left\langle N_{k}\right\rangle _{H_{\Lambda
}}=\rho \left( \beta ,\mu \right) -\rho _{c}^{BZ}\left( \theta \right) =%
\frac{1}{\lambda }\left( \mu -\mu _{c}^{BZ}\left( \theta \right) \right) >0.
\label{newdiagmodel40}
\end{equation}
Here $\rho \left( \beta ,\mu \right) $ is defined by (\ref{newdiagmodel21bis}).
If $\varepsilon _{0}<0$, then this condensation coexists with the
non-conventional condensation in the mode $k=0$ (see Theorem \ref
{newdiagmodelth7}).
\end{theorem}
%%%%%%%%%%%%%%%%%%%%%%%%%%%%%%%%%%%%%%%%%%%%%%%%%%%%%%%%%%%%%%%%%%%%%%%%%%%%%%%%%%%%%%%%%%%%%%%%%%%%
\textit{\textit{Proof.}} Let $g>0$ and $\Delta g>0$ be such that $g-\Delta
g>0$. Then by the {Bogoliubov convexity inequality (see e.g. \cite{BBZKT84})}, one gets:
\begin{equation}
0\leq \frac{\Delta g}{2V^{2}}\stackunder{k\in \Lambda ^{*}\backslash \left\{
0\right\} }{\sum }\left\langle N_{k}^{2}\right\rangle _{H_{\Lambda }}\leq
p_{\Lambda }[ H_{\Lambda }-\frac{\Delta g}{2V}\stackunder{k\in \Lambda
^{*}\backslash \left\{ 0\right\} }{\sum }N_{k}^{2}] - p_{\Lambda
}\left[ H_{\Lambda }\right] .  \label{newdiagmodel41}
\end{equation}
Notice that by Theorems \ref{newdiagmodelth1} and \ref{newdiagmodelth2} the
thermodynamic limits of pressures for two models (\ref{newdiagmodel2}) with
parameters $g>0$ and $g-\Delta g>0$ coincide with (\ref{newdiagmodel11}),
i.e. one has
\begin{equation}
\stackunder{\Lambda }{\lim }\{ p_{\Lambda }[ H_{\Lambda }-\frac{%
\Delta g}{2V}\stackunder{k\in \Lambda ^{*}\backslash \left\{ 0\right\} }{%
\sum }N_{k}^{2}] -p_{\Lambda }\left[ H_{\Lambda }\right] \} =0.
\label{newdiagmodel42}
\end{equation}
Since for any $k\in \Lambda ^{*}\backslash \left\{ 0\right\} $ we have the
estimate
\[
0\leq \left( \frac{\left\langle N_{k}\right\rangle _{H_{\Lambda }}}{V}%
\right) ^{2}\leq \frac{\left\langle N_{k}^{2}\right\rangle _{H_{\Lambda }}}{%
V^{2}}\leq \frac{1}{V^{2}}\stackunder{k\in \Lambda ^{*}\backslash \left\{
0\right\} }{\sum }\left\langle N_{k}^{2}\right\rangle _{H_{\Lambda }},
\]
its combination with (\ref{newdiagmodel41}) and (\ref{newdiagmodel42}) gives
(\ref{newdiagmodel39}).\newline
Let $\delta >0,$ then we have
\begin{equation}
\frac{1}{V}\stackunder{\left\{ k\in \Lambda ^{*}:\text{ }0<\left\| k\right\|
<\delta \right\} }{\sum }\left\langle N_{k}\right\rangle _{H_{\Lambda
}}=\rho _{\Lambda }\left( \beta ,\mu \right) -\left\langle \frac{%
a_{0}^{*}a_{0}}{V}\right\rangle _{H_{\Lambda }}-\frac{1}{V}\stackunder{%
\left\{ k\in \Lambda ^{*}:\left\| k\right\| \geq \delta \right\} }{\sum }%
\left\langle N_{k}\right\rangle _{H_{\Lambda }}.  \label{newdiagmodel43}
\end{equation}
Now we can follow the same line of reasoning as in proofs of Theorems \ref
{newdiagmodelth6} and \ref{newdiagmodelth7}: we have the set $\left\{
p_{\Lambda }\left( \beta ,\mu ,\gamma \right) \right\} _{\Lambda }$ of
convex functions of $\gamma \in \left( -\infty ,\varepsilon _{\delta
}\right] $ with
\[
\frac{1}{V}\stackunder{\left\{ k\in \Lambda ^{*}:\left\| k\right\| \geq
\delta \right\} }{\sum }\left\langle N_{k}\right\rangle _{H_{\Lambda ,\gamma
}}=\partial _{\gamma }p_{\Lambda }\left( \beta ,\mu ,\gamma \right) ,
\]
which by the Griffiths lemma and (\ref{newdiagmodel55}), (\ref
{newdiagmodel56}) implies for $\gamma =0$:
\begin{equation}
\stackunder{\Lambda }{\lim }\frac{1}{V}\stackunder{\left\{ k\in \Lambda
^{*}:\left\| k\right\| \geq \delta \right\} }{\sum }\left\langle
N_{k}\right\rangle _{H_{\Lambda }}=\partial _{\gamma }p\left( \beta ,\mu
,\gamma =0\right) .  \label{newdiagmodel57}
\end{equation}
Then by definitions (\ref{newdiagmodel54bis}), (\ref{newdiagmodel55}) and
Theorem \ref{newdiagmodelth6}, see (\ref{newdiagmodel21}), (\ref
{newdiagmodel21bis}), together with explicit formula (\ref{diagmodel42.2})
we get for $\mu <\mu _{c}^{BZ}\left( \theta \right) $:
\begin{equation}
\partial _{\gamma }p\left( \beta ,\mu ,\gamma =0\right) =\frac{1}{\left(
2\pi \right) ^{d}}\stackunder{\left\| k\right\| \geq \delta }{\int }\frac{%
d^{d}k}{e^{\beta \left( \varepsilon _{k}-\widetilde{\alpha }\left( \beta
,\mu \right) \right) }-1}  \label{newdiagmodel58}
\end{equation}
and
\begin{equation}
\partial _{\gamma }p\left( \beta ,\mu ,\gamma =0\right) =\frac{1}{\left(
2\pi \right) ^{d}}\stackunder{\left\| k\right\| \geq \delta }{\int }\frac{%
d^{d}k}{e^{\beta \varepsilon _{k}}-1}  \label{newdiagmodel59}
\end{equation}
for $\mu \geq \mu _{c}^{BZ}\left( \theta \right) $. Now, by virtue of (\ref
{newdiagmodel21bis}), (\ref{newdiagmodel28}) and definition (\ref
{diagmodel45.1}) we obtain (\ref{newdiagmodel40}) from (\ref{newdiagmodel43}%
), (\ref{newdiagmodel57}) and (\ref{newdiagmodel59}) by taking first the
thermodynamic limit and then the limit $\delta \rightarrow +0$. \hfill $\square$

%%%%%%%%%%%%%%%%%%%%%%%%%%%%%%%%%%%%%%%%%%%%%%%%%%%%%%%%%%%%%%%%%%%%%%%%%%%%%%%%%%%%%%%%%%%%%%%%
\setcounter{equation}{0}
%%%%%%%%%%%%%%%%%%%%%%%%%%%%%%%%%%%%%%%%%%%%%%%%%%%%%%%%%%%%%%%%%%%%%%%%%%%%%%%%%%%%%%%%%%%%%%%%
\section{Comments}\label{sec:4}
%%%%%%%%%%%%%%%%%%%%%%%%%%%%%%%%%%%%%%%%%%%%%%%%%%%%%%%%%%%%%%%%%%%%%%%%%%%%%%%%%%%%%%%%%%%%%%%%
We have presented a new exactly soluble model (\ref{newdiagmodel2}) which is
inspired by the MSV model \cite{MiVer} and our model \cite{BZ99Ph}. Due to
\textit{attractive}-type interaction in the mode $k=0$ it belongs to the family
of models which manifest two kinds of condensations: \textit{non-conventional} one in the mode
$k=0$ and \textit{conventional} (generalised)
BEC in modes $k\neq 0$. These condensations coexist
for large total particle densities $\rho >\rho _{c}^{BZ}\left( \theta
\right) $, or $\mu \geq \mu _{c}^{BZ}\left( \theta \right) =\lambda \rho
_{c}^{BZ}\left( \theta \right) $. This model demonstrates the richness of
the notion of Bose-condensation. It gives also a better understanding of the
difference between \textit{non-conventional} and \textit{conventional}
condensations.

\hspace{0.4cm}
First, in spite of superstability of the model, which implies
\[
\stackunder{\mu \in \Bbb{R}^{1}}{\sup }\rho \left( \beta ,\mu \right)
=+\infty ,
\]
the conventional condensation is due to a mechanism of saturation. Since,
after saturation of the non-conventional condensation, the kinetic-energy
density attains its maximal value at the critical density $\rho
_{c}^{BZ}\left( \theta \right) $, the further growth of the total energy
density for $\rho >\rho _{c}^{BZ}\left( \theta \right) $ is caused by a
macroscopic amount of particles with almost zero momenta.

\hspace{0.4cm}
The second important feature of models (\ref{newdiagmodel2}) (similar to
\cite{BZ99Ph} and in contrast to \cite{BZ98JP}) is that the repulsion between
bosons with $k\neq 0$ is strong enough to produce a generalized type III
(i.e. non-extensive) BEC. Notice that in the
Bogoliubov Weakly Imperfect Bose Gas \cite{BZ98JP,BZ00}, the BEC is of type I.

\setcounter{proposition}{0}
%
%EndExpansion

\subsection{Classification of the Bose-condensation types}\label{classification des condensations}

\subsubsection{The van den Berg-Lewis-Pul\`{e} classification: condensation
of type I, II and III}

For reader's convenience we remind a nomenclature of (generalized\textit{)}
Bose-Einstein condensations according to \cite{vdBLP86}:\newline
- the condensation is called the \textit{type I} when a finite number of
single-particle levels are macroscopically occupied;\newline
- it is of \textit{type II} when an infinite number of the levels are
macroscopically occupied;\newline
- it is called the \textit{type III}, or the \textit{non-extensive}
condensation, when no of the levels are macroscopically occupied whereas one
has
\[
\stackunder{\delta \rightarrow 0^{+}}{\lim }\stackunder{\Lambda }{\lim }%
\frac{1}{V}\stackunder{\left\{ k\in \Lambda ^{*},0<\left\| k\right\| \leq
\delta \right\} }{\sum }\left\langle N_{k}\right\rangle =\rho -\rho
_{c}\left( \theta \right) .
\]

\hspace{0.4cm}
An example of these different condensations is given in \cite{vdBL82}.
This paper demonstrates that three types of BEC can
be realised in the case of the PBG in an \textit{anisotropic} rectangular
box $\Lambda \subset \Bbb{R}^{3}$ of volume $V=\left| \Lambda \right|
=L_{x} \cdot L_{y} \cdot L_{z}$ and with Dirichlet boundary conditions. Let $%
L_{x}=V^{\alpha _{x}}$, $L_{y}=V^{\alpha _{y}}$, $L_{z}=V^{\alpha _{z}}$ for
$\alpha _{x}+\alpha _{y}+\alpha _{z}=1$ and $\alpha _{x}\leq \alpha _{y}\leq
\alpha _{z}$. If $\alpha _{z}<1/2$, then for sufficient large density $\rho $%
, we have the BEC of type I in the fundamental mode $%
k=\left( \frac{2\pi }{L_{x}},\frac{2\pi }{L_{y}},\frac{2\pi }{L_{z}}\right) $%
. For $\alpha _{z}=1/2$ one gets a condensation of type II characterized by
a macroscopic occupation of infinite package of modes $k=\left( \frac{2\pi }{%
L_{x}},\frac{2\pi }{L_{y}},\frac{2\pi n}{L_{z}}\right) $, $n\in \Bbb{N}$,
whereas for $\alpha _{z}>1/2$ we obtain a condensation of type III. In \cite
{Pule83} it was shown that the type III condensation can be caused in the PBG
by a weak external potential or (see \cite{vdB82}) by a specific choice of
boundary conditions and geometry. Another example of the \textit{%
non-extensive} condensation is given in \cite{MiVer,BZ99Ph} for bosons in an \textit{%
isotropic} box $\Lambda $ with \textit{repulsive interactions} which spread
out the \textit{conventional} BEC of type I into
Bose-Einstein\textit{\ }condensation of type III.

\subsubsection{Non-conventional versus conventional Bose condensation}

Here we classify Bose condensations by their mechanisms of formation. In the
most of papers (cf.\cite{vdBL82,vdB82,Pule83,MiVer}), the condensation is due to
\textit{saturation} of the total particle density, originally discovered by
Einstein \cite{Ein25} in the Bose gas without interaction (PBG). We call it
\textit{conventional} BEC \cite{ZUK}.

The existence of condensations, which is induced by \textit{interaction}, is
pointed out in papers \cite{BZ99Ph,BZ98JP,BZ00}. It is also the case of
Huang-Yang-Luttinger model \cite{vdBLP88} since it contains attractive interactions.
In particular, this is the case of the Bogoliubov Weakly Imperfect Bose-Gas \cite{BZ98JP}.
We call it \textit{non-conventional} Bose condensation.

$\left( i\right) $ As it is shown above, the non-conventional condensation does not
exclude the appearance of the BEC when total density of particles grows and exceeds
some saturation limit $\rho _{c}^{BZ}\left( \theta \right) $.

$\left( ii\right) $ To appreciate the notion of non-conventional
condensation let us remark that in models (\ref{diagmodel1}) and (\ref
{newdiagmodel2}) for $d=1,2$, there exists only one kind of condensation,
namely the \textit{non-conventional}.
%%%%%%%%%%%%%%%%%%%%%%%%%%%%%%%%%%%%%%%%%%%%%%%%%%%%%%%%%%%%%%%%%%%%%%%%%%%%%%%%%%%%%%%%%%%%%%%%%%%%%
\begin{remark}
A non-conventional BEC can always be characterized by its
type. Therefore, formally one obtains six kinds of condensations: a
non-conventional versus conventional of types I, II, or III.
\end{remark}
%%%%%%%%%%%%%%%%%%%%%%%%%%%%%%%%%%%%%%%%%%%%%%%%%%%%%%%%%%%%%%%%%%%%%%%%%%%%%%%%%%%%%%%%%%%%%%%%%%%%%
\setcounter{proposition}{0}

\subsection{The Griffiths lemma}\label{Griffiths lemma}
%%%%%%%%%%%%%%%%%%%%%%%%%%%%%%%%%%%%%%%%%%%%%%%%%%%%%%%%%%%%%%%%%%%%%%%%%%%%%%%%%%%%%%%%%%%%%%%%%%%%%
\begin{lemma}\rm{\cite{Grif}}
Let $\left\{ f_{n}\left( x\right) \right\} _{n\geq 1}$ be a sequence of
convex functions on a compact $I\subset \Bbb{R}$. If there exists a
pointwise limit
\begin{equation}
\stackunder{n\rightarrow \infty }{\lim }f_{n}\left( x\right) =f\left(
x\right) ,\text{ }x\in I,  \label{eq condC1}
\end{equation}
then
\begin{equation}
\begin{array}{l}
\stackunder{n\rightarrow \infty }{\lim }{\inf }\text{ }\partial
_{x}f_{n}\left( x-0\right) \geq \partial _{x}f\left( x-0\right) , \\
\stackunder{n\rightarrow \infty }{\lim }{\sup }\text{ }\partial
_{x}f_{n}\left( x+0\right) \leq \partial _{x}f\left( x+0\right) .
\end{array}
\label{first equality pour griffits}
\end{equation}
\end{lemma}
%%%%%%%%%%%%%%%%%%%%%%%%%%%%%%%%%%%%%%%%%%%%%%%%%%%%%%%%%%%%%%%%%%%%%%%%%%%%%%%%%%%%%%%%%%%%%%%%%%%%
\textit{\textit{Proof.}} By convexity one has
\begin{equation}
\begin{array}{l}
\partial _{x}f_{n}\left( x+0\right) \leq \frac{1}{l}\left[ f_{n}\left(
x+l\right) -f_{n}\left( x\right) \right] , \\
\partial _{x}f_{n}\left( x-0\right) \geq \frac{1}{l}\left[ f_{n}\left(
x\right) -f_{n}\left( x-l\right) \right] ,
\end{array}
\label{toto}
\end{equation}
for $l>0$. Then taking the limit $n\rightarrow \infty $ in (\ref{toto}), by (%
\ref{eq condC1}) we obtain:
\begin{equation}
\begin{array}{l}
\stackunder{n\rightarrow \infty }{\lim }{\sup }\text{ }\partial
_{x}f_{n}\left( x+0\right) \leq \frac{1}{l}\left[ f\left( x+l\right)
-f\left( x\right) \right] , \\
\stackunder{n\rightarrow \infty }{\lim }{\inf }\text{ }\partial
_{x}f_{n}\left( x-0\right) \geq \frac{1}{l}\left[ f\left( x\right) -f\left(
x-l\right) \right] .
\end{array}
\label{totobis}
\end{equation}
Now taking the limit $l\rightarrow +0,$ in (\ref{totobis}), one gets (\ref
{first equality pour griffits}). \hfill $\square$

\begin{remark}
In particular, if $x_{0}\in I$ is such that $\partial _{x}f_{n}\left(
x_{0}-0\right) =\partial _{x}f_{n}\left( x_{0}+0\right) $ and $\partial
_{x}f\left( x_{0}-0\right) =\partial _{x}f\left( x_{0}+0\right) $, then
\[
\stackunder{n\rightarrow \infty }{\lim }\partial _{x}f_{n}\left(
x_{0}\right) =\partial _{x}f\left( x_{0}\right) .
\]
\end{remark}

\setcounter{proposition}{0}

%TCIMACRO{\TeXButton{appendixpersofin}{\let\section\oldsect }}
%BeginExpansion
%\let\section\oldsect %
%EndExpansion

%%%%%%%%%%%%%%%%%%%%%%%%%%%%%%%%%%%%%%%%%%%%%%%%%%%%%%%%%%%%%%%%%%%%%%%%%%%%%%%%%%%%%%%%%%%%%%%%
\setcounter{equation}{0}
%%%%%%%%%%%%%%%%%%%%%%%%%%%%%%%%%%%%%%%%%%%%%%%%%%%%%%%%%%%%%%%%%%%%%%%%%%%%%%%%%%%%%%%%%%%%%%%%
\section{Effect of non-diagonal interaction in the \\
Weakly Imperfect Bose-Gas} \label{section 0}
%%%%%%%%%%%%%%%%%%%%%%%%%%%%%%%%%%%%%%%%%%%%%%%%%%%%%%%%%%%%%%%%%%%%%%%%%%%%%%%%%%%%%%%%%%%%%%%%
Note that neither in the toy model (\ref{newdiagmodel2}), nor in the HYL model
(\ref{HYL}) there is no two-body
potential in direct space, which is responsible for attraction between bosons. Instead,
the attraction instability there is mimicked by potentials in the dual (momentum) space. They are
favouring the accumulation of bosons in zero-mode $k =0$ enhancing the entropy/kinetic-energy
mechanism existed for conventional BEC in the perfect Bose-gas.
%%%%%%%%%%%%%%%%%%%%%%%%%%%%%%%%%%%%%%%%%%%%%%%%%%%%%%%%%%%%%%%%%%%%%%%%%%%%%%%%%%%%%%%%%%%%%%%%%%%%

In this section we present quantum mechanics arguments in order to explain conditions on the
two-body particle interaction potential that ensure a non-trivial
thermodynamic behaviour and non-conventional (dynamical) BEC manifested by the
Weakly Imperfect Bose-Gas (WIBG).
They based on the Fr\"{o}hlich transformation of the Bogoliubov truncated Hamiltonian
(known also as the WIBG model \cite{ZagBru01}), which is aimed to produce a partial
diagonalisation of the Hamiltonian.
%%%%%%%%%%%%%%%%%%%%%%%%%%%%%%%%%%%%%%%%%%%%%%%%%%%%%%%%%%%%%%%%%%%%%%%%%%%%%%%%%%%%%%%%%%%%%%%%%%%%
\smallskip

\noindent
\textbf{6.1} In \cite{Bog47, Bog47a}  Bogoliubov proposed a model of the Weakly Imperfect
Bose-Gas by truncation of the full Hamiltonian for bosons with two-body interaction.
In the grand-canonical ensemble this truncation yields
\begin{equation}
H_{\Lambda }^{B}\left( \mu \right) =T_{\Lambda }\left( \mu \right)
+U_{\Lambda }^{D}+U_{\Lambda }\, ,  \label{eq 1}
\end{equation}
where
\begin{eqnarray*}
T_{\Lambda }\left( \mu \right)  &=&\stackunder{k\in \Lambda ^{*}}{\sum }%
\left( \varepsilon _{k}-\mu \right) a_{k}^{*}a_{k}, \\
U_{\Lambda }^{D} &=&\frac{v\left( 0\right) }{V}a_{0}^{*}a_{0}\stackunder{%
k\in \Lambda ^{*},k\neq 0}{\sum }a_{k}^{*}a_{k}+\frac{1}{2V}\stackunder{k\in
\Lambda ^{*},k\neq 0}{\sum }v\left( k\right) a_{0}^{*}a_{0}\left(
a_{k}^{*}a_{k}+a_{-k}^{*}a_{-k}\right)  \\
&&+\frac{v\left( 0\right) }{2V}a_{0}^{*^{2}}a_{0}^{2}, \\
U_{\Lambda } &=&\frac{1}{2V}\stackunder{k\in \Lambda ^{*},k\neq 0}{\sum }%
v\left( k\right) \left(
a_{k}^{*}a_{-k}^{*}a_{0}^{2}+a_{0}^{*^{2}}a_{k}a_{-k}\right) ,
\end{eqnarray*}
with $\mu $ the chemical potential. Here $\{ a_{k}^{\#}\} _{k\in
\Lambda ^{*}}$ are the boson creation and annihilation operators
corresponding to the second quantisation in the cubic box $\Lambda =L\times
L\times L\subset \Bbb{R}^{3}$ with periodic boundary conditions on $\partial
\Lambda $, $\varepsilon _{k}=\hbar ^{2}k^{2}/2m,$
\[
\Lambda ^{*}= \{ k\in \Bbb{R}^{3}:\alpha =1,2,3\text{ , }k_{\alpha }=%
\frac{2\pi n_{\alpha }}{L} \ \text{ et } \ n_{\alpha }=0,\pm 1,\pm
2,\ldots  \, .\} ,
\]
$v\left( k\right) = \ \stackunder{\Bbb{R}^{3}}{\int }d^{3}x\varphi \left(
x\right) e^{-ikx},$ and $V=L^{3}$. We can remark that $H_{\Lambda
}^{B}\left( \mu \right) $ is defined in the boson Fock space over $L^{2}\left(
\Lambda \right) ,$ $\mathcal{F}_{\Lambda }\approx \mathcal{F}_{0\Lambda
}\otimes \mathcal{F}_{\Lambda }^{\prime }$ where $\mathcal{F}_{0\Lambda }$
and $\mathcal{F}_{\Lambda }^{\prime }$ are the boson Fock spaces constructed
out of $\mathcal{H}_{0\Lambda }$ (the one-dimensional subspace generated by $%
\psi _{k=0}$) and of its orthogonal complement $\mathcal{H}_{0\Lambda
}^{\bot }$ respectively. Note that for any complex $c\in \Bbb{C}$, we can
define in $\mathcal{F}_{0\Lambda }$ a coherent vector
\begin{equation}
\vspace{1pt}\psi _{0\Lambda }\left( c\right) =e^{-V\left| c\right| ^{2}/2}%
\stackrel{\infty }{\stackunder{k=0}{\sum }}\frac{1}{k!}\left( \sqrt{V}%
c\right) ^{k}\left( a_{0}^{*}\right) ^{k}\Omega _{0},  \label{eq 1bis}
\end{equation}
\vspace{1pt}where $\Omega _{0}$ is the vacuum of $\mathcal{F}_{\Lambda }$
and then $a_{0}\psi _{0\Lambda }\left( c\right) =c\sqrt{V}\psi _{0\Lambda
}\left( c\right) $.

Below we suppose that :

(A) the pair potential $\varphi \left( x\right) $ is
absolutely integrable (to ensure the existence of $v\left( k\right) $);

(B) the Fourier-transform $0\leq v\left( k\right) =v\left( -k\right) \leq
v\left( 0\right) $ and $v\left( 0\right) >0$.

%%%%%%%%%%%%%%%%%%%%%%%%%%%%%%%%%%%%%%%%%%%%%%%%%%%%%%%%%%%%%%%%%%%%%%%%%%%%%%%%%%%%%%%%%%%%%%%%%%%%
\smallskip

\noindent
\textbf{6.2} It is known \cite{AVZ92} that the WIBG is thermodynamically stable: the pressure
$p^{B}(\beta ,\mu) = \ \stackunder{\Lambda }{\lim }p_{\Lambda }\left[ H_{\Lambda }^{B}\right]$
is bounded, only for $\mu \leq
0$. If one considers only the \textit{diagonal} part of the Bogoliubov
Hamiltonian \vspace{1pt}$H_{\Lambda }^{BD}\left( \mu \right) =T_{\Lambda
}\left( \mu \right) +U_{\Lambda }^{D}$, one can show (cf. \cite{BZ-PL98}) that
\begin{equation}
p^{BD}\left( \beta ,\mu \leq 0\right) = \ \stackunder{\Lambda }{\lim }%
p_{\Lambda }\left[ H_{\Lambda }^{BD}\right] = p^{PBG}\left( \beta ,\mu \leq
0\right) ,  \label{eq 2}
\end{equation}
i.e. that thermodynamics of the diagonal part of the Bogoliubov Hamiltonian
and that of the PBG coincide, including the Bose-condensation which
occurs at $k=0$. This means in particular that the thermodynamic
non-equivalence between the Bogoliubov Hamiltonian and PBG, i.e.,
\begin{equation}
p^{B}\left( \beta ,\mu \leq 0\right) \neq p^{PBG}\left( \beta ,\mu \leq 0\right) ,
\label{eq 2bis}
\end{equation}
is due to \textit{non-diagonal} terms of interaction $U_{\Lambda }$.
The formal proof of (\ref{eq 2bis}) is given in Section \ref{section 2},
Lemma \ref{borne inf de p bogo}, cf. \cite{BZ98JP,BZ00}.

\hspace{0.4cm}
Note that $U_{\Lambda}$ corresponds to the interaction between bosons in the mode $k = 0$ and
those with $k \neq 0$. We aim to give an evidence that it is effective
attraction between particles with $k = 0$, which is responsible for non-conventional
(\textit{dynamical}) condensation of particles at $k=0$ for $\mu_{0} < \mu < 0$ discovered
in {{\cite{BZ-PL98}}}.
%%%%%%%%%%%%%%%%%%%%%%%%%%%%%%%%%%%%%%%%%%%%%%%%%%%%%%%%%%%%%%%%%%%%%%%%%%%%%%%%%%%%%%%%%%%%%%%%%%%%
%%%%%%%%%%%%%%%%%%%%%%%%%%%%%%%%%%%%%%%%%%%%%%%%%%%%%%%%%%%%%%%%%%%%%%%%%%%%%%%%%%%%%%%%%%%%
\begin{figure}[p]

\vspace{-24cm}

\centering
\includegraphics[width=1.9 \linewidth]{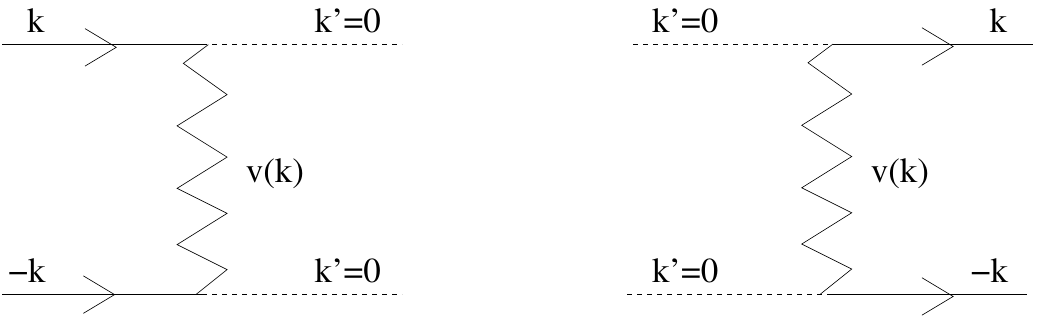}

%\end{figure}

\vspace{1.0cm}

{\bf Figure 1.}
Vertices corresponding to the non-diagonal part $U_{\Lambda }$ of the Bogoliubov Hamiltonian
for WIBG.
	
\end{figure}
%%%%%%%%%%%%%%%%%%%%%%%%%%%%%%%%%%%%%%%%%%%%%%%%%%%%%%%%%%%%%%%%%%%%%%%%%%%%%%%%%%%%%%%%%%%%%
\smallskip

\noindent
\textbf{6.3} The \textit{non-diagonal} part $U_{\Lambda}$ of the Bogoliubov Hamiltonian
(\ref{eq 1}) can be represented in term of vertices (see Figure 1). In order to
understand the r\^{o}le of non-diagonal part of the Bogoliubov Hamiltonian, one has to evaluate
the \textit{effective} interactions, which is induced by $U_{\Lambda}$ between particles
\textit{in} the zero-mode $k=0$ and between particle \textit{outside} the zero-mode $k=0$,
see Figure 2 and Figure 3.

The simplest way to calculate the corresponding
two coupling constants $g_{\Lambda ,pq}$ and $g_{\Lambda ,00}$ is to use the
Fr\"{o}hlich transformation \cite{Fro52} originally proposed to deduce the
Bardeen-Cooper-Schriffer-Bogoliubov (BCS-Bogoliubov)
model interaction in the theory of superconductivity {{\cite{Bog58}, \cite{BTSh58}}}.
This is unitary transformation of a Hamiltonian $H$:
%%%%%%%%%%%%%%%%%%%%%%%%%%%%%%%%%%%%%%%%%%%%%%%%%%%%%%%%%%%%%%%%%%%%%%%%%%%%%%%%%%%%%%%%%%%%
\begin{figure}[p]

\vspace{-20cm}

\centering
\includegraphics[width=1.8\linewidth]{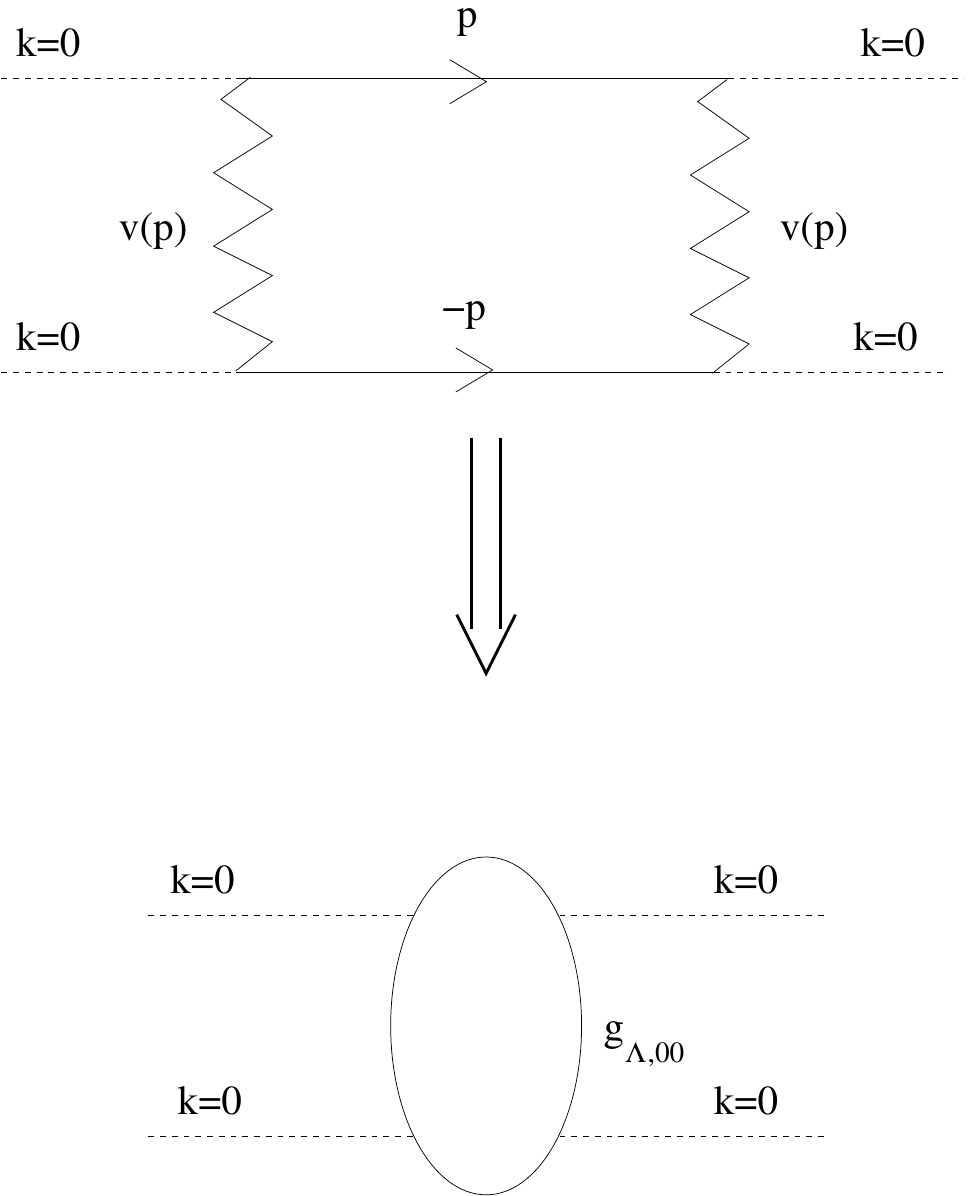}

\vspace{1.0cm}

{\bf Figure 2.}
Illustration to the effective vertex, generated by $U_{\Lambda }$, for interaction between
particles in the condensate (zero-mode).
	
\end{figure}
%%%%%%%%%%%%%%%%%%%%%%%%%%%%%%%%%%%%%%%%%%%%%%%%%%%%%%%%%%%%%%%%%%%%%%%%%%%%%%%%%%%%%%%%%%%%
%%%%%%%%%%%%%%%%%%%%%%%%%%%%%%%%%%%%%%%%%%%%%%%%%%%%%%%%%%%%%%%%%%%%%%%%%%%%%%%%%%%%%%%%%%%%
\begin{equation}
\widetilde{H}=e^{-iS}He^{iS},  \label{eq 3}
\end{equation}
with self-adjoint generator $S=S^{*}$. By developing $e^{iS}$ and $e^{-iS}$, one obtains
commutator series :
\begin{equation}
\widetilde{H}=H+i\left[ H,S\right] -\frac{1}{2}\left[ \left[ H,S\right]
,S\right] +...  \label{eq 4}
\end{equation}

In the case of the Bogoliubov Hamiltonian $H_{\Lambda }^{B}\left( \mu
\right) $, we have to search for such operator $S$ that the non-diagonal part
$U_{\Lambda }$ will be canceled producing instead two diagonal terms with \textit{vertices}
of the form:
%%%%%%%%%%%%%%%%%%%%%%%%%%%%%%%%%%%%%%%%%%%%%%%%%%%%%%%%%%%%%%%%%%%%%%%%
\begin{equation}\label{eq 4a}
\frac{g_{\Lambda ,00}}{V}\, a_{0}^{*}a_{0}^{*}a_{0}a_{0} \ \ \ {\rm{and}} \ \ \
\frac{g_{\Lambda,pq}}{V}\, a_{p}^{*}a_{-p}^{*}a_{-q}a_{q}\, .
\end{equation}
%%%%%%%%%%%%%%%%%%%%%%%%%%%%%%%%%%%%%%%%%%%%%%%%%%%%%%%%%%%%%%%%%%%%%%%%
To this end, we define the
self-adjoint operator $S$ as follows:
\begin{equation}
S:= \ \stackunder{k\in \Lambda ^{*},k\neq 0}{\sum }\left( \Phi
(k)a_{k}^{*}a_{-k}^{*}a_{0}^{2}+\overline{\Phi (k)}a_{0}^{*^{2}}a_{k}a_{-k}%
\right) ,  \label{eq 5}
\end{equation}
where $\Phi (k)$ have to be determined in such a way that to cancel $U_{\Lambda }$.
Thus, by analogy with perturbation theory, to evaluate $%
g_{\Lambda ,00}$ and $g_{\Lambda ,kq}$, we have to calculate (\ref{eq 4}) up to the
second order in $v(k)$.
%%%%%%%%%%%%%%%%%%%%%%%%%%%%%%%%%%%%%%%%%%%%%%%%%%%%%%%%%%%%%%%%%%%%%%%%%%%%%%%%%%%%%%%%%%%%%%%%%%%%
\begin{figure}[p]

\vspace{-20cm}

\centering
\includegraphics[width=1.8\linewidth]{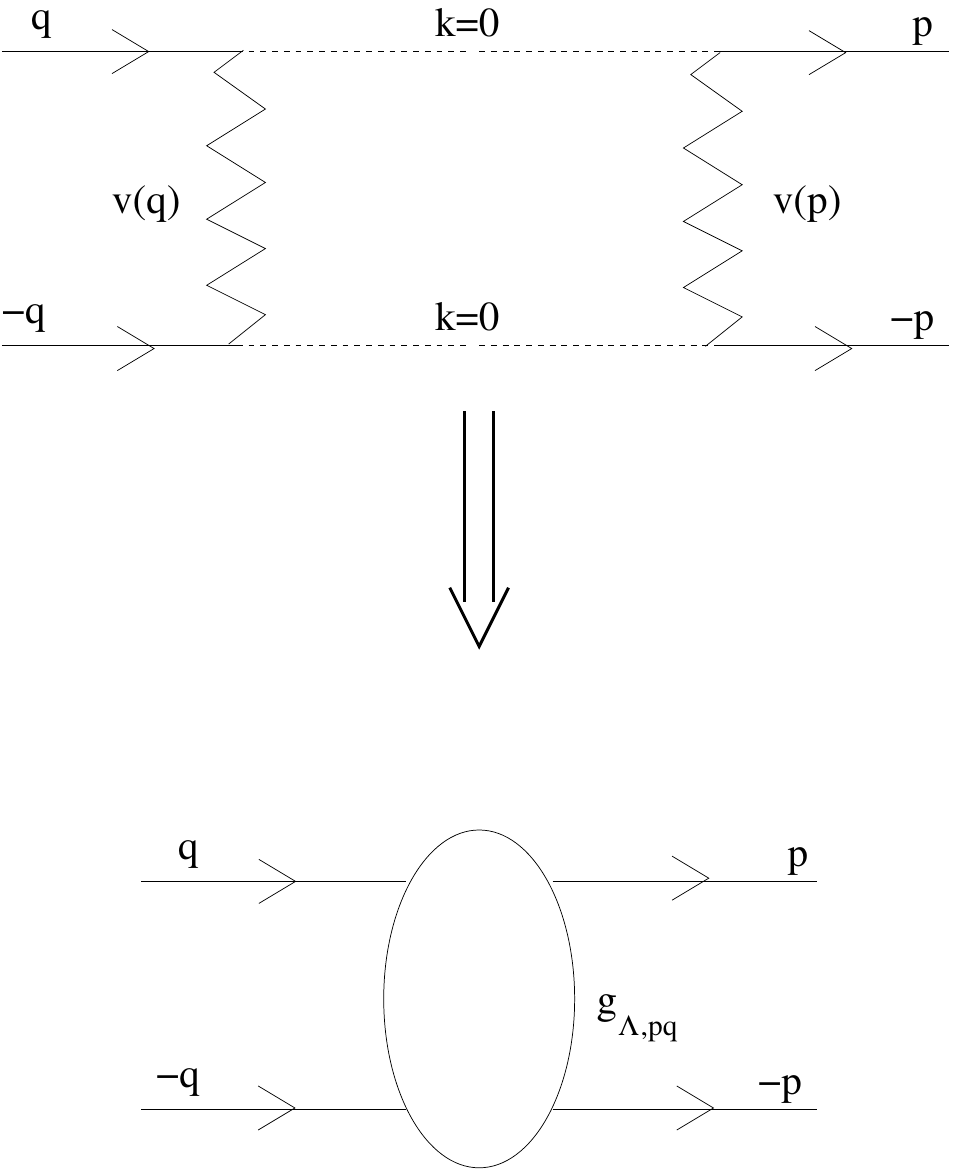}

\vspace{1.0cm}

{\bf Figure 3.}
Illustration to the effective vertex, generated by $U_{\Lambda }$, for interaction between
particles outside the zero-mode condensate.
	
\end{figure}
%%%%%%%%%%%%%%%%%%%%%%%%%%%%%%%%%%%%%%%%%%%%%%%%%%%%%%%%%%%%%%%%%%%%%%%%%%%%%%%%%%%%%%%%%%%%
\smallskip

\noindent
\textbf{6.4} Therefore, we obtain
\begin{eqnarray}
\widetilde{H}_{\Lambda }^{B} &=&e^{-iS}H_{\Lambda }^{B}\left( \mu \right)
e^{iS} =  \widetilde{H}_{\Lambda ,1}^{B}+ \widetilde{H}_{\Lambda ,2}^{B}+ ...=\nonumber \\
&=&H_{\Lambda }^{B}\left( \mu \right) +i\left[ H_{\Lambda }^{B}\left( \mu
\right) ,S\right] -\frac{1}{2}\left[ \left[ H_{\Lambda }^{B}\left( \mu
\right) ,S\right] ,S\right] +....  \label{eq 6}
\end{eqnarray}
Here the \textit{first-order} term in $v(k)$ is equal to
\begin{equation}
 \widetilde{H}_{\Lambda ,1}^{B}=U_{\Lambda }^{D}+U_{\Lambda
}+i\left[ T_{\Lambda }\left( \mu \right) ,S\right] .  \label{eq 7}
\end{equation}
Thus, to calculate $\Phi (k)$ we have the equation
\begin{equation}
U_{\Lambda }+i\left[ T_{\Lambda }\left( \mu \right) ,S\right] =0.
\label{eq 8}
\end{equation}
After direct calculations, one obtains that
\[
i\left[ T_{\Lambda }\left( \mu \right) ,S\right] =2i\stackunder{k\in \Lambda
^{*},k\neq 0}{\sum }\left\{ \varepsilon _{k}\Phi
(k)a_{-k}^{*}a_{k}^{*}a_{0}^{2}-\varepsilon _{k}\overline{\Phi (k)}%
a_{0}^{*^{2}}a_{-k}a_{k}\right\}\, .
\]
So, to satisfy (\ref{eq 8}) we have to define $\Phi (k)$ as
\begin{equation}
\Phi (k):= \frac{iv(k)}{4V\varepsilon _{k}} \, ,  \label{eq 9}
\end{equation}
for every $k\in \Lambda ^{*},k\neq 0$.

By (\ref{eq 6}) and  (\ref{eq 9}), the \textit{second-order} term in $v(k)$ is equal to
\begin{equation}
 \widetilde{H}_{\Lambda ,2}^{B}=i\left[ U_{\Lambda
}^{D}+U_{\Lambda }+\frac{i}{2}\left[ T_{\Lambda }\left( \mu \right)
,S\right] ,S\right] .  \label{eq 10}
\end{equation}
By virtue of equation (\ref{eq 8}) we obtain that
\[
U_{\Lambda }+\frac{i}{2}\left[ T_{\Lambda }\left( \mu \right) ,S\right] = \frac{1}{2}U_{\Lambda },
\]
and hence (\ref{eq 10}) yields
\begin{equation}
\widetilde{H}_{\Lambda ,2}^{B}=i\left[ U_{\Lambda
}^{D},S\right] +\frac{i}{2}\left[ U_{\Lambda },S\right] .  \label{eq 11}
\end{equation}

Note that straightforward calculations allow to check that $i\left[ U_{\Lambda}^{D},S\right]$
does not give any term with vertices of the forms (\ref{eq 4a}) for
$p,q\in \Lambda ^{*}\backslash \{0\}$, which is not surprising. Indeed, one can realise that
vertices corresponding to diagonal interaction $U_{\Lambda }^{D}$ and $S$ can not produce terms
of these forms.

\hspace{0.4cm}
On the other hand, direct calculation of the second term of (\ref{eq 11}) gives :
\begin{eqnarray}
\frac{i}{2}\left[ U_{\Lambda },S\right]  &=&\stackunder{q\in \Lambda
^{*},q\neq 0}{\stackunder{k\in \Lambda ^{*},k\neq 0}{\sum }}\frac{iv(q)}{4V}\
{\Large \{}\Phi (k)[
a_{0}^{*^{2}}a_{q}a_{-q},a_{k}^{*}a_{-k}^{*}a_{0}^{2}]   \nonumber \\
&& + \ \overline{\Phi (k)}[
a_{-q}^{*}a_{q}^{*}a_{0}^{2},a_{0}^{*^{2}}a_{-k}a_{k}]{\Large \}} \, ,
\label{eq 12}
\end{eqnarray}
with commutator
\begin{eqnarray}
[ a_{0}^{*^{2}}a_{-q}a_{q},a_{k}^{*}a_{-k}^{*}a_{0}^{2}]
&=&2a_{0}^{*^{2}}a_{0}^{2}\left( 1+a_{-k}^{*}a_{-k}+a_{k}^{*}a_{k}\right)
\delta _{q,k}  \nonumber \\
&&-2\left( 1+2a_{0}^{*}a_{0}\right) a_{k}^{*}a_{-k}^{*}a_{-q}a_{q}  \nonumber
\\
&=&- [ a_{-q}^{*}a_{q}^{*}a_{0}^{2},a_{0}^{*^{2}}a_{-k}a_{k}] ^{*}.
\label{eq 13}
\end{eqnarray}
Therefore, (\ref{eq 12}) and (\ref{eq 13}) yield only one vertex of type (\ref{eq 4a})
for effective interaction $\widetilde{H}_{\Lambda ,2}^{B,I}$ of bosons \textit{all} in the
zero-mode:
\begin{equation}
\widetilde{H}_{\Lambda ,2}^{B,I} =
-\frac{1}{V}\stackunder{k\in \Lambda ^{*},k\neq 0}{\sum }\frac{\left[ v(k)\right]
^{2}}{4 V \varepsilon _{k}} \ \
a_{0}^{*^{2}}a_{0}^{2}\, . \label{eq 14}
\end{equation}
Thus, the coupling constant $g_{\Lambda,00}$ corresponding to the effective vertex of
Figure 2 is negative and equal to
\begin{equation}
g_{\Lambda ,00}=- \stackunder{k\in \Lambda ^{*},k\neq 0}{\sum }%
\frac{\left[ v(k)\right] ^{2}}{4 V \, \varepsilon _{k}}\, ,  \label{eq 15}
\end{equation}
or, in the thermodynamic limit, to
\begin{equation}
g_{00}:= \lim_{\Lambda }g_{\Lambda ,00}=-\frac{1}{4\left( 2\pi \right)
^{3}}\int_{\Bbb{R}^{3}}d^{3}k\frac{\left[ v(k)\right] ^{2}}{\varepsilon _{k}}%
<0 \, .  \label{eq 15bisbis}
\end{equation}

\hspace{0.4cm}
On the other hand, the commutator (\ref{eq 12}) contains also terms corresponding to effective
vertex for bosons \textit{all} interacting outside the zero-mode:
\begin{eqnarray}
 \widetilde{H}_{\Lambda ,2}^{B,II} &=&\frac{1%
}{V}\stackunder{q\in \Lambda ^{*},q\neq 0}{\stackunder{k\in \Lambda
^{*},k\neq 0}{\sum }}\frac{v(k)v(q)}{8}\left( \frac{1}{\varepsilon _{k}}+%
\frac{1}{\varepsilon _{q}}\right)   \nonumber \\
&&\times \frac{\left( 1+2a_{0}^{*}a_{0}\right) }{V}%
a_{k}^{*}a_{-k}^{*}a_{-q}a_{q}\, .  \label{eq 15bis}
\end{eqnarray}

\hspace{0.4cm}
Recall that in order to find the effective BCS-interaction between two electrons mediated by the
phonons exchange, one has to project the result of the Fr\"{o}hlich
transformation on the quantum state which is vacuum for phonons, i.e. on the
\textit{fundamental} phonon-state, see e.g. {{\cite{Kv10} Ch.5,\S1, \cite{Dav73} Ch.XI,\S88}}.
Since effective interaction
between bosons with $k,p \neq 0$ is mediated by exchange via zero-mode condensate (see Figure 3.), we
project $\widetilde{H}_{\Lambda ,2}^{B,II}$ on the coherent state for the mode $k=0$, i.e.,
on the state $\left|C\right\rangle :=  \psi _{0}(c) \otimes \psi ^{\prime }$
with a given amount of condensate, where $\psi _{0}\left( c\right) \in \mathcal{F}_{0\Lambda }$
is defined by (\ref{eq 1bis}) and $\psi ^{\prime }\in \mathcal{F}_{\Lambda }^{\prime }$. Then for
condensate density $\left| c\right| ^{2} := \left\langle a_{0}^{*}a_{0}\right\rangle
_{\psi _{0}(c)}/V$, the operator of effective two-body interaction of bosons
outside of condensate is defined in $\mathcal{F}_{\Lambda }^{\prime }$ by sesquilinear forms
parameterised by $c \ $:
\begin{eqnarray*}
\left\langle C_{1}\right|  \widetilde{H}_{\Lambda ,2}^{B,II} \left| C_{2}\right\rangle  &=&
\frac{1}{V}\stackunder{q\in
\Lambda ^{*},q\neq 0}{\stackunder{k\in \Lambda ^{*},k\neq 0}{\sum }}\frac{%
v(k)v(q)}{8}\left( \frac{1}{\varepsilon _{k}}+\frac{1}{\varepsilon _{q}}%
\right) \times \\
&&\times \frac{\left( 1+2\left\langle a_{0}^{*}a_{0}\right\rangle
_{\psi _{0}(c)}\right) }{V}\left(\psi^{\prime}_{1},a_{k}^{*}a_{-k}^{*}
a_{-q}a_{q}\psi^{\prime}_{2} \right).
\end{eqnarray*}
Hence, we obtain that the coupling constant $g_{\Lambda ,pq}$ which corresponds
to effective interaction presented on Figure 3 is equal to
\begin{equation}
g_{\Lambda ,pq} =\frac{1}{8 V}v(p)v(q)\left( \frac{1}{\varepsilon _{p}}+%
\frac{1}{\varepsilon _{q}}\right) \
{\left( 1+2\left\langle a_{0}^{*}a_{0}\right\rangle
_{\psi _{0}(c)}\right)}\ .  \label{eq 16}
\end{equation}

\hspace{0.4cm}
Since in general, for the Gibbs state corresponding to Hamiltonian $H_{\Lambda }^{B}$,
\begin{equation}
\rho _{0}^{B}\left( \beta ,\mu \right) = \stackunder{\Lambda }{\lim }\frac{%
\left\langle a_{0}^{*}a_{0}\right\rangle _{H_{\Lambda }^{B}}}{V} \, ,
\label{eq 17}
\end{equation}
is density of the condensate in the mode $k=0$, we get that effective interaction
between particles outside the zero-mode is proportional to density (\ref{eq 17}) and repulsive
if $v(k) \geq 0$:
\begin{equation}
g_{pq}:= \stackunder{\Lambda }{\lim }g_{\Lambda ,pq}=\frac{v(p)v(q)}{4}%
\rho _{0}^{B}\left( \frac{1}{\varepsilon _{p}}+\frac{1}{\varepsilon _{q}}%
\right) \geq 0.  \label{eq 18}
\end{equation}
%%%%%%%%%%%%%%%%%%%%%%%%%%%%%%%%%%%%%%%%%%%%%%%%%%%%%%%%%%%%%%%%%%%%%%%%%%%%%%%%%%%%%%%%%%%%%%%%%%%%
\begin{remark}\label{rem:PH1} We recall that there is
another way to truncate the full Hamiltonian (\ref{hamiltonien total}), which is
less severe than Bogoliubov's \textit{ansatz} that gives the WIBG model (\ref{eq 1}).
This kind of truncation was proposed by Zubarev and Tserkovnikov in {\rm{\cite{ZubarevTserkovnikov}}}.
They were inspired by idea of generalisation the Bogoliubov WIBG Hamiltonian and
by studies of the BCS-Bogoliubov model, where producing the Cooper pairs coupling
plays a central r\^{o}le.
Later this idea was developed in several papers, e.g., {\rm{\cite{Luban,PuleZagrebnov1}}}
and review {\rm{\cite{PuleVerbZag08}}} with references there.
\end{remark}
%%%%%%%%%%%%%%%%%%%%%%%%%%%%%%%%%%%%%%%%%%%%%%%%%%%%%%%%%%%%%%%%%%%%%%%%%%%%%%%%%%%%%%%%%%

\hspace{0.4cm}
The Hamiltonian {truncated} according to the Zubarev-Tserkovnikov \textit{ansatz} is known
as the \textit{Pair Hamiltonian}. It has the form:
\begin{equation}
H_{\Lambda }^{Pair}=T_{\Lambda }+U_{\Lambda }^{PairD}+U_{\Lambda
}^{PairN}\equiv T_{\Lambda }+U_{\Lambda }^{Pair}.  \label{Physreport203}
\end{equation}
Here $U_{\Lambda }^{PairD}$ is the \textit{diagonal} part of the interaction
in $H_{\Lambda }^{Pair}$:
\begin{eqnarray}
U_{\Lambda }^{PairD} &=&\frac{1}{2V}\stackunder{k_{1},k_{2}\in \Lambda ^{*}}{%
\sum }v(0)a_{k_{1}}^{*}a_{k_{2}}^{*}a_{k_{2}}a_{k_{1}}  \nonumber \\
&&+\frac{1}{2V}\stackunder{k_{1}\in \Lambda ^{*},k_{2}\in \Lambda
^{*}\backslash \left\{ \pm k_{1}\right\} }{\sum }%
v(k_{1}-k_{2})a_{k_{2}}^{*}a_{k_{2}}a_{k_{1}}^{*}a_{k_{1}},
\label{Physreport204}
\end{eqnarray}
and $U_{\Lambda }^{PairN}$ is the corresponding \textit{nondiagonal} part:
\begin{equation}
U_{\Lambda }^{PairN}=\frac{1}{2V}\stackunder{k\in \Lambda ^{*},k^{\prime
}\in \Lambda ^{*}\backslash \left\{ k\right\} }{\sum }v(k-k^{\prime
})a_{k}^{*}a_{-k}^{*}a_{-k^{\prime }}a_{k^{\prime }}.  \label{Physreport205}
\end{equation}
From (\ref{Physreport204}) and (\ref{Physreport205}) it is clear that the
full interaction $U_{\Lambda }$ in (\ref{hamiltonien total}) is truncated in
the following way: first, put $q=0$ or $q=k_{1}-k_{2}$ and then $%
k_{1}=k^{\prime },$ $k_{2}=-k^{\prime },$ $q=k-k^{\prime }.$ Another evident
remark is that interaction $U_{\Lambda }^{Pair}$ contains the Bogoliubov
interacting terms $U_{\Lambda }^{BD}$  and $U_{\Lambda }^{B}$, see (\ref{eq 1}).
To obtain $U_{\Lambda }^{BD}$ one has to truncate the both of the sums in (\ref{Physreport204}) by
constraints: $k_{1}=0,$ $k_{2}\neq 0$ or $k_{1}\neq 0,$ $k_{2}=0$.
Similarity, one obtains $U_{\Lambda }^{B}$ via truncation (\ref{Physreport205}) by
conditions $k=0,$ $k^{\prime }\neq 0$ or $k\neq 0,$ $%
k^{\prime }=0$.

\hspace{0.4cm}
There are few rigorous results about the model $H_{\Lambda }^{Pair}$
(\ref{Physreport203}), see, e.g., {\rm{\cite{PuleZagrebnov1,PuleZagrebnov2,PuleVerbZag08}}}
and references there.
Inspired by the success in the rigorous study of the BCS-Bogoliubov model, the
papers {\rm{\cite{ZubarevTserkovnikov,Luban,PuleZagrebnov1,PuleZagrebnov2}}} used either the
BCS-Bogoliubov variational principle or the Approximating Hamiltonian Method.

\hspace{0.4cm}
An important conclusion of the rigorous analysis was that if $U_{\Lambda }^{PairN}$ reveals
a \textit{moderate} attractive interaction equilibrated for stability by repulsion in
$U_{\Lambda }^{PairD}$, then the model (\ref{Physreport203}) manifests a conventional and a
boson-pairs condensations which may coexist. On the other hand, in {\rm{\cite{PuleZagrebnov1}}}
Section 3.C, it was proved that by tuning parameters of attraction and repulsion one can produce
a regime, where \textit{only} pair condensate is possible with a similar to the WIBG model
discontinuous. This scenario is quite different from behaviour of condensates in WIBG,
see Section \ref{section 4} and Section \ref{section 5}, where conventional condensation
always follows \textit{after} dynamical condensation, see Figure 4 and Figure 5.
%%%%%%%%%%%%%%%%%%%%%%%%%%%%%%%%%%%%%%%%%%%%%%%%%%%%%%%%%%%%%%%%%%%%%%%%%%%%%%%%%%%%%%%%%%%%%%%%%%%%
\smallskip

\noindent
\textbf{6.5} Thus, the non-diagonal part $U_{\Lambda }$ of the Bogoliubov Hamiltonian \textit{itself}
yields an effective \textit{attraction} between particles \textit{in} the mode $k=0$ and effective
\textit{repulsion} of particles \textit{outside} of the condensate mode $k=0$. Therefore,
$U_{\Lambda }$ favours creation of the non-conventional condensation of bosons in the
mode $k=0$ due to effective attraction between them. We call it non-conventional \textit{dynamical}
condensation in contrast to conventional Bose-condensation, which is due to a simple saturation
of occupation numbers at modes $k \neq 0$.

\hspace{0.4cm}
To estimate for WIBG $H_{\Lambda }^{B}$ the effective two-body particle interaction
at zero-mode, one has to take into account besides (\ref{eq 14}) also
a direct repulsion corresponding to the last term in the
diagonal part $U_{\Lambda }^{D}$, see (\ref{eq 7}). Thus, we have to evaluate for $v(k) \geq 0$
and $d=3$ the sign of
\begin{equation}
\left( \frac{v(0)}{2}+g_{00}\right) =\frac{v(0)}{2}-\frac{1}{4\left( 2\pi
\right) ^{3}}\int_{\Bbb{R}^{3}}d^{3}k\frac{\left[ v(k)\right] ^{2}}{%
\varepsilon _{k}}\, .  \label{eq 19}
\end{equation}
In the next Sections \ref{section 2} and \ref{section 4} we show that inequality
\begin{equation}
\left( \frac{v(0)}{2}+g_{00}\right) <0,  \label{eq 20}
\end{equation}
gives a sufficient condition for existence in WIBG of the non-conventional {dynamical}
condensation at zero-mode, cf. \cite{BZ98QPL}.

\hspace{0.4cm}
In fact, in Sections \ref{section 2}-\ref{section 4} we shall rigorously show that the condition
(\ref{eq 20}), (interpreted here as a competition between $U_{\Lambda }^{D}$ and $U_{\Lambda }$)
is sufficient and necessary for non-equivalence between WIBG and
PBG and that under (\ref{eq 20}) (for dimensions $d \geq 1$) a non-conventional condensation
$\rho _{0}^{B} \neq 0$ occurs at $k=0$ in the WIBG  for negative chemical potentials
$\mu \in \left( \mu _{0},0\right) $. Notice that if
\begin{equation}
\rho _{0}^{B}\left( \beta ,\mu \right) =0,  \label{eq 21}
\end{equation}
then $g_{pq}=0$ (\ref{eq 18}), and the non-diagonal part $U_{\Lambda }$ seems to has no
influence on the thermodynamics of the WIBG. In fact, we shall show that condition
(\ref{eq 21}) implies thermodynamic equivalence between WIBG and PBG.
%%%%%%%%%%%%%%%%%%%%%%%%%%%%%%%%%%%%%%%%%%%%%%%%%%%%%%%%%%%%%%%%%%%%%%%%%%%%%%%%%%%%%%%%%%%%%%%%%%%%

\hspace{0.4cm}
In Section \ref{section 4} we show that in the limit
$\mu \rightarrow 0$, for densities $\rho> \rho^{B}_{c}(\beta):=
\rho^{B}( \beta ,\mu =0)$
(where $\rho^{B}(\beta ,\mu)$ is the particle density in WIBG), one observes
a conventional (\textit{generalised}) Bose-condensation.
Therefore, the WIBG manifests \textit{two} different types of Bose-condensation:

- the first, for $\mu_{0} < \mu < 0$, due to attraction between bosons in the mode $k=0$
(non-conventional {dynamical} condensation);

- the second, for $\mu=0$, due to the conventional saturation mechanism
(\textit{generalised} Bose-condensation \`{a} la van den Berg-Lewis-Pul\`{e} \cite{vdBLP86}).
Moreover, in this large-density regime the non-conventional condensate and the conventional BEC
in WIBG \textit{coexist}.

\hspace{0.4cm}
For the first time this was established in \cite{BZ98JP} Theorem 4.9, \cite{BZ00} Remark 2.5.
These results were summarised in review \cite{ZagBru01}.
%%%%%%%%%%%%%%%%%%%%%%%%%%%%%%%%%%%%%%%%%%%%%%%%%%%%%%%%%%%%%%%%%%%%%%%%%%%%%%%%%%%%%%%%%%%%%%%%%%%%
\smallskip

\noindent
\textbf{6.6} In conclusion of this section we note that recently the attraction
(\ref{eq 19}) and non-conventional {dynamical} condensation were discussed in the framework
of \textit{ansatz} of formation in condensate the bound atoms pairs \cite{Blag20}.

\hspace{0.4cm}
First, this idea of pairing allows to avoid a discrepancy between
experiment-assistant estimate of $g_{00}$ in \cite{BBKKKPPSY94} and calculations based on
formula (\ref{eq 15bisbis}). The renormalised (due to the pairing) integrand in (\ref{eq 15bisbis})
gives a correct estimate for the bounding energy (\ref{eq 19}).

\hspace{0.4cm}
Second, the pairing induced double-mass helium atom scaling : $m \rightarrow 2m$ in the
\textit{van Hove} structure factor fits well with experimental data. The Feynman formula
demonstrate the excellent agreement with experimental elementary excitations Bogoliubov spectrum
\cite{BBKKKPPSY94}.

\hspace{0.4cm}
Third, the \textit{coexistence} of {dynamical} condensate and the conventional BEC in the WIBG
model bolsters the basic assumption of \cite{Blag20} that helium atoms
participate in both the single atom–atom and pair–pair motions, thus possessing the independent
relaxation times for ground state of liquid helium.

\hspace{0.4cm}
Note that {dynamical} condensate is \textit{saturated}
by the value $\rho^{B}_{0}(\beta, 0)$ at $\mu =0$ (Figure 4.) and the critical total particle density
$\rho =\rho^{B}_{c}(\beta):= \rho^{PBG}_{c}(\beta) + \rho^{B}_{0}(\beta, 0)$, see Figure 5.
Then further increasing of the particle density produces conventional BEC:
$\rho -\rho^{B}_{c}(\beta) > 0$, see \cite{BZ00} Remark 2.5 and Corollary 2.6. Therefore,
at zero temperature the totality of particles are in condensate, which is a \textit{mixture} of
dynamical and conventional condensates, as in scenario assumed in \cite{Blag20}.

\hspace{0.4cm}
We shall return to discussion of these properties of condensate in the WIBG model
below in Section \ref{section 4} and Section \ref{section 5}.

\hspace{0.4cm}
To my knowledge the first attempt to understand a possible r\^{o}le of the non-conventional
condensate in superfluid $\rm{^4He}$ comes back to a very complete review \cite{VYIC}.
In contract to microscopic (cf.\textit{Cooper pairs} in the BCS-Bogoliubov theory of
superconductivity, the \textit{boson-pairing} in Remark \ref{rem:PH1} or in WIBG \cite{Blag20})
the authors claimed a boson soliton-soliton pairing in WIBG via mesoscopic Gross-Pitayevskii
description. This might be an interesting direction in understanding one-particle versus pair
condensate which could be appropriate for bosons in trap, but out of the scope of the present paper.
%%%%%%%%%%%%%%%%%%%%%%%%%%%%%%%%%%%%%%%%%%%%%%%%%%%%%%%%%%%%%%%%%%%%%%%%%%%%%%%%%%%%%%%%%%%%%%%
%\setcounter{proposition}{0}
\setcounter{equation}{0}
%%%%%%%%%%%%%%%%%%%%%%%%%%%%%%%%%%%%%%%%%%%%%%%%%%%%%%%%%%%%%%%%%%%%%%%%%%%%%%%%%%%%%%%%%%%%%%%
\section{The Weakly Imperfect Bose-Gas: set up \\ of the problem\label{section 1}}
\textbf{7.1} A pragmatic procedure for the description of the properties of superfluids,
e.g. derivation of the experimentally observed spectra, was initiated in
Bogoliubov's classical papers \cite{Bog47,Bog47a}, where he considered a
Hamiltonian with truncated interaction, giving rise to what will be called
the Bogoliubov Hamiltonian for a Weakly Imperfect Bose-Gas (WIBG) \cite{ZagBru01}.

\hspace{0.4cm}
Consider a system of bosons of mass $m$ in a cubic box $\Lambda \subset \Bbb{%
R}^{3}$ of the volume $V=L^{3}$, with \textit{periodic boundary conditions} on $\partial\Lambda$.
If $\varphi \left( x\right) $ denotes an integrable two-body interaction potential and
\begin{equation}
v\left( q\right) = \ \stackunder{\Bbb{R}^{3}}{\int }d^{3}x\varphi \left(
x\right) e^{-iqx},\text{ }q\in \Bbb{R}^{3},  \label{potentiel fourier}
\end{equation}
then its second-quantised Hamiltonian acting in the boson Fock space $\mathcal{F}_{\Lambda }$
can be written as
\begin{equation}
H_{\Lambda }= \ \stackunder{k}{\sum }\varepsilon _{k}a_{k}^{*}a_{k}+\frac{1}{2V}%
\stackunder{k_{1},k_{2},q}{\sum }v\left( q\right)
a_{k_{1}+q}^{*}a_{k_{2}-q}^{*}a_{k_{1}}a_{k_{2}}  \label{hamiltonien total}
\end{equation}
where all sums run over the set $\Lambda ^{*}$ defined by
\begin{equation}\label{dual-Lambda}
\Lambda ^{*}=\{ k\in \Bbb{R}^{3}:\alpha =1,2,3\text{, }k_{\alpha }=%
\frac{2\pi n_{\alpha }}{L}\text{ et }n_{\alpha }=0,\pm 1,\pm 2,\ldots \, . \}
\end{equation}
Here $\varepsilon _{k}=\hbar ^{2}k^{2}/2m$ is the kinetic energy, and $%
a_{k}^{\#}=\left\{ a_{k}^{*},a_{k}\right\} $ are the usual boson creation
and annihilation operator in the one-particle states
$\{\psi_{k}(x) = e^{ikx}/\sqrt{V}\}_{k\in \Lambda^{*}, x \in \Lambda}$:
\[
a_{k}^{*}:= a^{*}(\psi_{k}) = \stackunder{\Lambda}{\int}dx \, \psi_{k}(x)
a^{*}(x)\, ,
\]
where $a^{\#}\left( x\right)$ are the basic boson operators in the Fock space over
$L^{2}\left( \Bbb{R}^{3}\right)$. If one expects that BEC, which occurs
in the ideal Bose-gas for $k=0$, persists for a weak interaction $\varphi
\left( x\right) $ then, according to Bogoliubov, the most important terms in
(\ref{hamiltonien total}) should be those in which\ at least \textit{two} operators $%
a_{0}^{*},$ $a_{0}$ appear. We are thus led to consider the following
truncated Hamiltonian (the Bogoliubov Hamiltonian for WIBG, see eq.(3.81) in \cite{Bog70},Part 3):
\begin{equation}
H_{\Lambda }^{B}=T_{\Lambda }+U_{\Lambda }^{D}+U_{\Lambda }
\label{hamiltonien de Bogoliubov}
\end{equation}
where
\begin{eqnarray}
T_{\Lambda } &=&\stackunder{k\in \Lambda ^{*}}{\sum }\varepsilon
_{k}a_{k}^{*}a_{k},  \label{perfect gas without mu} \\
U_{\Lambda }^{D} &=&\frac{v\left( 0\right) }{V}a_{0}^{*}a_{0}\stackunder{%
k\in \Lambda ^{*},k\neq 0}{\sum }a_{k}^{*}a_{k}+\frac{1}{2V}\stackunder{k\in
\Lambda ^{*},k\neq 0}{\sum }v\left( k\right) a_{0}^{*}a_{0}\left(
a_{k}^{*}a_{k}+a_{-k}^{*}a_{-k}\right)  \nonumber \\
&&+\frac{v\left( 0\right) }{2V}a_{0}^{*^{2}}a_{0}^{2},
\label{interaction diag} \\
U_{\Lambda } &=&\frac{1}{2V}\stackunder{k\in \Lambda ^{*},k\neq 0}{\sum }%
v\left( k\right) \left(
a_{k}^{*}a_{-k}^{*}a_{0}^{2}+a_{0}^{*^{2}}a_{k}a_{-k}\right) .
\label{interaction non diag}
\end{eqnarray}
$H_{\Lambda }^{BD}:= \left( T_{\Lambda }+U_{\Lambda }^{D}\right) $
represents the diagonal part of the Bogoliubov Hamiltonian $H_{\Lambda }^{B}$
while $U_{\Lambda }$ represents the non-diagonal part.
%%%%%%%%%%%%%%%%%%%%%%%%%%%%%%%%%%%%%%%%%%%%%%%%%%%%%%%%%%%%%%%%%%%%%%%%%%%%%%%%%%%%%%%%%%%%%%%%%%%%
\begin{remark}
Below the following assumptions on the interaction potential $\varphi $ are
imposed:

{\rm{(A)}} $\varphi \in L^{1}\left( \Bbb{R}^{3}\right)$ and  $\varphi(x) = \varphi(\|x\|)$;

{\rm{(B)}} $k \mapsto v( k) \in \Bbb{R}$ is continuous, such that $v\left(0\right) >0$ and
$0\leq v\left( k\right) \leq v\left(0\right)$ for $k\in \Bbb{R}^{3}$.
\end{remark}
%%%%%%%%%%%%%%%%%%%%%%%%%%%%%%%%%%%%%%%%%%%%%%%%%%%%%%%%%%%%%%%%%%%%%%%%%%%%%%%%%%%%%%%%%%%%%%%%%%%%%

\hspace{0.4cm}
It is known \cite{ZagBru01} that under these (and in fact, even weaker) conditions pair potential
$\varphi$ is \textit{superstable} and hence that the grand-canonical
partition function associated with the full Hamiltonian (\ref{hamiltonien total})
\begin{equation}
\Xi _{\Lambda }\left( \beta ,\mu \right) ={\rm{Tr}}_{\mathcal{F}_{\Lambda }}\left(
e^{-\beta \left( H_{\Lambda }-\mu N_{\Lambda }\right) }\right)
\end{equation}
and the finite-volume pressure
\begin{equation}
p_{\Lambda }\left[ H_{\Lambda }\right] := p_{\Lambda }\left( \beta ,\mu
\right) =\frac{1}{\beta V}\ln \Xi _{\Lambda }\left( \beta ,\mu \right)
\end{equation}
are finite for all real $\mu $\ and all $\beta >0$.

However, it is \textit{not} true for the Bogoliubov Hamiltonian (\ref{hamiltonien de
Bogoliubov}) :
%%%%%%%%%%%%%%%%%%%%%%%%%%%%%%%%%%%%%%%%%%%%%%%%%%%%%%%%%%%%%%%%%%%%%%%%%%%%%%%%%%%%%%%%%%%%%%%%%%%%%
\begin{proposition}\label{proposition de stabilite}{\rm{\cite{AVZ92}}}
Let $\Xi _{\Lambda}^{B}\left( \beta ,\mu \right) $ be the grand-canonical partition function
associated for the Hamiltonian (\ref{hamiltonien de Bogoliubov}). Then :

$\left( a\right) $ the Bogoliubov model of WIBG is stable ( $\Xi _{\Lambda
}^{B}\left( \beta ,\mu \right) <+\infty $ ) for $\mu \leq 0$ and unstable
for $\mu >0$.

$\left( b\right) $ for $\mu \leq \mu ^{*}=-\frac{1}{2}\varphi \left(
0\right) $ the pressure
\begin{equation}
p^{B}\left( \beta ,\mu \right) = \stackunder{\Lambda }{\lim }p_{\Lambda
}\left[ H_{\Lambda }^{B}\right]   \label{eq 1.10}
\end{equation}

coincides with the pressure of the perfect Bose-gas (PBG)
\begin{equation}
p^{PBG}\left( \beta ,\mu \right) = \stackunder{\Lambda }{\lim }p_{\Lambda
}\left[ T_{\Lambda }\right] .
\end{equation}
\end{proposition}
%%%%%%%%%%%%%%%%%%%%%%%%%%%%%%%%%%%%%%%%%%%%%%%%%%%%%%%%%%%%%%%%%%%%%%%%%%%%%%%%%%%%%%%%%%%%%%%%%%%%
Note that the proof is a corollary of (\ref{inegality 2 between PBD and P0}) and (\ref{inequality 2}),
whereas the proof of (b) follows from Remark \ref{remark 2.4} and Corollary \ref{cor 2.5}.
Moreover, the following conjecture was formulated in \cite{AVZ92} :
%%%%%%%%%%%%%%%%%%%%%%%%%%%%%%%%%%%%%%%%%%%%%%%%%%%%%%%%%%%%%%%%%%%%%%%%%%%%%%%%%%%%%%%%%%%%%%%%%%%%
\begin{conjecture}\label{conjecture}\ The Bogoliubov Hamiltonian $H_{\Lambda
}^{B}$\ is exactly soluble in the sense that it is thermodynamically
equivalent, in the grand-canonical ensemble, to the PBG for all chemical
potential $\mu \leq 0$; which means precisely that
\begin{equation}
p^{B}\left( \beta ,\mu \leq 0\right) = p^{PBG}\left( \beta ,\mu \leq 0\right) .
\end{equation}
\end{conjecture}
%%%%%%%%%%%%%%%%%%%%%%%%%%%%%%%%%%%%%%%%%%%%%%%%%%%%%%%%%%%%%%%%%%%%%%%%%%%%%%%%%%%%%%%%%%%%%%%%%%%%
%\noindent
\textbf{7.2} The aim of the next Sections \ref{section 2}--\ref{section 5} is twofold :

- first, to show that the phase diagram of the Bogoliubov model (\ref
{hamiltonien de Bogoliubov}) is less trivial than it is expressed by
Conjecture \ref{conjecture};

- second, to calculate exactly $p^{B}\left( \beta ,\mu \right) $ in domain
where it does not coincide with $p^{PBG}\left( \beta ,\mu \right) $.

The results of Sections \ref{section 2}--\ref{section 5} are organized as follows: \\
In the next Section \ref{section 2}, we show that
\begin{equation}
p^{BD}\left( \beta ,\mu \leq 0\right) = \ \stackunder{\Lambda }{\lim }%
p_{\Lambda }\left[ H_{\Lambda }^{BD}\right] =p^{PBG}\left( \beta ,\mu \leq
0\right) ,
\end{equation}
i.e. that thermodynamics of the diagonal part of the Bogoliubov Hamiltonian
and that of the ideal gas coincide including Bose-condensation which occurs
at $k=0$. This means in particular that the thermodynamic non-equivalence
between Bogoliubov Hamiltonian and PBG is due to non-diagonal terms of
interaction (\ref{interaction non diag}). Moreover, in this section, we
study the Conjecture \ref{conjecture}. First, we show that for any
interaction which satisfies (A) and (B) there is a domain $\Gamma $ of the
phase diagram (plane $Q=\left( \mu \leq 0,\theta =\beta ^{-1}\geq 0\right) $
) where indeed
\begin{equation}
p^{B}\left( \beta ,\mu \leq 0\right) =p^{PBG}\left( \beta ,\mu \leq 0\right) .
\end{equation}
Then we formulate a sufficient condition on the interaction $v\left(
k\right) $ to ensure the existence of domain $D_{0}\subset Q$ where
\begin{equation}
p^{B}\left( \beta ,\mu \right) \neq p^{PBG}\left( \beta ,\mu \right) .
\end{equation}
In fact we show that this is equivalent to the statement that the system $%
H_{\Lambda }^{B}$ manifests in this domain a Bose-condensation.

Thermodynamic limit of the pressure (\ref{eq 1.10}) of the system $%
H_{\Lambda }^{B}$ in domain $D\supseteq D_{0}$ defined by
\begin{equation}
p^{B}\left( \beta ,\mu \right) \neq p^{PBG}\left( \beta ,\mu \right) .
\end{equation}
is studied on Section \ref{section 3}. We give an exact formula for $%
p^{B}\left( \beta ,\mu \right) $ showing its relation to the concept of the
Bogoliubov approximation \`{a} la Ginibre \cite{Gin68}. As a corollary we
get that $D=D_{0}$. In Section \ref{section 4} we study the breaking of the
gauge symmetry and the behaviour of the Bose-condensate, i.e., the phase
diagram of the WIBG. We reserve Section \ref{section 5} for discussions and
concluding remarks.

%%%%%%%%%%%%%%%%%%%%%%%%%%%%%%%%%%%%%%%%%%%%%%%%%%%%%%%%%%%%%%%%%%%%%%%%%%%%%%%%%%%%%%%%%%%%%%
\setcounter{proposition}{0}
\setcounter{equation}{0}
%%%%%%%%%%%%%%%%%%%%%%%%%%%%%%%%%%%%%%%%%%%%%%%%%%%%%%%%%%%%%%%%%%%%%%%%%%%%%%%%%%%%%%%%%%%%%%
\section{The Bogoliubov weakly imperfect gas versus \\ the perfect Bose-gas}\label{section 2}

\subsection{Diagonal part of the Bogoliubov Hamiltonian}

The diagonal part of the Bogoliubov Hamiltonian \vspace{1pt}$H_{\Lambda
}^{BD}=\left( T_{\Lambda }+U_{\Lambda }^{D}\right) $ (\ref{perfect gas
without mu})-(\ref{interaction diag}) can be rewritten using the
occupation-number operators for modes $k\in \Lambda
^{*},n_{k}=a_{k}^{*}a_{k} $. So, the Hamiltonian $H_{\Lambda }^{BD}\left(
\mu \right) := H_{\Lambda }^{BD}-\mu N_{\Lambda }$ has the form
\begin{eqnarray}
H_{\Lambda }^{BD}\left( \mu \right) &=&\stackunder{k\in \Lambda ^{*}}{\sum }%
\left( \varepsilon _{k}-\mu \right) a_{k}^{*}a_{k}+\frac{v\left( 0\right) }{V%
}a_{0}^{*}a_{0}\stackunder{k\in \Lambda ^{*},k\neq 0}{\sum }a_{k}^{*}a_{k}
\nonumber \\
&&+\frac{1}{2V}\stackunder{k\in \Lambda ^{*},k\neq 0}{\sum }v\left( k\right)
a_{0}^{*}a_{0}\left( a_{k}^{*}a_{k}+a_{-k}^{*}a_{-k}\right) +\frac{v\left(
0\right) }{2V}a_{0}^{*^{2}}a_{0}^{2},  \nonumber \\
&&  \label{hamiltonien de Bogoliubov que diagonal}
\end{eqnarray}
where $N_{\Lambda }= \ \stackunder{k\in \Lambda ^{*}}{\sum }a_{k}^{*}a_{k}$. If
$v\left( k\right) $ satisfies (B) then one obviously gets :
\begin{eqnarray}
H_{\Lambda }^{BD}\left( \mu \right) &=&\stackunder{k\in \Lambda ^{*}}{\sum }%
\left( \varepsilon _{k}-\mu \right) n_{k}+\frac{v\left( 0\right) }{V}%
n_{0}N_{\Lambda }-\frac{v\left( 0\right) }{2V}\left( n_{0}^{2}+n_{0}\right)
\nonumber \\
&&+\frac{1}{V}\stackunder{k\in \Lambda ^{*},k\neq 0}{\sum }v\left( k\right)
n_{0}n_{k}  \label{eqHBnd} \\
&\geq &T_{\Lambda }\left( \mu \right) := T_{\Lambda }-\mu N_{\Lambda }.
\label{ineqHBnd}
\end{eqnarray}

\begin{theorem}
$\label{pressiondiag}$Let $v\left( k\right) $ satisfy (A) and (B). Then

$\left( a\right) $ for any $\mu \leq 0$ and $\beta >0$ one has
\begin{equation}
p^{BD}\left( \beta ,\mu \right) := \stackunder{\Lambda }{\lim }%
p_{\Lambda }\left[ H_{\Lambda }^{BD}\right] =p^{PBG}\left( \beta ,\mu \right)
\text{ ,}  \label{egality between PBD and P0}
\end{equation}

$\left( b\right) $ for any $\beta >0$ one has
\[
p^{BD}\left( \beta ,\mu >0\right) =+\infty
\]
\end{theorem}
%%%%%%%%%%%%%%%%%%%%%%%%%%%%%%%%%%%%%%%%%%%%%%%%%%%%%%%%%%%%%%%%%%%%%%%%%%%%%%%%%%%%%%%%%%%%%%%%%%%%
\emph{Proof.}
$\left( a\right) $ In virtue of representation (\ref{eqHBnd}) and inequality
(\ref{ineqHBnd}) we get that partition function

\[
\Xi _{\Lambda }^{BD}\left( \beta ,\mu \right) ={\rm{Tr}}_{\mathcal{F}_{\Lambda
}}e^{-\beta H_{\Lambda }^{BD}\left( \mu \right) }\leq {\rm{Tr}}_{\mathcal{F}%
_{\Lambda }}e^{-\beta T_{\Lambda }\left( \mu \right) }=\Xi _{\Lambda
}^{PBG}\left( \beta ,\mu \right) .
\]
Hence, for any $\mu <0$%
\begin{equation}
p_{\Lambda }\left[ H_{\Lambda }^{BD}\right] \leq p_{\Lambda }\left[
T_{\Lambda }\right] .  \label{inegality 1 between PBD and P0}
\end{equation}
By (\ref{eqHBnd}), we can calculate ${\rm{Tr}}_{%
\mathcal{F}_{\Lambda }}$ in the basis of occupation-number operators :
\[
\Xi _{\Lambda }^{BD}\left( \beta ,\mu \right) =\vspace{1pt}\stackunder{%
n_{0}=0}{\stackrel{\infty }{\sum }}\left\{ e^{\left[ -\beta \left({
v\left( 0\right) }\left( n_{0}^{2}-n_{0}\right)/{2V} -\mu n_{0}\right)
\right] }\stackunder{k\in \Lambda ^{*},k\neq 0}{\prod }\left( 1-e^{\left[
-\beta \left( \varepsilon _{k}-\mu + {\left[ v\left( 0\right) +v\left(
k\right) \right] } n_{0}/{V} \right) \right] }\right) ^{-1}\right\} ,
\]
which gives estimate
\[
\Xi _{\Lambda }^{BD}\left( \beta ,\mu \right) \geq \vspace{1pt}\stackunder{%
k\in \Lambda ^{*},k\neq 0}{\prod }\left( 1-e^{\left[ -\beta \left(
\varepsilon _{k}-\mu \right) \right] }\right) ^{-1}.
\]
Therefore,
\begin{equation}
p_{\Lambda }\left[ H_{\Lambda }^{BD}\right] \geq \widetilde{p}_{\Lambda
}^{PBG}\left( \beta ,\mu \right) := \frac{1}{\beta V}\stackunder{k\in
\Lambda ^{*},k\neq 0}{\sum }\ln \left[ \left( 1-e^{\left[ -\beta \left(
\varepsilon _{k}-\mu \right) \right] }\right) ^{-1}\right]
\label{inegality 2 between PBD and P0}
\end{equation}
Note that $\widetilde{p}_{\Lambda }^{PBG}\left( \beta ,\mu \right) $ is the
pressure of an ideal Bose-gas with excluded mode $k=0$ and $\widetilde{p}%
_{\Lambda }^{PBG}\left( \beta ,\mu \right) <+\infty $ for $\mu <\stackunder{%
k\neq 0}{\inf }\varepsilon _{k}$. Hence, for any $\mu <0$ one gets
\begin{equation}
\stackunder{\Lambda }{\lim }\widetilde{p}_{\Lambda }^{PBG}\left( \beta ,\mu
\right) = \ \stackunder{\Lambda }{\lim }p_{\Lambda }\left[ T_{\Lambda }\right]
=p^{PBG}\left( \beta ,\mu \right)   \label{limit for P0 and P0 without k=0}
\end{equation}
Therefore, taking thermodynamic limit in (\ref{inegality 1 between PBD and
P0}), (\ref{inegality 2 between PBD and P0}), by (\ref{limit for P0 and P0
without k=0}) we obtain (\ref{egality between PBD and P0}) for $\mu <0$.
Then taking limit $\mu \rightarrow - 0$ one gets (\ref{egality between PBD
and P0}) for $\mu =0$.

$\left( b\right) $ Follows directly from the estimate (\ref{inegality 2
between PBD and P0}). \hfill $\square$

\begin{corollary}\label{densitediag}
Since functions $\left\{ p_{\Lambda }^{BD}\left( \beta,\mu \right) := p_{\Lambda }
\left[ H_{\Lambda }^{BD}\right] \right\}_{\Lambda \subset \Bbb{R}^{d}}$ are convex for
$\mu \leq 0$ and the limit $p^{PBG}\left( \beta ,\mu \right) $ is differentiable for $\mu <0$,
the Griffiths lemma {\rm{\cite{Grif}}} yields
\[
\stackunder{\Lambda }{\lim }\partial _{\mu }p_{\Lambda }^{BD}\left( \beta
,\mu \right) =\partial _{\mu }p^{PBG}\left( \beta ,\mu \right) \text{, }
\]
i.e, the particle-density for the system (\ref{hamiltonien de Bogoliubov que
diagonal}) coincides with that for the ideal gas :
\begin{equation}
\rho ^{BD}\left( \beta ,\mu \right) := \stackunder{\Lambda }{\lim }%
\left\langle \frac{N}{V}\right\rangle _{H_{\Lambda }^{BD}}\left( \beta ,\mu
\right) =\partial _{\mu }p^{PBG}\left( \beta ,\mu \right) :=
\rho ^{PBG}\left( \beta ,\mu \right)   \label{egality between roBD and ro0}
\end{equation}
Here $\left\langle -\right\rangle _{H_{\Lambda }}\left( \beta ,\mu \right) $
corresponds to the grand-canonical Gibbs state for Hamiltonian $H_{\Lambda }$.
Taking in (\ref{egality between roBD and ro0}) the limit $\mu \rightarrow - 0$ we extend this
equality to $\mu \in \left( -\infty ,0\right] $.
\end{corollary}

\hspace{0.4cm}
Resuming (\ref{egality between PBD and P0}) and (\ref{egality between roBD
and ro0}) we see that diagonal part of the Bogoliubov Hamiltonian $%
H_{\Lambda }^{BD}$ is \textit{thermodynamically equivalent} to $T_{\Lambda }$. The
Bose condensate in the system (\ref{hamiltonien de Bogoliubov que diagonal})
has the same properties as in the PBG. Below we show that it is
non-diagonal interaction (\ref{interaction non diag}) that makes the
essential difference between $H_{\Lambda }^{B}$ and $T_{\Lambda }$.

\subsection{Domain $\Gamma $ : $p^{B}\left( \beta ,\mu \right) = p^{PBG}\left(
\beta ,\mu \right) $}

Similar to PBG the Bogoliubov WIBG exists only for $\mu \leq 0$, see
Proposition \ref{proposition de stabilite}. In fact we can claim more.
%%%%%%%%%%%%%%%%%%%%%%%%%%%%%%%%%%%%%%%%%%%%%%%%%%%%%%%%%%%%%%%%%%%%%%%%%%%%%%%%%%%%%%%%%%%%%%%%%%%%
\begin{lemma} \label{borne inf de p bogo}
For any $\mu \leq 0$, one has
\begin{equation}
p^{PBG}\left( \beta ,\mu \right) \leq p^{B}\left( \beta ,\mu \right) .
\label{inequality 1}
\end{equation}
\end{lemma}
%%%%%%%%%%%%%%%%%%%%%%%%%%%%%%%%%%%%%%%%%%%%%%%%%%%%%%%%%%%%%%%%%%%%%%%%%%%%%%%%%%%%%%%%%%%%%%%%%%%%
\emph{Proof}:
By the {Bogoliubov convexity inequality (see, e.g., \cite{BBZKT84})}, one gets :
\begin{equation}
\frac{1}{V}\left\langle U_{\Lambda }\right\rangle _{H_{\Lambda }^{B}}\leq
p_{\Lambda }\left[ H_{\Lambda }^{BD}\right] -p_{\Lambda }\left[ H_{\Lambda
}^{B}\right] \leq \frac{1}{V}\left\langle U_{\Lambda }\right\rangle
_{H_{\Lambda }^{BD}}.  \label{inequality 2}
\end{equation}
Since $\left\langle U_{\Lambda }\right\rangle _{H_{\Lambda}^{BD}}=0$,
combining (\ref{inegality 2 between PBD and P0}), (\ref{limit for P0 and P0 without k=0}) and
(\ref{inequality 2}) we obtain in the thermodynamic limit (\ref{inequality 1}).
\hfill $\square $
%%%%%%%%%%%%%%%%%%%%%%%%%%%%%%%%%%%%%%%%%%%%%%%%%%%%%%%%%%%%%%%%%%%%%%%%%%%%%%%%%%%%%%%%%%%%%%%%%%%
\begin{remark}\label{remark 2.4}
Let $v\left( k\right) $ satisfy (A) and (B). Then
regrouping terms in (\ref{interaction diag}), (\ref{interaction non diag})
one gets
\begin{equation}
H_{\Lambda }^{B}=\widetilde{H}_{\Lambda }+\frac{1}{2V}\stackunder{k\in
\Lambda ^{*},k\neq 0}{\sum }v\left( k\right) \left(
a_{0}^{*}a_{k}+a_{-k}^{*}a_{0}\right) ^{*}\left(
a_{0}^{*}a_{k}+a_{-k}^{*}a_{0}\right) \geq \widetilde{H}_{\Lambda }\text{ },
\label{equation for stability}
\end{equation}
where
\begin{equation}
\widetilde{H}_{\Lambda }= \ \stackunder{k\in \Lambda ^{*},k\neq 0}{\sum }\left(
\varepsilon _{k}-\frac{v\left( k\right) }{2V}+\frac{v\left( 0\right) }{V}%
n_{0}\right) n_{k}+\frac{v\left( 0\right) }{2V}n_{0}^{2}-\frac{1}{2}\varphi
\left( 0\right) n_{0}.  \label{Hamiltonien pour stabilite}
\end{equation}
\end{remark}

\hspace{0.4cm}
Hence, by the Bogoliubov inequality for (\ref{equation for stability}) and for
its diagonal part (\ref{Hamiltonien pour stabilite}) we obtain in the
thermodynamic limit for $\mu \leq 0$%
\begin{equation}
p^{B}\left( \beta ,\mu \right) \leq \stackunder{\Lambda }{\lim }p_{\Lambda
}\left[ \widetilde{H}_{\Lambda }\right] = \ \stackunder{\rho _{0}\geq 0}{\sup }%
G\left( \beta ,\mu ;\rho _{0}\right) .
\label{equation for sufficient condition}
\end{equation}
Here
\begin{equation}
G\left( \beta ,\mu ;\rho _{0}\right) := \left[ -\frac{v\left( 0\right) }{%
2}\rho _{0}^{2}+\left( \mu +\frac{1}{2}\varphi \left( 0\right) \right) \rho
_{0}+p^{PBG}\left( \beta ,\mu -v\left( 0\right) \rho _{0}\right) \right] .
\label{equation de G}
\end{equation}

\begin{corollary}\label{cor 2.5} {\rm{\cite{AVZ92}}}
If $\mu \leq -\frac{1}{2}\varphi \left( 0\right) $, then $%
\stackunder{\rho _{0}\geq 0}{\sup }G\left( \beta ,\mu ;\rho _{0}\right)
=p^{PBG}\left( \beta ,\mu \right) $. Therefore, by Lemma \ref{borne inf de p
bogo} and inequality (\ref{equation for sufficient condition}) we get
\begin{equation}
p^{B}\left( \beta ,\mu \right) =p^{PBG}\left( \beta ,\mu \right) ,\text{ for }%
\Gamma _{\mu _{*}}=\left\{ \theta \geq 0,\text{ }\mu \leq -\frac{1}{2}%
\varphi \left( 0\right) := \mu _{*}\right\} .
\label{old domain for p bogo}
\end{equation}
\end{corollary}

\hspace{0.4cm}
The next statement extends the domain $\Gamma _{\mu _{*}}$.
\begin{theorem}
\label{theorem 2.2}Let $v\left( k\right) $ satisfy {\rm{(A)}} and {\rm{(B)}} and let
\begin{equation}
h\left( z,\beta \right) := z+\frac{v\left( 0\right) }{\left( 2\pi
\right) ^{3}}\stackunder{\Bbb{R}^{3}}{\int }d^{3}k\left( e^{ \beta
\left( \varepsilon _{k}+z\right) }-1\right) ^{-1},   \label{eq 2.16}
\end{equation}
for $d = 3$. Then we have
\begin{equation}
p^{B}\left( \beta ,\mu \right) = p^{PBG}\left( \beta ,\mu \right) ,\text{ for }%
\left( \theta ,\mu \right) \in \Gamma ,
\label{equation from sufficient condition}
\end{equation}
where
\begin{equation}
\Gamma =\left\{ \left( \theta ,\mu \right) :\frac{1}{2}\varphi \left(
0\right) \leq \stackunder{z\geq -\mu }{\inf }h\left( z,\beta \right)
\right\} \subset Q.  \label{domaine gamma}
\end{equation}
\end{theorem}
%%%%%%%%%%%%%%%%%%%%%%%%%%%%%%%%%%%%%%%%%%%%%%%%%%%%%%%%%%%%%%%%%%%%%%%%%%%%%%%%%%%%%%%%
\emph{Proof.}
By virtue of (\ref{inequality 1}) and (\ref{equation for sufficient
condition}), (\ref{equation de G}), the equality (\ref{equation from
sufficient condition}) is insured by
\begin{equation}
\stackunder{\rho _{0}\geq 0}{\sup }G\left( \beta ,\mu ;\rho _{0}\right)
=G\left( \beta ,\mu ;\rho _{0}=0\right) .  \label{sufficient condition2}
\end{equation}
If $\partial _{\rho _{0}}G\left( \beta ,\mu ;\rho _{0}\right) \leq 0$ or
equivalently $\frac{1}{2}\varphi \left( 0\right) \leq h\left( v\left(
0\right) \rho _{0}-\mu ,\beta \right) $ for $\rho _{0}\geq 0$, then
sufficient condition (\ref{domaine gamma}) guarantees (\ref{sufficient
condition2}) and hence, (\ref{equation from sufficient condition}).
\hfill $\square $
%%%%%%%%%%%%%%%%%%%%%%%%%%%%%%%%%%%%%%%%%%%%%%%%%%%%%%%%%%%%%%%%%%%%%%%%%%%%%%%%%%%%%%%%%%%%%%%%%%%
\begin{corollary}
\label{corrolarry fo gamma prime}Since $h\left( z,\beta \right) $ is a
convex function of $z\geq 0$ and $h\left( z,\beta \right) \geq z$, then by (%
\ref{eq 2.16}) we get
\begin{equation}
\lambda \left( \theta \right) \leq \stackunder{z\geq -\mu }{\inf }h\left(
z,\beta \right) ,  \label{condition suffisante 1}
\end{equation}
where
\begin{equation}
\lambda \left( \theta \right) := \stackunder{z\geq 0}{\inf }h\left(
z,\beta \right) .
\end{equation}
Therefore, by (\ref{condition suffisante 1}) we get a sufficient condition
independent of $\mu \leq 0$ (high-temperature domain) :
\begin{equation}
\Gamma _{\theta _{*}}=\left\{ \left( \theta ,\mu \leq 0\right) :\frac{1}{2}%
\varphi \left( 0\right) \leq \lambda \left( \theta \right) \right\} ,
\label{domaine gamma prime}
\end{equation}
which insures (\ref{equation from sufficient condition}).
\end{corollary}

\begin{remark}
Note that the inequality $h\left( z,\beta \right) \geq z$ and (\ref{domaine
gamma}), for $-\mu \geq \frac{1}{2}\varphi \left( 0\right) $ implies (\ref
{old domain for p bogo}), i.e., $\Gamma _{\mu _{*}}\subset \Gamma $. On the
other hand, (\ref{domaine gamma}) for $\mu =0$ implies (\ref{domaine gamma
prime}), i.e., $\Gamma _{\theta _{*}}\subset \Gamma $.
\end{remark}

\begin{remark}
Since $\partial _{v(0)}\lambda \left( \theta \right) \geq 0$, one can always
insure (\ref{domaine gamma prime}) for a fixed temperature $\theta $, by
increasing $v(0)$ without changing $\varphi \left( 0\right) $.
\end{remark}
%%%%%%%%%%%%%%%%%%%%%%%%%%%%%%%%%%%%%%%%%%%%%%%%%%%%%%%%%%%%%%%%%%%%%%%%%%%%%%%%%%%%%%%%%%%%%%%%%%
\begin{remark}
Note that $p^{PBG}\left( \beta =+\infty ,\mu \right) =0$ and $\lambda \left(
\theta =0\right) =0$. Therefore, at zero temperature the sufficient
condition (\ref{domaine gamma}) reduces to (\ref{old domain for p bogo}).
In fact this part of $\Gamma $ is known since  {\rm{ \cite{AVZ92}}}.
Theorem \ref{theorem 2.2} shows that Conjecture \ref{conjecture} formulated
there can be extended at least to domain $\Gamma $ (\ref{domaine gamma}).
\end{remark}

\hspace{0.4cm}
Below we show that this conjecture is not valid in the complement $%
Q\backslash \Gamma $.

\subsection{Domain $D$ : $p^{B}\left( \beta ,\mu \right) \neq p^{PBG}\left(
\beta ,\mu \right) $}

Let $\mathcal{H}_{0\Lambda }\subset L^{2}\left( \Lambda \right) $ be the
one-dimensional subspace generated by $\psi _{k=0}$, see Section 1. Then $%
\mathcal{F}_{\Lambda }\approx \mathcal{F}_{0\Lambda }\otimes \mathcal{F}%
_{\Lambda }^{\prime }$ where $\mathcal{F}_{0\Lambda }$ and $\mathcal{F}%
_{\Lambda }^{\prime }$ are the boson Fock spaces constructed out of $%
\mathcal{H}_{0\Lambda }$ and of its orthogonal complement $\mathcal{H}%
_{0\Lambda }^{\bot }$ respectively. For any complex $c\in \Bbb{C}$, we can
define in $\mathcal{F}_{0\Lambda }$ a coherent vector

\begin{equation}
\vspace{1pt}\psi _{0\Lambda }\left( c\right) =e^{-V\left| c\right| ^{2}/2}%
\stackrel{\infty }{\stackunder{k=0}{\sum }}\frac{1}{k!}\left( \sqrt{V}%
c\right) ^{k}\left( a_{0}^{*}\right) ^{k}\Omega _{0},
\end{equation}
\vspace{1pt}where $\Omega _{0}$ is the vacuum of $\mathcal{F}_{\Lambda }$.
Then $a_{0}\psi _{0\Lambda }\left( c\right) =c\psi _{0\Lambda }\left(
c\right) $.
%%%%%%%%%%%%%%%%%%%%%%%%%%%%%%%%%%%%%%%%%%%%%%%%%%%%%%%%%%%%%%%%%%%%%%%%%%%%%%%%%%%%%%%%%%%%%%%%%%%%%
\begin{definition}\label{definition of Bogo approx}
The Bogoliubov approximation to a Hamiltonian $H_{\Lambda }\left( \mu \right) := H_{\Lambda }-\mu
N_{\Lambda }$ in $\mathcal{F}_{\Lambda }$ is the operator $H_{\Lambda
}\left( c^{\#},\mu \right) $ defined in $\mathcal{F}_{\Lambda }^{\prime }$
by its sesquilinear form:
\begin{equation}
\left( \psi _{1}^{\prime },H_{\Lambda }\left( c^{\#}\right) \psi
_{2}^{\prime }\right) _{\mathcal{F}_{\Lambda }^{\prime }}=\left( \psi
_{0\Lambda }\left( c\right) \otimes \psi _{1}^{\prime },H_{\Lambda }\psi
_{0\Lambda }\left( c\right) \otimes \psi _{2}^{\prime }\right) _{\mathcal{F}%
_{\Lambda }}  \label{definition de approx de bogoliubov}
\end{equation}
for $\psi _{0\Lambda }\left( c\right) \otimes \psi_{1,2}^{\prime }$ in the form domain
of $H_{\Lambda }\left( \mu \right) $, where $c^{\#}=\left( c,\overline{c}\right) $.
\end{definition}

\hspace{0.4cm}
This formulation of the Bogoliubov approximation \cite{BZ98JP,ZagBru01} provides an
estimate for the pressure $p_{\Lambda }\left[ H_{\Lambda }^{B}\right] $ from
below which allows to refine (\ref{inequality 1}).
%%%%%%%%%%%%%%%%%%%%%%%%%%%%%%%%%%%%%%%%%%%%%%%%%%%%%%%%%%%%%%%%%%%%%%%%%%%%%%%%%%%%%%%%%%%%%%%%%%%
\begin{proposition}\label{pression inf} {\rm{\cite{AVZ92}}}
For any $\left( \theta ,\mu \right)\in Q$ one has
\begin{equation}
\stackunder{c\in \Bbb{C}}{\sup }\widetilde{p}_{\Lambda }^{B}\left( \beta
,\mu ;c^{\#}\right) \leq p_{\Lambda }\left[ H_{\Lambda }^{B}\right] ,
\label{equation inf for p bogo}
\end{equation}
where
\begin{equation}
\widetilde{p}_{\Lambda }^{B}\left( \beta ,\mu ;c^{\#}\right) := \frac{1}{%
\beta V}\ln {\rm{Tr}}_{\mathcal{F}_{\Lambda }^{\prime }}e^{-\beta H_{\Lambda
}^{B}\left( c^{\#},\mu \right) }.  \label{defintion of p de c}
\end{equation}
\end{proposition}
%%%%%%%%%%%%%%%%%%%%%%%%%%%%%%%%%%%%%%%%%%%%%%%%%%%%%%%%%%%%%%%%%%%%%%%%%%%%%%%%%%%%%%%%%%%%%%%%%%
\begin{remark}
By Definition \ref{definition of Bogo approx} we get from (\ref{hamiltonien
de Bogoliubov})-(\ref{interaction non diag}) that
\begin{eqnarray}
H_{\Lambda }^{B}\left( c^{\#},\mu \right)  &=&\stackunder{k\in \Lambda
^{*},k\neq 0}{\sum }\left[ \varepsilon _{k}-\mu +v\left( 0\right) \left|
c\right| ^{2}\right] a_{k}^{*}a_{k}  \nonumber \\
&&+\stackunder{k\in \Lambda ^{*},k\neq 0}{\frac{1}{2}\sum }v\left( k\right)
\left| c\right| ^{2}\left[ a_{k}^{*}a_{k}+a_{-k}^{*}a_{-k}\right]   \nonumber
\\
&&+\stackunder{k\in \Lambda ^{*},k\neq 0}{\frac{1}{2}\sum }v\left( k\right)
\left[ c^{2}a_{k}^{*}a_{-k}^{*}+\overline{c}^{2}a_{k}a_{-k}\right]
\nonumber \\
&&-\mu \left| c\right| ^{2}V+\frac{1}{2}v\left( 0\right) \left| c\right|
^{4}V.  \label{hamiltonien de Bogoliubov sans fluctuation}
\end{eqnarray}
Therefore, after diagonalization one can calculate (\ref{defintion of p de c})
in the explicit form:
\begin{equation}
\begin{array}{l}
\widetilde{p}_{\Lambda }^{B}\left( \beta ,\mu ;c^{\#}\right) =\xi _{\Lambda
}\left( \beta ,\mu ;x\right) +\eta _{\Lambda }\left( \mu ;x\right) , \\
\xi _{\Lambda }\left( \beta ,\mu ;x\right) =\frac{1}{\beta V}\stackunder{%
k\in \Lambda ^{*},k\neq 0}{\sum }\ln \left( 1-e^{-\beta E_{k}}\right) ^{-1},
\\
\eta _{\Lambda }\left( \mu ;x\right) =-\frac{1}{2V}\stackunder{k\in \Lambda
^{*},k\neq 0}{\sum }\left( E_{k}-f_{k}\right) +\mu x-\frac{1}{2}v\left(
0\right) x^{2},
\end{array}
\label{eq 2.28}
\end{equation}
where $E_{k}$ and $f_{k}$ are functions of $x=\left| c\right| ^{2}\geq 0$
and $\mu \leq 0$ :
\begin{equation}
\begin{array}{l}
f_{k}=\varepsilon _{k}-\mu +x\left[ v\left( 0\right) +v\left( k\right)
\right] , \\
h_{k}=xv\left( k\right) , \\
E_{k}=\sqrt{f_{k}^{2}-h_{k}^{2}}.
\end{array}
\label{eq 2.29}
\end{equation}
\end{remark}
%%%%%%%%%%%%%%%%%%%%%%%%%%%%%%%%%%%%%%%%%%%%%%%%%%%%%%%%%%%%%%%%%%%%%%%%%%%%%%%%%%%%%%%%%%%%%%

\hspace{0.4cm}
Now the strategy for localisation of domain $D$ gets clear: by
virtue of (\ref{inequality 1}) and (\ref{equation inf for p bogo}) one has
to $\left( \theta ,\mu \right) \in Q$ such that
\begin{equation}
p^{PBG}\left( \beta ,\mu \right) <\stackunder{\Lambda }{\lim }\left[
\stackunder{c\in \Bbb{C}}{\sup }\widetilde{p}_{\Lambda }^{B}\left( \beta
,\mu ;c^{\#}\right) \right] .  \label{eq 2.30}
\end{equation}
%%%%%%%%%%%%%%%%%%%%%%%%%%%%%%%%%%%%%%%%%%%%%%%%%%%%%%%%%%%%%%%%%%%%%%%%%%%%%%%%%%%%%%%%%%%%%%%%%%
\begin{proposition}\label{proposition P(x) inf} {\rm{\cite{AVZ92}}}
Let $v\left( k\right) $ satisfy {\rm{(A), (B)}} and
\begin{equation}
v(0)\geq \frac{1}{2\left( 2\pi \right) ^{3}}\int_{\Bbb{R}^{3}}d^{3}k\frac{%
\left[ v(k)\right] ^{2}}{\varepsilon _{k}}.
\label{condition sur le potentiel}
\end{equation}
Then, cf. (\ref{inegality 2 between PBD and P0}),
\[
\stackunder{c\in \Bbb{C}}{\sup }\widetilde{p}_{\Lambda }^{B}\left( \beta
,\mu ;c^{\#}\right) =\widetilde{p}_{\Lambda }^{B}\left( \beta ,\mu ;0\right)
=\widetilde{p}_{\Lambda }^{PBG}\left( \beta ,\mu \right) .
\]
Therefore, in the thermodynamic limit (see (\ref{limit for P0 and P0 without
k=0})) we get
\begin{equation}
\stackunder{\Lambda }{\lim }\left[ \stackunder{c\in \Bbb{C}}{\sup }%
\widetilde{p}_{\Lambda }^{B}\left( \beta ,\mu ;c^{\#}\right) \right]
=p^{PBG}\left( \beta ,\mu \right) .  \label{limit de px qd trivial}
\end{equation}
\end{proposition}
%%%%%%%%%%%%%%%%%%%%%%%%%%%%%%%%%%%%%%%%%%%%%%%%%%%%%%%%%%%%%%%%%%%%%%%%%%%%%%%%%%%%%%%%%%%%%%%%%%
\begin{lemma}
\label{lemma pour etha}Let $v\left( k\right) $ satisfy {\rm{(A), (B)}} and
\begin{equation}
{\rm{(C):}} \hspace{1cm} v(0)<\frac{1}{2\left( 2\pi \right) ^{3}}
\int_{\Bbb{R}^{3}}d^{3}k\frac{\left[v(k)\right] ^{2}}{\varepsilon _{k}}\, .
\label{condition sur le potentiel 2}
\end{equation}
Then, there is $\mu _{0}<0$ such that
\begin{equation}
\stackunder{\Lambda }{\lim }\left( \stackunder{x\geq 0}{\sup }\eta _{\Lambda
}\left( \mu ;x\right) \right) =\eta \left( \mu ;\overline{x}\left( \mu
\right) >0\right) >0\text{ for }\mu \in \left( \mu _{0},0\right] .
\label{equation pour etha}
\end{equation}
\end{lemma}
\emph{Proof.}
By the explicit formulae (\ref{eq 2.28}) and (\ref{eq 2.29}) we readily get
that for $\mu \leq 0$:

$\left( a\right) $ $\eta _{\Lambda }\left( \mu ;x=0\right) =0$ and $\eta
_{\Lambda }\left( \mu ;x\right) \leq const - v\left( 0\right)x^{2}/{2}; $

$\left( b\right) $ $\partial _{x}\eta _{\Lambda }\left( \mu ;x=0\right) =\mu
$ and
\[
\partial _{x}^{2}\eta _{\Lambda }\left( \mu ;x=0\right) =\frac{1}{2V}%
\stackunder{k\in \Lambda ^{*},k\neq 0}{\sum }\frac{\left[ v(k)\right] ^{2}}{%
\left( \varepsilon _{k}-\mu \right) }-v\left( 0\right) .
\]
Since
\[
\stackunder{\Lambda }{\lim }\frac{1}{2V}\stackunder{k\in \Lambda ^{*},k\neq 0%
}{\sum }\frac{\left[ v(k)\right] ^{2}}{\left( \varepsilon _{k}-\mu \right) }=%
\frac{1}{2\left( 2\pi \right) ^{3}}\int_{\Bbb{R}^{3}}d^{3}k\frac{\left[
v(k)\right] ^{2}}{\left( \varepsilon _{k}-\mu \right) },
\]
the condition (\ref{condition sur le potentiel 2}) implies the existence of $%
\widetilde{\mu }_{0}<0$ such that
\[
\stackunder{\Lambda }{\lim } \partial _{x}^{2}\eta _{\Lambda }\left(
\mu >\widetilde{\mu }_{0};x=0\right) >0\text{ }.
\]
By virtue of $\left( a\right) $, $\left( b\right) $, and $\stackunder{%
\Lambda }{\lim } \partial _{x}\eta _{\Lambda }\left( \mu =0;x=0\right)
 =0$ this means that
\begin{equation}
\stackunder{\Lambda }{\lim }\left( \stackunder{x\geq 0}{\sup }\eta _{\Lambda
}\left( \mu =0;x\right) \right) =\eta \left( \mu =0;\overline{x}\left( \mu
=0\right) >0\right) >0\text{ .}  \label{equation 2.36}
\end{equation}
Therefore, by continuity of (\ref{equation 2.36}) on the interval $\left(
\widetilde{\mu }_{0},0\right] $ it follows the existence of $\mu _{0}$ : $%
\widetilde{\mu }_{0}\leq \mu _{0}<0,$ such that one has (\ref{equation pour etha}).
\hfill $\square$
%%%%%%%%%%%%%%%%%%%%%%%%%%%%%%%%%%%%%%%%%%%%%%%%%%%%%%%%%%%%%%%%%%%%%%%%%%%%%%%%%%%%%%%%%%%%%%%%%%
\begin{theorem}
\label{th for not equality with P0}Let $v\left( k\right) $ satisfy {\rm{(A), (B)}}
and {\rm{(C)}}. Then, for any $\mu \in \left( \mu _{0},0\right] $, there is $\theta
_{0}\left( \mu \right) >0$ such that one has (see Fig. 1) :
\begin{equation}
p^{PBG}\left( \beta ,\mu \right) <p^{B}\left( \beta ,\mu \right) ,\text{ in }%
D_{0}=\left\{ \left( \theta ,\mu \right) :\mu _{0}<\mu \leq 0,0\leq \theta
<\theta _{0}\left( \mu \right) \right\} ,
\label{eq for not equality with P0}
\end{equation}
where $\mu _{0}$ is defined in Lemma \ref{lemma pour etha} and domain $D_{0}$
coincides in fact with
\[
D_{0}:= \left\{ \left( \theta ,\mu \right) :\text{ }\stackunder{\Lambda
}{\lim }\text{ }\stackunder{c\in \Bbb{C}}{\sup }\widetilde{p}_{\Lambda
}^{B}\left( \beta ,\mu ;c^{\#}\right) >p^{PBG}\left( \beta ,\mu \right)
\right\} .
\]
\end{theorem}
\emph{Proof.}
First we note that by (\ref{eq 2.28})-(\ref{eq 2.29}) one has $\xi _{\Lambda
}\left( \beta ,\mu ;x=0\right) =\widetilde{p}_{\Lambda }^{PBG}\left( \beta,\mu \right) $ and
\begin{equation}
\begin{array}{c}
\left( i\right) \text{ }\partial _{x}\xi _{\Lambda }\left( \beta ,\mu
;x\right) \leq 0\text{ and }\stackunder{x\rightarrow +\infty }{\lim }\xi
_{\Lambda }\left( \beta ,\mu ;x\right) =0,\text{ for any }\Lambda ; \\
\left( ii\right) \text{ }\partial _{\theta }\xi _{\Lambda }\left( \beta ,\mu
;x\right) \geq 0\text{ and }\stackunder{\theta \rightarrow 0}{\lim }\xi
_{\Lambda }\left( \beta ,\mu ;x\right) =0,\text{ for any }\Lambda .
\end{array}
\label{eq 2.38}
\end{equation}
Next, by Lemma \ref{lemma pour etha} for $\mu =\mu _{0}<0$ we have
\begin{equation}
\stackunder{\Lambda }{\lim }\left( \stackunder{x\geq 0}{\sup }\eta _{\Lambda
}\left( \mu _{0};x\right) \right) =\eta \left( \mu _{0};0\right) =\eta
\left( \mu _{0};\overline{x}\left( \mu _{0}\right) >0\right) =0\text{ .}
\label{equation pour etha 1}
\end{equation}
Hence, according to (\ref{eq 2.38}) and (\ref{equation pour etha 1}) one
obtains:
\begin{equation}
\begin{array}{c}
\left( iii\right) \text{ }\theta >0:\text{ }\stackunder{\Lambda }{\lim }%
\left[ \stackunder{c\in \Bbb{C}}{\sup }\widetilde{p}_{\Lambda }^{B}\left(
\beta ,\mu _{0};c^{\#}\right) \right] = \ \stackunder{x\geq 0}{\sup }\left[ \xi
\left( \beta ,\mu _{0};x\right) +\eta \left( \mu _{0};x\right) \right]  \\
\text{ }=\widetilde{p}^{B}\left( \beta ,\mu _{0};c^{\#}=0\right)
=p^{PBG}\left( \beta ,\mu \right) ,
\end{array}
\label{eq 2.40}
\end{equation}
and by (\ref{eq 2.38}), $\left( ii\right) $ and (\ref{equation pour etha 1})
we get:
\[
\begin{array}{l}
\left( i\nu \right) \text{ }\theta =0:\text{ }\stackunder{\Lambda }{\lim }%
\left[ \stackunder{c\in \Bbb{C}}{\sup }\widetilde{p}_{\Lambda }^{B}\left(
\beta =\infty ,\mu _{0};c^{\#}\right) \right] =\widetilde{p}^{B}\left( \beta
=\infty ,\mu _{0};c^{\#}=0\right)  \\
\text{ }=\widetilde{p}^{B}\left( \beta =\infty ,\mu _{0};c^{\#}\right) |%
\stackunder{\left| c\right| ^{2}=\overline{x}\left( \mu _{0}\right) >0}{}=0.
\end{array}
\]

\hspace{0.4cm}
Now by (\ref{eq 2.28}), (\ref{eq 2.38}) and Lemma \ref{lemma pour etha} one
gets that for $\mu _{0}<\mu \leq 0$
\begin{equation}
\stackunder{\Lambda }{\lim }\left[ \stackunder{c\in \Bbb{C}}{\sup }%
\widetilde{p}_{\Lambda }^{B}\left( \beta ,\mu >\mu _{0};c^{\#}\right)
\right] \geq \eta \left( \mu >\mu _{0};\overline{x}\left( \mu \right)
>0\right) >0.  \label{eq 2.41}
\end{equation}
Since by (\ref{eq 2.38}), $\left( ii\right) $ the pressure $p^{PBG}\left(
\beta ,\mu \leq 0\right) $ is monotonously decreasing for $\theta \searrow 0$%
, there is a temperature $\theta _{0}\left( \mu \right) \ $such that for $%
\theta <\theta _{0}\left( \mu >\mu _{0}\right) $ we get from (\ref{eq 2.41})
\begin{eqnarray}
p^{PBG}\left( \beta >\beta _{0}\left( \mu \right) ,\mu >\mu _{0}\right)
&<&\eta \left( \mu >\mu _{0};\overline{x}\left( \mu \right) >0\right)
\label{e.q.2.42}\\
&<&\stackunder{\Lambda }{\lim }\left[ \stackunder{c\in \Bbb{C}}{\sup }%
\widetilde{p}_{\Lambda }^{B}\left( \beta >\beta _{0}\left( \mu \right) ,\mu
>\mu _{0};c^{\#}\right) \right] \nonumber  .
\end{eqnarray}
Then (\ref{equation inf for p bogo}) and (\ref{e.q.2.42}) imply (\ref{eq for
not equality with P0}) for $\left( \theta ,\mu \right) \in
D_{0}$. \hfill $\square$
%%%%%%%%%%%%%%%%%%%%%%%%%%%%%%%%%%%%%%%%%%%%%%%%%%%%%%%%%%%%%%%%%%%%%%%%%%%%%%%%%%%%%%%%%%%%%%%
\begin{corollary}
\label{corollary 2.17}Let
\begin{equation}
D:= \left\{ \left( \theta ,\mu \right) :p^{B}\left( \beta ,\mu \right)
>p^{PBG}\left( \beta ,\mu \right) \right\} .  \label{definition de D}
\end{equation}
Then by (\ref{equation inf for p bogo}) and (\ref{eq for not equality with
P0}) one obviously gets
\[
D\supseteq D_{0}=\left\{ \left( \theta ,\mu \right) :\mu _{0}<\mu \leq
0,0\leq \theta <\theta _{0}\left( \mu \right) \right\} .
\]
Here $\mu _{0}<0$ is defined in Lemma \ref{lemma pour etha} and $\theta
_{0}\left( \mu \right) $ in Theorem \ref{th for not equality with P0}.
\end{corollary}
%%%%%%%%%%%%%%%%%%%%%%%%%%%%%%%%%%%%%%%%%%%%%%%%%%%%%%%%%%%%%%%%%%%%%%%%%%%%%%%%%%%%%%%%%%%%%%%%%
\begin{remark}
\label{remark 2.18}The condition (\ref{condition sur le potentiel 2}) is
sufficient to guarantee that $\mu _{0}<0$, i.e., $D\supseteq D_{0}\neq
\left\{ \emptyset \right\} $. On the other hand, the contrary condition (\ref
{condition sur le potentiel}) implies only the triviality (\ref{limit de px
qd trivial}) of the lower bound (\ref{equation inf for p bogo}) for $%
p^{B}\left( \beta ,\mu \right) $ but not $D=\left\{ \emptyset \right\} $,
see Lemma \ref{borne inf de p bogo}.
\end{remark}

\hspace{0.4cm}
Therefore, for the moment we do not know whether condition (C) is necessary
for $D\neq \left\{ \emptyset \right\} $, we postpone the answer to this
question to Section 3. Here we discuss a relation of the conditions (\ref
{condition sur le potentiel}),(\ref{condition sur le potentiel 2}) which
result from the rather restricted analysis of convexity and monotonicity of
the $\widetilde{p}_{\Lambda }^{B}\left( \beta ,\mu ;c^{\#}\right) $ in the
vicinity of $x=0$ and the condition (\ref{old domain for p bogo}) which
gives triviality of the upper bound (\ref{equation for sufficient condition}%
) for $p^{B}\left( \beta ,\mu \right) $ for all temperatures.
%%%%%%%%%%%%%%%%%%%%%%%%%%%%%%%%%%%%%%%%%%%%%%%%%%%%%%%%%%%%%%%%%%%%%%%%%%%%%%%%%%%%%%%%%%%%%%%%%%%
\begin{remark}
Let $v\left( k\right) $ satisfy {\rm{(A), (B)}} and {\rm{(C)}}. Then there is $\widetilde{%
\mu }_{0}<0$ such that for $\mu \leq \widetilde{\mu }_{0}$ one has
\begin{equation}
v\left( 0\right) \geq \frac{1}{2\left( 2\pi \right) ^{3}}\int_{\Bbb{R}^{3}}%
\frac{\left[ v\left( k\right) \right] ^{2}}{\left( \varepsilon _{k}-\mu
\right) }d^{3}k,  \label{eq 2.44}
\end{equation}
and by consequence $\partial _{x}^{2}\eta \left( \mu \leq \widetilde{\mu }%
_{0};x=0\right) \leq 0$, see Lemma \ref{lemma pour etha}. One can represent
the inequality (\ref{eq 2.44}) as
\begin{equation}
\int_{\Bbb{R}^{3}}\frac{d^{3}k}{\left( 2\pi \right) ^{3}}v\left( k\right)
\left\{ \frac{v\left( k\right) }{2\left( \varepsilon _{k}-\mu \right) }-%
\frac{v\left( 0\right) }{\varphi \left( 0\right) }\right\} \leq 0.
\label{eq 2.45}
\end{equation}
Since by (B) and $\mu \leq 0$%
\[
\frac{v\left( k\right) }{2\left( \varepsilon _{k}-\mu \right) }\leq \frac{%
v\left( 0\right) }{\left( -2\mu \right) }\text{ },
\]
the condition $\mu <-\frac{1}{2}\varphi \left( 0\right) := \mu _{*}$ (%
\ref{old domain for p bogo}) implies (\ref{eq 2.45}), i.e., $\mu _{*}\leq
\widetilde{\mu }_{0}$. Therefore, a local convexity condition (\ref{eq 2.44}) is intimately
related to condition insuring $p^{B}\left(
\beta ,\mu \right) =p^{PBG}\left( \beta ,\mu \right) $. In particular, one has
to note that for condition (\ref{condition sur le potentiel}) the inequality
(\ref{eq 2.44}) is valid for any $\mu <0$.
\end{remark}

\hspace{0.4cm}
We conclude this section by a new simple and important for below
characterisation of domain $D$ (cf. (\ref{definition de D})).
%%%%%%%%%%%%%%%%%%%%%%%%%%%%%%%%%%%%%%%%%%%%%%%%%%%%%%%%%%%%%%%%%%%%%%%%%%%%%%%%%%%%%%%%%%%%%%%%%%%
\begin{theorem}
\label{theorem 2.20}Let
\begin{equation}
\rho _{0}^{B}\left( \beta ,\mu \right) := \stackunder{\Lambda }{\lim }%
\left\langle \frac{a_{0}^{*}a_{0}}{V}\right\rangle _{_{H_{\Lambda
}^{B}}}\left( \beta ,\mu \right)
\label{definition de densite de condensant}
\end{equation}
be density of the Bose-condensate for the Bogoliubov Hamiltonian (\ref
{hamiltonien de Bogoliubov}). Then
\begin{equation}
D=\left\{ \left( \theta ,\mu \right) \in Q:\text{ }\rho _{0}^{B}\left( \beta
,\mu \right) >0\right\} .  \label{definition de Dbis}
\end{equation}
\end{theorem}
\emph{Proof.}
Put
\begin{equation}
\widehat{H}_{\Lambda }^{B}:= H_{\Lambda }^{B}+\frac{1}{2}\varphi \left(
0\right) a_{0}^{*}a_{0}.  \label{eq 2.48}
\end{equation}
Then by Remark \ref{remark 2.4} we get
\begin{equation}
\stackunder{\Lambda }{\lim }p_{\Lambda }\left[ \widehat{H}_{\Lambda
}^{B}\right] \leq \stackunder{\rho _{0}\geq 0}{\sup }\left\{ G\left( \rho
_{0},\mu \right) -\frac{1}{2}\varphi \left( 0\right) \rho _{0}\right\}
=p^{PBG}\left( \beta ,\mu \right) .  \label{eq 2.49}
\end{equation}
By the Bogoliubov inequality for $H_{\Lambda }^{B}$ and $\widehat{H}_{\Lambda
}^{B}$ one has
\begin{equation}
p_{\Lambda }\left[ H_{\Lambda }^{B}\right] -\frac{\varphi \left( 0\right) }{2%
}\left\langle \frac{a_{0}^{*}a_{0}}{V}\right\rangle _{_{H_{\Lambda
}^{B}}}\leq p_{\Lambda }\left[ \widehat{H}_{\Lambda }^{B}\right] .
\label{eq 2.50}
\end{equation}
Now by virtue of (\ref{inequality 1}), (\ref{eq 2.49}) and (\ref{eq 2.50})
we get in the thermodynamic limit that
\[
p^{PBG}\left( \beta ,\mu \right) -\frac{\varphi \left( 0\right) }{2}\rho
_{0}^{B}\left( \beta ,\mu \right) \leq p^{B}\left( \beta ,\mu \right) -\frac{%
\varphi \left( 0\right) }{2}\rho _{0}^{B}\left( \beta ,\mu \right) \leq
p^{PBG}\left( \beta ,\mu \right) .
\]
Therefore, $p^{B}\left( \beta ,\mu \right) =p^{PBG}\left( \beta ,\mu \right) $
if and only if $\rho _{0}^{B}\left( \beta ,\mu \right) =0$, which gives (\ref
{definition de Dbis}). \hfill $\square$
%%%%%%%%%%%%%%%%%%%%%%%%%%%%%%%%%%%%%%%%%%%%%%%%%%%%%%%%%%%%%%%%%%%%%%%%%%%%%%%%%%%%%%%%%%%%%%%%%%
\begin{remark}
The fact that $p^{B}\left( \beta ,\mu \right) \neq p^{PBG}\left( \beta ,\mu
\right) $ only when $\rho _{0}^{B}\left( \beta ,\mu \right) \neq 0$ is very
similar to what is known since Bogoliubov model for superfluidity {\rm{\cite{ZagBru01}}} Sec.2.2.
An essential difference is that in the Bogoliubov model the gapless spectrum
occurs for positive chemical potential $\mu =v\left( 0\right) \rho $, where
the system corresponding to the Bogoliubov Hamiltonian for WIBG is unstable,
for further discussion see {\rm{\cite{ZagBru01,BZ08}}} and Section \ref{section 5}.
\end{remark}

%%%%%%%%%%%%%%%%%%%%%%%%%%%%%%%%%%%%%%%%%%%%%%%%%%%%%%%%%%%%%%%%%%%%%%%%%%%%%%%%%%%%%%%%%%%%%%
\setcounter{proposition}{0}
\setcounter{equation}{0}
%%%%%%%%%%%%%%%%%%%%%%%%%%%%%%%%%%%%%%%%%%%%%%%%%%%%%%%%%%%%%%%%%%%%%%%%%%%%%%%%%%%%%%%%%%%%%%
\section{Exactness of the Bogoliubov approximation}\label{section 3}
%%%%%%%%%%%%%%%%%%%%%%%%%%%%%%%%%%%%%%%%%%%%%%%%%%%%%%%%%%%%%%%%%%%%%%%%%%%%%%%%%%%%%%%%%%%%%%
Since the pressure $p^{B}\left( \beta ,\mu \right) \neq p^{PBG}\left( \beta
,\mu \right) $ only in domain $D$, where the Bose condensate $\rho
_{0}^{B}\left( \beta ,\mu \right) >0$, the aim of this section is to
identify $p^{B}\left( \beta ,\mu \right) $ in this domain. Below we shall
show that
\begin{eqnarray}
p^{B}\left( \beta ,\mu \right)  &=&\stackunder{\Lambda }{\lim }\left[
\stackunder{c\in \Bbb{C}}{\sup }\widetilde{p}_{\Lambda }^{B}\left( \beta
,\mu ;c^{\#}\right) \right]   \nonumber \\
&=&\stackunder{c\in \Bbb{C}}{\sup }\widetilde{p}^{B}\left( \beta ,\mu
;c^{\#}\right) ,  \label{equality of p de c et pbogo}
\end{eqnarray}
and that in fact (cf. (\ref{eq for not equality with P0}), (\ref{definition
de D})) one has
\begin{equation}
D=D_{0}.  \label{eq  3.2}
\end{equation}
Therefore, the condition (C) is necessary for $D\neq \left\{ \emptyset
\right\} $, cf. Remark \ref{remark 2.18}. By definition of $\widetilde{p}%
^{B}\left( \beta ,\mu ;c^{\#}\right) $, see (\ref{equation inf for p bogo})-(\ref{eq 2.28}),
the statement (\ref{equality of p de c et pbogo}) means that
the Bogoliubov approximation for the WIBG is exact. Since
$\widetilde{p}_{\Lambda }^{B}\left( \beta ,\mu ;c^{\#}\right) $ is known explicitly the
statement (\ref{equality of p de c et pbogo}) gives exact solution of this model on the
\textit{thermodynamic level}.

\hspace{0.4cm}
According to results of Section \ref{section 2} it is non-diagonal part $U_{\Lambda }$ (\ref
{interaction non diag}) of the Bogoliubov Hamiltonian (\ref{hamiltonien de
Bogoliubov}) that makes that $p^{B}\left( \beta ,\mu \right) \neq
p^{PBG}\left( \beta ,\mu \right) $ in domain $D\neq \left\{ \emptyset \right\}$
for (\ref{condition sur le potentiel 2}). Since for condition (C) (\ref{condition sur le potentiel 2})
the interaction $U_{\Lambda }$ is known to be effectively attractive (\ref{eq 20}), to prove
(\ref{equality of p de c et pbogo}) we use the Approximation Hamiltonian Method originally
invented for quantum systems with attractive interactions, see e.g. {\rm{\cite{BBZKT84}}}.
%%%%%%%%%%%%%%%%%%%%%%%%%%%%%%%%%%%%%%%%%%%%%%%%%%%%%%%%%%%%%%%%%%%%%%%%%%%%%%%%%%%%%%%%%%%%%
\begin{remark}
This method was adapted by Ginibre {\rm{\cite{Gin68}}} to prove the exactness of
the Bogoliubov approximation for non-ideal Bose-gas (\ref{hamiltonien total}%
) with superstable interaction, which is the case if $v\left( q\right) $
satisfies {\rm{(B)}}. But after truncation of (\ref{hamiltonien total}) the
Hamiltonian $H_{\Lambda }^{B}$ (\ref{hamiltonien de Bogoliubov}) is no more
superstable. The system (\ref{hamiltonien de Bogoliubov}) is unstable for $%
\mu >0$, Proposition \ref{proposition de stabilite}. Below we follow the
Approximation Hamiltonian Method \`{a} la Ginibre adapted for the WIBG model.
\end{remark}

\hspace{0.4cm}
Since in the approximating Hamiltonian $H_{\Lambda }^{B}\left( c^{\#},\mu
\right) $ (\ref{hamiltonien de Bogoliubov sans fluctuation}) the gauge
symmetry is broken, we introduce
\begin{eqnarray}
H_{\Lambda }^{B}\left( \nu ^{\#}\right)  &=&H_{\Lambda }^{B}-\sqrt{V}\left(
\overline{\nu }a_{0}+\nu a_{0}^{*}\right) ,  \nonumber \\
H_{\Lambda }^{B}\left( \mu ,\nu ^{\#}\right)  &=&H_{\Lambda }^{B}\left( \nu
^{\#}\right) -\mu N_{\Lambda }  \label{definition de Hbogo avec source}
\end{eqnarray}
with sources $\nu \in \Bbb{C}$ breaking the symmetry of $H_{\Lambda }^{B}$%
, here $\nu ^{\#}=\left( \nu ,\overline{\nu }\right) $. Then by Proposition
\ref{pression inf} and the Bogoliubov inequality for Hamiltonians: $H_{\Lambda }^{B}\left(
\mu ,\nu ^{\#}\right) $ and $H_{\Lambda }^{B}\left( c^{\#},\mu ,\nu
^{\#}\right)$, one gets :
\begin{eqnarray}
0 &\leq &\Delta _{\Lambda }\left( \beta ,\mu ;c^{\#},\nu ^{\#}\right) :=
p_{\Lambda }\left[ H_{\Lambda }^{B}\left( \nu ^{\#}\right) \right] -%
\widetilde{p}_{\Lambda }^{B}\left( \beta ,\mu ;c^{\#},\nu ^{\#}\right) \leq
\nonumber \\
&\leq &\frac{1}{V}\left\langle H_{\Lambda }^{B}\left( c^{\#},\mu ,\nu
^{\#}\right) -H_{\Lambda }^{B}\left( \mu ,\nu ^{\#}\right) \right\rangle
_{H_{\Lambda }^{B}\left( \nu ^{\#}\right) }.  \label{eq 3.4}
\end{eqnarray}
Let $A:= a_{0}-\sqrt{V}c,A^{*}:= a_{0}^{*}-\sqrt{V}\overline{c}$.
Then Taylor expansion around $a_{0}^{\#}$ gives :
\begin{eqnarray}
&&H_{\Lambda }^{B}\left( c^{\#},\mu ,\nu ^{\#}\right) -H_{\Lambda }^{B}\left(
\mu ,\nu ^{\#}\right)  = - A^{*}\left[ a_{0},H_{\Lambda }^{B}\left( \mu ,\nu
^{\#}\right) \right] +h.c.  \nonumber \\
&&+\frac{1}{2}A^{*^{2}}\left[ a_{0},\left[ a_{0},H_{\Lambda }^{B}\left( \mu
,\nu ^{\#}\right) \right] \right] +h.c.
+A^{*}\left[ a_{0},\left[ H_{\Lambda }^{B}\left( \mu ,\nu ^{\#}\right)
,a_{0}^{*}\right] \right] A  \nonumber \\
&&-\frac{1}{2}A^{*^{2}}\left[ a_{0},\left[ a_{0},\left[ H_{\Lambda
}^{B}\left( \mu ,\nu ^{\#}\right) ,a_{0}^{*}\right] \right] \right] A+h.c.
\nonumber \\
&&+\frac{1}{4}A^{*^{2}}\left[ a_{0},\left[ a_{0},\left[ \left[ H_{\Lambda
}^{B}\left( \mu ,\nu ^{\#}\right) ,a_{0}^{*}\right], a_{0}^{*}\right] \right]
\right] A^{2}.  \label{dev de taylor de hbogo}
\end{eqnarray}
%%%%%%%%%%%%%%%%%%%%%%%%%%%%%%%%%%%%%%%%%%%%%%%%%%%%%%%%%%%%%%%%%%%%%%%%%%%%%%%%%%%%%%%%%%%%%%%%
\begin{remark}
\label{remark 3.2}Explicit calculations show that the third and the fourth
other terms in (\ref{dev de taylor de hbogo}) are bounded from above :
\begin{eqnarray}
-\frac{v\left( 0\right) }{\sqrt{V}}\left( \overline{c}A^{*}AA+cA^{*}A^{*}A%
\right) -\frac{v\left( 0\right) }{2V}A^{*^{2}}A^{2} &=&2v\left( 0\right)
\left| c\right| ^{2}A^{*}A-  \nonumber \\
-\frac{v\left( 0\right) }{2V}\left( A^{2}+2\sqrt{V}cA\right) ^{*}\left(
A^{2}+2\sqrt{V}cA\right)  &\leq &2v\left( 0\right) \left| c\right|
^{2}A^{*}A.  \label{troisieme + quatrieme terme}
\end{eqnarray}
\end{remark}
%%%%%%%%%%%%%%%%%%%%%%%%%%%%%%%%%%%%%%%%%%%%%%%%%%%%%%%%%%%%%%%%%%%%%%%%%%%%%%%%%%%%%%%%%%%%%%%%%
\begin{remark}
After some manipulations the terms of the first and the second order in (\ref
{dev de taylor de hbogo}) can be combined in
\begin{eqnarray}
&&-\frac{1}{2}\left[ A^{*}A,\left[ H_{\Lambda }^{B}\left( \mu ,\nu
^{\#}\right) ,A^{*}A\right] \right] +2A^{*}\left[ A,\left[ H_{\Lambda
}^{B}\left( \mu ,\nu ^{\#}\right) ,A^{*}\right] \right] A  \nonumber \\
&&-\frac{3}{2}A^{*}\left[ A,H_{\Lambda }^{B}\left( \mu ,\nu ^{\#}\right)
\right] -\frac{3}{2}\left[ H_{\Lambda }^{B}\left( \mu ,\nu ^{\#}\right)
,A^{*}\right] A.  \label{premier + deuxieme terme}
\end{eqnarray}
\end{remark}
%%%%%%%%%%%%%%%%%%%%%%%%%%%%%%%%%%%%%%%%%%%%%%%%%%%%%%%%%%%%%%%%%%%%%%%%%%%%%%%%%%%%%%%%%%%%%%%%%
\begin{lemma}
One has the following inequality :
\begin{equation}
\left\langle \left[ A^{*}A,\left[ H_{\Lambda }^{B}\left( \mu ,\nu
^{\#}\right) ,A^{*}A\right] \right] \right\rangle _{H_{\Lambda }^{B}\left(
\nu ^{\#}\right) }\geq 0.  \label{inegualite du prd scalaire de bogo}
\end{equation}
\end{lemma}
\emph{Proof.}
Denote by $\left(\cdot , \cdot \right)_{H_{\Lambda }}$ a positive semi-definite
scalar product with respect to a Hamiltonian $H_{\Lambda }$:
\begin{equation}
\left( X,Y\right) _{H_{\Lambda }}:= \frac{1}{\beta \ \Xi _{\Lambda }\left(
\beta ,\mu \right) }\stackunder{0}{\stackrel{\beta }{\int }}d\tau {\rm{Tr}}_{%
\mathcal{F}_{\Lambda }}\left( e^{-\left( \beta -\tau \right) H_{\Lambda
}\left( \mu \right) }X^{*}e^{-\tau H_{\Lambda }\left( \mu \right) }Y\right) .
\label{eq 3.9}
\end{equation}
Then $\left( \mathbf{1},Y\right) _{H_{\Lambda }}=\left\langle Y\right\rangle
_{H_{\Lambda }}$ and
\begin{equation}
\beta \left( \left[ X,H_{\Lambda }\left( \mu \right) \right] ,\left[
X,H_{\Lambda }\left( \mu \right) \right] \right) _{H_{\Lambda
}}=\left\langle \left[ X,\left[ H_{\Lambda }\left( \mu \right) ,X^{*}\right]
\right] \right\rangle _{H_{\Lambda }}.  \label{eq 3.10}
\end{equation}
Applying (\ref{eq 3.10}) to $H_{\Lambda }\left( \mu \right) =H_{\Lambda
}^{B}\left( \mu ,\nu ^{\#}\right) $ and $X=A^{*}A$ one gets (\ref{inegualite
du prd scalaire de bogo}). \hfill $\square$
%%%%%%%%%%%%%%%%%%%%%%%%%%%%%%%%%%%%%%%%%%%%%%%%%%%%%%%%%%%%%%%%%%%%%%%%%%%%%%%%%%%%%%%%%%%%%%%%
\begin{lemma}
One has the following estimate :
\begin{eqnarray}
-2\left\langle A^{*}\left[ A,H_{\Lambda }^{B}\left( \mu ,\nu ^{\#}\right)
\right] \right\rangle _{H_{\Lambda }^{B}\left( \nu ^{\#}\right) } &\leq
&\left\langle \left[ A^{*},\left[ H_{\Lambda }^{B}\left( \mu ,\nu
^{\#}\right) ,A\right] \right] \right\rangle _{H_{\Lambda }^{B}\left( \nu
^{\#}\right) }  \nonumber \\
&&+\left\langle \left[ A^{*},\left[ H_{\Lambda }^{B}\left( \mu ,\nu
^{\#}\right) ,A\right] \right] ^{*}\right\rangle _{H_{\Lambda }^{B}\left(
\nu ^{\#}\right) }  \nonumber \\
&&+2\beta ^{-1}\left\langle \left\{ A,A^{*}\right\} \right\rangle
_{H_{\Lambda }^{B}\left( \nu ^{\#}\right) },  \label{eq 3.11}
\end{eqnarray}
where $\left\{ X,Y\right\} =XY+YX.$
\end{lemma}
\emph{Proof.}
By the spectral decomposition of the Hamiltonian $\left( H_{\Lambda
}^{B}\left( \mu ,\nu ^{\#}\right) \psi _{n}=E_{n}\psi _{n}\right) $ one gets
\begin{eqnarray}
&&\left\langle \left\{ A^{*},\left[ H_{\Lambda }^{B}\left( \mu ,\nu
^{\#}\right) ,A\right] \right\} \right\rangle _{H_{\Lambda }^{B}\left( \nu
^{\#}\right) } =   \label{eq 3.12}\\
&&\frac{1}{\Xi _{\Lambda }^{B}\left( \beta ,\mu ,\nu
^{\#}\right) }\stackunder{m,n}{\sum }\left| \left( \psi _{m},A\psi
_{n}\right) \right| ^{2}
\left( E_{m}-E_{n}\right) \left( e^{-\beta E_{n}}+e^{-\beta
E_{m}}\right). \nonumber
\end{eqnarray}
Since
\begin{equation}
\frac{1}{2}\left( e^{x}+e^{y}\right) -\frac{1}{2}\left| e^{x}-e^{y}\right|
\leq \frac{e^{x}-e^{y}}{x-y}\leq \frac{1}{2}\left( e^{x}+e^{y}\right) ,
\label{eq 3.13}
\end{equation}
one gets
\begin{eqnarray}
&& \beta \left( E_{m}-E_{n}\right) \left( e^{-\beta E_{n}}+e^{-\beta
E_{m}}\right) \leq \nonumber \\
&& 2\left( e^{-\beta E_{n}}-e^{-\beta E_{m}}\right)
+\beta \left( E_{m}-E_{n}\right) \left| e^{-\beta E_{n}}-e^{-\beta
E_{m}}\right|  \leq \nonumber \\
&&2\left( e^{-\beta E_{n}}+e^{-\beta E_{m}}\right)
+\beta \left( E_{m}-E_{n}\right) \left( e^{-\beta E_{n}}-e^{-\beta
E_{m}}\right) .  \label{extimate 3.14}
\end{eqnarray}
Inserting the estimate (\ref{extimate 3.14}) into (\ref{eq 3.12}) we obtain
\begin{eqnarray}
&&\left\langle \left\{ A^{*},\left[ H_{\Lambda }^{B}\left( \mu ,\nu
^{\#}\right) ,A\right] \right\} \right\rangle _{H_{\Lambda }^{B}\left( \nu
^{\#}\right) } \leq \label{eq 3.15} \\
&&2\beta ^{-1}\left\langle AA^{*}+A^{*}A\right\rangle_{H_{\Lambda }^{B}\left( \nu ^{\#}\right) }
+\left\langle \left[ A^{*},\left[ H_{\Lambda }^{B}\left( \mu ,\nu
^{\#}\right) ,A\right] \right] \right\rangle _{H_{\Lambda }^{B}\left( \nu
^{\#}\right) }.  \nonumber
\end{eqnarray}
Note that
\begin{eqnarray}
&&-2\left\langle A^{*}\left[ A,H_{\Lambda }^{B}\left( \mu ,\nu ^{\#}\right)
\right] \right\rangle _{H_{\Lambda }^{B}\left( \nu ^{\#}\right) } = \label{eq 3.16} \\
&&\left\langle \left[ A^{*},\left[ H_{\Lambda }^{B}\left( \mu ,\nu
^{\#}\right) ,A\right] \right] \right\rangle_{H_{\Lambda }^{B}\left( \nu
^{\#}\right) }
+\left\langle \left\{ A^{*},\left[ H_{\Lambda }^{B}\left( \mu ,\nu
^{\#}\right) ,A\right] \right\} \right\rangle _{H_{\Lambda }^{B}\left( \nu
^{\#}\right) }.  \nonumber
\end{eqnarray}
Then combining (\ref{eq 3.15}) and (\ref{eq 3.16}) one gets (\ref{eq 3.11}).
\hfill $\square$
%%%%%%%%%%%%%%%%%%%%%%%%%%%%%%%%%%%%%%%%%%%%%%%%%%%%%%%%%%%%%%%%%%%%%%%%%%%%%%%%%%%%%%%%%%%%%%%%%%
\begin{corollary}
Since
\[
\left\langle A^{*}\left[ A,H_{\Lambda }^{B}\left( \mu ,\nu ^{\#}\right)
\right] \right\rangle _{H_{\Lambda }^{B}\left( \nu ^{\#}\right)
}=\left\langle \left[ H_{\Lambda }^{B}\left( \mu ,\nu ^{\#}\right)
,A^{*}\right] A\right\rangle _{H_{\Lambda }^{B}\left( \nu ^{\#}\right) },
\]
by (\ref{eq 3.11}) the mean value of the last two terms of (\ref{premier +
deuxieme terme}) is bounded from above:
\begin{eqnarray}
-3\left\langle A^{*}\left[ A,H_{\Lambda }^{B}\left( \mu ,\nu ^{\#}\right)
\right] \right\rangle _{H_{\Lambda }^{B}\left( \nu ^{\#}\right) } &\leq &%
\frac{3}{2}\left\langle \left[ A^{*},\left[ H_{\Lambda }^{B}\left( \mu ,\nu
^{\#}\right) ,A\right] \right] +h.c.\right\rangle _{H_{\Lambda }^{B}\left(
\nu ^{\#}\right) }+  \nonumber \\
&&+3\beta ^{-1}\left\langle AA^{*}+A^{*}A\right\rangle _{H_{\Lambda
}^{B}\left( \nu ^{\#}\right) }.  \label{eq 3.17}
\end{eqnarray}
\end{corollary}

\hspace{0.4cm}
Since we are looking for the estimate of (\ref{dev de taylor de hbogo}) (and
consequently by of (\ref{premier + deuxieme terme})) from above,
inequalities (\ref{inegualite du prd scalaire de bogo}) and (\ref{eq 3.17})
show that it rests to estimate the mean value only of the second term in (%
\ref{premier + deuxieme terme}). Here we formulate the result, see proof in \cite{BZ98JP},
Appendix A.
%%%%%%%%%%%%%%%%%%%%%%%%%%%%%%%%%%%%%%%%%%%%%%%%%%%%%%%%%%%%%%%%%%%%%%%%%%%%%%%%%%%%%%%%%%%%%%%%%%
\begin{theorem}
\label{theorem 3.7}Let $\left( \theta ,\mu \right) \in D$ Then in $D$ there
are non-negative functions
\begin{equation}
\begin{array}{l}
a=a\left( \theta ,\mu ,\nu ^{\#}\right) , \\
b=b\left( \theta ,\mu ,\nu ^{\#}\right) ,
\end{array}
\label{eq 3.18}
\end{equation}
such that for $\left| \nu \right| \leq r_{0}$, $r_{0}>0$, one has :
\begin{equation}
\left\langle A^{*}\left[ A,\left[ H_{\Lambda }^{B}\left( \mu ,\nu
^{\#}\right) ,A^{*}\right] \right] A\right\rangle _{H_{\Lambda }^{B}\left(
\nu ^{\#}\right) }\leq a\left\langle A^{*}A\right\rangle _{H_{\Lambda
}^{B}\left( \nu ^{\#}\right) }+b.  \label{eq 3.19}
\end{equation}
\end{theorem}

\hspace{0.4cm}
Now, to prove the main assertion of this section (Theorem \ref{theorem 3.16}) we also need the
following two lemmata.
%%%%%%%%%%%%%%%%%%%%%%%%%%%%%%%%%%%%%%%%%%%%%%%%%%%%%%%%%%%%%%%%%%%%%%%%%%%%%%%%%%%%%%%%%%%%%%%%%%%
\begin{lemma}\label{lemma 3.8}
For $\left( \theta ,\mu \right) \in Q$ and $\nu \in \Bbb{C}$ we have
\begin{equation}
p_{\Lambda }\left[ H_{\Lambda }^{B}\left( \nu ^{\#}\right) \right] \leq
\widetilde{p}_{\Lambda }^{PBG}\left( \beta ,\mu \right) +\left\{ \frac{1}{%
\beta V}\stackunder{n_{0}=0}{\stackrel{\infty }{\sum }}e^{\frac{\beta }{2}%
\left[ \left( \varphi \left( 0\right) +2\right) n_{0}-v\left( 0\right)
n_{0}^{2}/V\right] }\right\} +\left| \nu \right| ^{2}.  \label{eq 3.20}
\end{equation}
\end{lemma}
%%%%%%%%%%%%%%%%%%%%%%%%%%%%%%%%%%%%%%%%%%%%%%%%%%%%%%%%%%%%%%%%%%%%%%%%%%%%%%%%%%%%%%%%%%%%%%%%%
\emph{Proof.}
By the inequality
\[
-\sqrt{V}\left( \overline{\nu }a_{0}+\nu a_{0}^{*}\right) \geq
-a_{0}^{*}a_{0}-\left| \nu \right| ^{2}V,
\]
it follows immediately from the estimate (cf. (\ref{equation for stability}%
), (\ref{Hamiltonien pour stabilite}))
\[
H_{\Lambda }^{B}\left( \nu ^{\#}\right) -\mu N_{\Lambda }\geq \stackunder{%
k\in \Lambda ^{*},k\neq 0}{\sum }\left( \varepsilon _{k}-\mu -\frac{v(k)}{2V}%
\right) n_{k} \ +
\]
\[
+ \ \frac{v(0)}{2V}n_{0}^{2}-\left( \mu +\frac{1}{2}\varphi \left(
0\right) +1\right) n_{0}-\left| \nu \right| ^{2}V.
\]
\hfill $\square$
%%%%%%%%%%%%%%%%%%%%%%%%%%%%%%%%%%%%%%%%%%%%%%%%%%%%%%%%%%%%%%%%%%%%%%%%%%%%%%%%%%%%%%%%%%%%%%%%%%
\begin{corollary}
By (\ref{eq 3.20}), in the thermodynamic limit, one gets
\begin{equation}
p^{B}\left( \beta ,\mu ;\nu ^{\#}\right) \leq p^{PBG}\left( \beta ,\mu \right)
+\frac{1}{2}\stackunder{\rho \geq 0}{\sup }\left[ \left( \varphi \left(
0\right) +2\right) \rho -v\left( 0\right) \rho ^{2}\right] +\left| \nu
\right| ^{2}  \label{eq 3.21}
\end{equation}
for $\left( \theta ,\mu \right) \in Q,\nu \in \Bbb{C}.$
\end{corollary}
%%%%%%%%%%%%%%%%%%%%%%%%%%%%%%%%%%%%%%%%%%%%%%%%%%%%%%%%%%%%%%%%%%%%%%%%%%%%%%%%%%%%%%%%%%%%%%%%%
\begin{lemma}
\label{lemma 3.10}For any $\mu <0$ and $\nu \in \Bbb{C}$ one has the
estimate
\begin{equation}
\left\langle \frac{N_{\Lambda }}{V}\right\rangle _{H_{\Lambda }^{B}\left(
\nu ^{\#}\right) }\leq g_{\Lambda }\left( \beta ,\mu ;\nu ^{\#}\right)
<\infty .  \label{eq 3.22}
\end{equation}
\end{lemma}
\emph{Proof.}
For any $\mu <0$ there is $\delta >0$ such that $\mu +\delta <0$. Then by
the Bogoliubov inequality we obtain
\begin{equation}
\delta \left\langle \frac{N_{\Lambda }}{V}\right\rangle _{H_{\Lambda
}^{B}\left( \nu ^{\#}\right) }\leq p_{\Lambda }\left[ H_{\Lambda }^{B}\left(
\nu ^{\#}\right) -\delta N_{\Lambda }\right] -p_{\Lambda }\left[ H_{\Lambda
}^{B}\left( \nu ^{\#}\right) \right] .  \label{eq 3.23}
\end{equation}
Therefore, by Lemma \ref{lemma 3.8} one gets (\ref{eq 3.22}) for
\begin{equation}
g_{\Lambda }\left( \beta ,\mu ;\nu ^{\#}\right) := \frac{1}{\delta }
\left( p_{\Lambda }^{B}\left( \beta ,\mu +\delta ;\nu ^{\#}\right)
-p_{\Lambda }^{B}\left( \beta ,\mu ;\nu ^{\#}\right) \right) .
\label{eq 3.24}
\end{equation}
\hfill $\square$
%%%%%%%%%%%%%%%%%%%%%%%%%%%%%%%%%%%%%%%%%%%%%%%%%%%%%%%%%%%%%%%%%%%%%%%%%%%%%%%%%%%%%%%%%%%%%%%%%%
\begin{corollary}
In the thermodynamic limit (\ref{eq 3.24}) gives
\begin{eqnarray}
\rho ^{B}\left( \beta ,\mu ;\nu ^{\#}\right)  &=&\stackunder{\Lambda }{\lim }%
\left\langle \frac{N_{\Lambda }}{V}\right\rangle _{H_{\Lambda }^{B}\left(
\nu ^{\#}\right) }\leq \frac{1}{\delta }\left( p^{B}\left( \beta ,\mu
+\delta ;\nu ^{\#}\right) -p^{B}\left( \beta ,\mu ;\nu ^{\#}\right) \right)
\nonumber \\
&:= &g\left( \beta ,\mu ;\nu ^{\#}\right) .  \label{eq 3.25}
\end{eqnarray}
\end{corollary}

\hspace{0.4cm}
In fact, by the Griffiths lemma (Section \ref{Griffiths lemma}) we get in domain $D$ that
\begin{equation}
\rho ^{B}\left( \beta ,\mu ;\nu ^{\#}\right) =\partial _{\mu }p^{B}\left(
\beta ,\mu ;\nu ^{\#}\right),  \ \ \mu <0, \ \nu \in \Bbb{C}.  \label{eq 3.26}
\end{equation}
%%%%%%%%%%%%%%%%%%%%%%%%%%%%%%%%%%%%%%%%%%%%%%%%%%%%%%%%%%%%%%%%%%%%%%%%%%%%%%%%%%%%%%%%%%%%%%%%%
\begin{corollary}\label{corollary 3.12} By virtue of (\ref{eq 3.22}) one obviously get :
\begin{eqnarray}
\left\langle \frac{a_{0}^{*}a_{0}}{V}\right\rangle _{H_{\Lambda }^{B}\left(
\nu ^{\#}\right) } &\leq &g_{\Lambda }\left( \beta ,\mu ;\nu ^{\#}\right) ,
\nonumber \\
\left| \left\langle \frac{a_{0}^{*}}{\sqrt{V}}\right\rangle _{H_{\Lambda
}^{B}\left( \nu ^{\#}\right) }\right|  &=&\left| \left\langle \frac{a_{0}^{*}%
}{\sqrt{V}}\right\rangle _{H_{\Lambda }^{B}\left( \nu ^{\#}\right) }\right|
\leq \sqrt{g_{\Lambda }\left( \beta ,\mu ;\nu ^{\#}\right) }\ .
\label{eq 3.27}
\end{eqnarray}
\end{corollary}
%%%%%%%%%%%%%%%%%%%%%%%%%%%%%%%%%%%%%%%%%%%%%%%%%%%%%%%%%%%%%%%%%%%%%%%%%%%%%%%%%%%%%%%%%%%%%%%%%
\begin{remark}
\label{remark 3.15} To optimise the estimate (\ref{eq 3.4}) we have to estimate
the value of $\stackunder{c\in \Bbb{C}}{\sup }\widetilde{p}_{\Lambda }^{B}\left(
\beta ,\mu ;c^{\#},\nu ^{\#}\right) $. Since by Definition \ref{definition
of Bogo approx} and (\ref{definition de Hbogo avec source})
\begin{equation}
H_{\Lambda }^{B}\left( c^{\#},\mu ,\nu ^{\#}\right) =H_{\Lambda }^{B}\left(
c^{\#},\mu \right) -V\left( \nu \overline{c}+\overline{\nu }c\right) \geq
H_{\Lambda }^{B}\left( c^{\#},\mu \right) -V\left( \left| \nu \right|
^{2}\left| c\right| ^{2}+1\right) ,  \label{eq 3.40}
\end{equation}
one gets by (\ref{eq 2.28}) that for any $\left( \theta ,\mu \right) \in Q$
and a fixed $\nu ^{\#}$ there exists $A\geq 0$ such that
\begin{equation}
\widetilde{p}_{\Lambda }^{B}\left( \beta ,\mu ;c^{\#},\nu ^{\#}\right) \leq
A-\frac{1}{2}v\left( 0\right) \left| c\right| ^{4}.  \label{eq 3.41}
\end{equation}
Therefore, the optimal value of $\left| c\right| $ is bounded by a positive
constant $M<\infty $.
\end{remark}
%%%%%%%%%%%%%%%%%%%%%%%%%%%%%%%%%%%%%%%%%%%%%%%%%%%%%%%%%%%%%%%%%%%%%%%%%%%%%%%%%%%%%%%%

\hspace{0.4cm}
Now we are in position to prove the main statement of this section (see (\ref
{equality of p de c et pbogo})) about exactness of the Bogoliubov
approximation for the WIBG.
%%%%%%%%%%%%%%%%%%%%%%%%%%%%%%%%%%%%%%%%%%%%%%%%%%%%%%%%%%%%%%%%%%%%%%%%%%%%%%%%%%%%%%%%
\begin{theorem}\label{theorem 3.16}
Let $\left( \theta ,\mu \right) \in D$. Then
\begin{equation}
\stackunder{\Lambda }{\lim }\left\{ p_{\Lambda }^{B}\left( \beta ,\mu ,\nu
^{\#}\right) -\stackunder{c\in \Bbb{C}}{\sup }\widetilde{p}_{\Lambda
}^{B}\left( \beta ,\mu ;c^{\#},\nu ^{\#}\right) \right\} =0  \label{eq 3.42}
\end{equation}
for $\left| \nu \right| \leq r_{0}$, $r_{0}>0$.
\end{theorem}
%%%%%%%%%%%%%%%%%%%%%%%%%%%%%%%%%%%%%%%%%%%%%%%%%%%%%%%%%%%%%%%%%%%%%%%%%%%%%%%%%%%%%%%%
\emph{Proof.}
By inequality (\ref{eq 3.4}) one gets
\begin{eqnarray}
0 &\leq &\stackunder{c\in \Bbb{C}}{\inf }\Delta _{\Lambda }\left( \beta ,\mu
;c^{\#},\nu ^{\#}\right) := \Delta _{\Lambda }\left( \beta ,\mu ;%
\widehat{c}_{\Lambda }^{\#}\left( \beta ,\mu ,\nu ^{\#}\right) ,\nu
^{\#}\right)   \nonumber \\
&\leq &\frac{1}{V}\left\langle H_{\Lambda }^{B}\left( c^{\#},\mu ,\nu
^{\#}\right) -H_{\Lambda }^{B}\left( \mu ,\nu ^{\#}\right) \right\rangle
_{H_{\Lambda }^{B}\left( \nu ^{\#}\right) }.  \label{eq 3.43}
\end{eqnarray}
In virtue of (\ref{dev de taylor de hbogo})-(\ref{premier + deuxieme terme}%
), estimates (\ref{troisieme + quatrieme terme}) (\ref{inegualite du prd
scalaire de bogo}), (\ref{eq 3.11}), (\ref{eq 3.17}), (\ref{eq 3.19}) and
Remark \ref{remark 3.15}, there are positive constants $u$ and $w$
independent of the volume $V$, such that
\begin{eqnarray}
&&\frac{1}{V}\left\langle H_{\Lambda }^{B}\left( c^{\#},\mu ,\nu ^{\#}\right)
-H_{\Lambda }^{B}\left( \mu ,\nu ^{\#}\right) \right\rangle _{H_{\Lambda
}^{B}\left( \nu ^{\#}\right) } \leq  \label{eq 3.44} \\
&& u+\frac{w}{2}
\left\langle \left\{ \left( a_{0}^{*}-\sqrt{V}c^{*}\right) ,\left( a_{0}-%
\sqrt{V}c\right) \right\} \right\rangle _{H_{\Lambda }^{B}\left( \nu
^{\#}\right) }.  \nonumber
\end{eqnarray}
Put $c=\left\langle a_{0}/\sqrt{V}\right\rangle _{H_{\Lambda }^{B}\left( \nu
^{\#}\right) }$ which is bounded, see (\ref{eq 3.27}). Then
\[
\Delta _{\Lambda }\left( \beta ,\mu ;\widehat{c}_{\Lambda }^{\#},\nu
^{\#}\right) \leq \Delta _{\Lambda }\left( \beta ,\mu ;\left\langle a_{0}^{\#}/%
\sqrt{V}\right\rangle_{H_{\Lambda }^{B}\left( \nu ^{\#}\right) },\nu^{\#}\right) ,
\]
and estimates (\ref{eq 3.43}), (\ref{eq 3.44}) give
\begin{equation}
0\leq \stackunder{c\in \Bbb{C}}{\inf }\Delta _{\Lambda }\left( \beta ,\mu
;c^{\#},\nu ^{\#}\right) \leq \frac{u}{V}+\frac{w}{2V}\left\langle \left\{
\left( a_{0}^{*}-\left\langle a_{0}^{*}\right\rangle \right) ,\left(
a_{0}-\left\langle a_{0}\right\rangle \right) \right\} \right\rangle
_{H_{\Lambda }^{B}\left( \nu ^{\#}\right) },  \label{eq 3.45}
\end{equation}
where for the shorthand $\left\langle a_{0}^{\#}\right\rangle :=
\left\langle a_{0}^{\#}\right\rangle _{H_{\Lambda }^{B}\left( \nu
^{\#}\right) }$. Let $\delta a_{0}^{\#}:= a_{0}^{\#}-\left\langle
a_{0}^{\#}\right\rangle $. Then, by the Harris inequality \cite{Harr67}
one gets
\begin{equation}
\frac{1}{2}\left\langle \left\{ \delta a_{0}^{*},\delta a_{0}\right\}
\right\rangle _{H_{\Lambda }^{B}\left( \nu ^{\#}\right) }\leq \left( \delta
a_{0}^{*},\delta a_{0}\right) _{H_{\Lambda }^{B}\left( \nu ^{\#}\right) }+%
\frac{\beta }{12}\left\langle \left[ \delta a_{0}^{*},\left[ H_{\Lambda
}^{B}\left( \mu ,\nu ^{\#}\right) ,\delta a_{0}\right] \right] \right\rangle
_{H_{\Lambda }^{B}\left( \nu ^{\#}\right) }.  \label{eq 3.46}
\end{equation}
Since by (B) and Lemma \ref{lemma 3.10} we have:
\begin{eqnarray}
\left\langle \left[ \delta a_{0}^{*},\left[ H_{\Lambda }^{B}\left( \mu ,\nu
^{\#}\right) ,\delta a_{0}\right] \right] \right\rangle _{H_{\Lambda
}^{B}\left( \nu ^{\#}\right) } &=&\left\langle \frac{v\left( 0\right) }{V}%
N_{\Lambda }-\mu +\frac{1}{V}\stackunder{k\in \Lambda ^{*}}{\sum }v\left(
k\right) a_{k}^{*}a_{k}\right\rangle _{H_{\Lambda }^{B}\left( \nu
^{\#}\right) } \nonumber \\
&\leq &2v\left( 0\right) g_{\Lambda }\left( \beta ,\mu ;\nu ^{\#}\right)
-\mu ,  \label{eq 3.34}
\end{eqnarray}
by (\ref{eq 3.25}) and the uniform boundedness of $g_{\Lambda }\left( \beta
,\mu ;\nu ^{\#}\right) $ on $D$ for $\left| \nu \right| \leq r_{0}$ by $g_{0}
$, the estimate (\ref{eq 3.45}) in this compact set gets the form:
\begin{equation}
0\leq \stackunder{c\in \Bbb{C}}{\inf }\Delta _{\Lambda }\left( \beta ,\mu
;c^{\#},\nu ^{\#}\right) \leq \frac{1}{V}\left[ \widetilde{u}+w\left( \delta
a_{0}^{*},\delta a_{0}\right) _{H_{\Lambda }^{B}\left( \nu ^{\#}\right)
}\right] .  \label{eq 3.47}
\end{equation}

\hspace{0.4cm}
Now we can proceed along the standard reasoning of the Approximation
Hamiltonian Method \cite{BBZKT84}. First we note that
\begin{equation}
\left( \delta a_{0}^{*},\delta a_{0}\right) _{H_{\Lambda }^{B}\left( \nu
^{\#}\right) }=\frac{1}{\beta }\partial _{\nu }\partial _{\overline{\nu }%
}p_{\Lambda }\left[ H_{\Lambda }^{B}\left( \nu ^{\#}\right) \right] .
\label{eq 3.48}
\end{equation}
By the (canonical) gauge transformation $a_{0}\rightarrow a_{0}e^{i\varphi }$%
, $\varphi =\arg \nu $, one gets that in fact
\[
p_{\Lambda }\left[ H_{\Lambda }^{B}\left( \nu ^{\#}\right) \right]
=p_{\Lambda }^{B}\left( \beta ,\mu ;\left| \nu \right| := r\right) .
\]
Then passing in (\ref{eq 3.48}) to polar coordinates $\left( r,\varphi
\right) $ we obtain :
\begin{equation}
\left( \delta a_{0}^{*},\delta a_{0}\right) _{H_{\Lambda }^{B}\left( \nu
^{\#}\right) }=\frac{1}{4\beta r}\partial _{r}\left( r\partial
_{r}p_{\Lambda }^{B}\right) .  \label{eq 3.49}
\end{equation}
Let $c=\left| c\right| e^{i\psi }$, $\psi =\arg c$. Then by (\ref{definition
de Hbogo avec source}), (\ref{eq 3.4}) one gets
\begin{eqnarray}
\stackunder{c\in \Bbb{C}}{\inf }\Delta _{\Lambda }\left( \beta ,\mu
;c^{\#},\nu ^{\#}\right)  &=&\stackunder{\left| c\right| ,\psi }{\inf }%
\Delta _{\Lambda }\left( \beta ,\mu ;\left| c\right| e^{\pm i\psi },re^{\pm
i\varphi }\right) = \ \stackunder{\left| c\right| }{\inf }\widehat{\Delta }%
_{\Lambda }\left( \beta ,\mu ;\left| c\right| e^{\pm i\varphi },r\right)
\nonumber \\
&:= & \stackunder{\left| c\right| }{\inf }\widetilde{\Delta }_{\Lambda }\left(
r\right) .  \label{eq 3.51bis}
\end{eqnarray}
Therefore, by (\ref{eq 3.47})
\begin{equation}
\stackunder{R}{\stackrel{R+\varepsilon }{\int }}r\stackunder{\left| c\right|
}{\inf }\widetilde{\Delta }_{\Lambda }\left( r\right) dr\leq \frac{1}{V}%
\left\{ \widetilde{u}\frac{\left( R+\varepsilon \right) ^{2}-R^{2}}{2}+\frac{%
w}{4\beta }
\left( r\partial _{r}p_{\Lambda }^{B}\right) \bigg|^{R+\varepsilon}_{R}\right\},
\label{eq 3.50}
\end{equation}
for $\left[ R,R+\varepsilon \right] \subset \left[ 0,r_{0}\right] $. Note
that by (\ref{eq 3.27}) we have
\begin{equation}
\partial _{r}p_{\Lambda }^{B}=2\left| \left\langle a_{0}/\sqrt{V}%
\right\rangle _{H_{\Lambda }^{B}\left( \nu ^{\#}\right) }\right| \leq
2g_{0}^{\frac{1}{2}},\text{ }\left( \theta ,\mu \right) \in D,\left| \nu
\right| \leq r_{0}.  \label{eq 3.52}
\end{equation}
Therefore, (\ref{eq 3.50}) gets the form
\begin{equation}
\stackunder{R}{\stackrel{R+\varepsilon }{\int }}r\stackunder{\left| c\right|
}{\inf }\widetilde{\Delta }_{\Lambda }\left( r\right) dr\leq \frac{1}{V}%
\left\{ \widetilde{u}\frac{\left( R+\varepsilon \right) ^{2}-R^{2}}{2}+\frac{%
w}{2\beta }g_{0}^{\frac{1}{2}}\left( 2R+\varepsilon \right) \right\} .
\label{eq 3.53}
\end{equation}
Since by Corollary \ref{corollary 3.12} and Remark \ref{remark 3.15}
\[
\left| \partial _{r}\stackunder{\left| c\right| }{\inf }\widetilde{\Delta }%
_{\Lambda }\left( r\right) \right| \leq 2g_{\Lambda }^{\frac{1}{2}}+2\left|
\widehat{c}_{\Lambda }\right| \leq 2\left( g_{0}^{\frac{1}{2}}+M\right) ,
\]
for $r\in \left[ R,R+\varepsilon \right] $ we get :
\[
\stackunder{\left| c\right| }{\inf }\widetilde{\Delta }_{\Lambda }\left(
R\right) \leq \stackunder{\left| c\right| }{\inf }\widetilde{\Delta }%
_{\Lambda }\left( r\right) +2\left( r-R\right) \left( g_{0}^{\frac{1}{2}%
}+M\right) .
\]
Hence,
\[
\stackunder{\left| c\right| }{\inf }\widetilde{\Delta }_{\Lambda }\left(
R\right) \frac{\left( R+\varepsilon \right) ^{2}-R^{2}}{2}\leq \stackrel{%
R+\varepsilon }{\stackunder{R}{\int }}r\stackunder{\left| c\right| }{\inf }%
\widetilde{\Delta }_{\Lambda }\left( r\right) dr+2\left( g_{0}^{\frac{1}{2}%
}+M\right) \left( \frac{r^{3}}{3}-R\frac{r^{2}}{2}\right) \bigg|^{R+\varepsilon}_{R}.
%\stackrel{%
%R+\varepsilon }{\stackunder{R}{\mid }}.
\]
Then by (\ref{eq 3.53}) we obtain
\begin{equation}
\stackunder{\left| c\right| }{\inf }\widetilde{\Delta }_{\Lambda }\left(
R\right) \leq \frac{1}{V}\left\{ \widetilde{u}+\frac{w}{\beta }g_{0}^{\frac{1%
}{2}}\varepsilon ^{-1}\right\} +\left( g_{0}^{\frac{1}{2}}+M\right)
\varepsilon \frac{R+\frac{2}{3}\varepsilon }{R+\frac{1}{2}\varepsilon }.
\label{eq 3.54}
\end{equation}
Note that $\varepsilon >0$ is still arbitrary. Minimising  the right-hand
side of (\ref{eq 3.54}) one obtains that for large $V$ the optimal value of $
\varepsilon \sim 1/\sqrt{V}$. Hence, for $V\rightarrow \infty $ one
gets from (\ref{eq 3.54}) the asymptotic estimate
\begin{equation}
0\leq \stackunder{c\in \Bbb{C}}{\inf }\Delta _{\Lambda }\left( \beta ,\mu
;c^{\#},\nu ^{\#}\right) \leq \delta _{\Lambda }=const. \ \frac{1}{\sqrt{V}}
\label{eq 3.55}
\end{equation}
valid for $\left( \theta ,\mu \right) \in D$ and $\left| \nu \right| \leq
r_{0}$. \hfill $\square$
%%%%%%%%%%%%%%%%%%%%%%%%%%%%%%%%%%%%%%%%%%%%%%%%%%%%%%%%%%%%%%%%%%%%%%%%%%%%%%%%%%%%%%%%%%%%%%%%%%%%
\begin{corollary}
\label{corollary 3.17} Since the variational pressure $\widetilde{p}_{\Lambda
}^{B}\left( \beta ,\mu ;c^{\#},\nu ^{\#}\right) $ is known in the explicit form
(see (\ref{hamiltonien de Bogoliubov sans fluctuation})) :
\begin{equation}
\widetilde{p}_{\Lambda }^{B}\left( \beta ,\mu ;c^{\#},\nu ^{\#}\right) =%
\widetilde{p}_{\Lambda }^{B}\left( \beta ,\mu ;c^{\#}\right) +2\left| \nu
\right| \left| c\right| ,  \label{eq 3.56}
\end{equation}
we get that the thermodynamic limits
\begin{eqnarray}
\widetilde{p}^{B}\left( \beta ,\mu ;c^{\#},\nu ^{\#}\right)  &=&\stackunder{%
\Lambda }{\lim }\widetilde{p}_{\Lambda }^{B}\left( \beta ,\mu ;c^{\#},\nu
^{\#}\right) ,  \nonumber \\
\widetilde{p}^{B}\left( \beta ,\mu ;\widehat{c}^{\#}\left( \beta ,\mu ;\nu
^{\#}\right) ,\nu ^{\#}\right)  &=&\stackunder{\Lambda }{\lim }\left[
\stackunder{c\in \Bbb{C}}{\sup }\widetilde{p}_{\Lambda }^{B}\left( \beta
,\mu ;c^{\#},\nu ^{\#}\right) \right]   \nonumber \\
&=&\stackunder{c\in \Bbb{C}}{\sup }\widetilde{p}^{B}\left( \beta ,\mu
;c^{\#},\nu ^{\#}\right)   \label{eq 3.57}
\end{eqnarray}
exist. Then by virtue of the uniform estimate (\ref{eq 3.55}) we get
\begin{equation}
p^{B}\left( \beta ,\mu ;\nu ^{\#}\right) =  \stackunder{\Lambda }{\lim }
p_{\Lambda }\left[ H_{\Lambda }^{B}\left( \nu ^{\#}\right) \right] =
\stackunder{c\in \Bbb{C}}{\sup }\widetilde{p}^{B}\left( \beta ,\mu
;c^{\#},\nu ^{\#}\right)   \label{eq 3.58}
\end{equation}
for $\left( \theta ,\mu \right) \in D$, $\left| \nu \right| \leq r_{0}$ and
(cf. (\ref{equality of p de c et pbogo})) the limit $\left| \nu \right|
\rightarrow 0$ :
\begin{equation}
p^{B}\left( \beta ,\mu \right) =  \stackunder{c\in \Bbb{C}}{\sup }\widetilde{p}%
^{B}\left( \beta ,\mu ;c^{\#}\right).   \label{eq 3.59}
\end{equation}
\end{corollary}
%%%%%%%%%%%%%%%%%%%%%%%%%%%%%%%%%%%%%%%%%%%%%%%%%%%%%%%%%%%%%%%%%%%%%%%%%%%%%%%%%%%%%%%%%%%%%%
\begin{corollary}
\label{corollary 3.18}Inequalities (\ref{equation inf for p bogo}) and (\ref
{eq 2.30}) give
\[
p^{PBG}\left( \beta ,\mu \right) \leq \stackunder{\Lambda }{\lim }\left[
\stackunder{c\in \Bbb{C}}{\sup }\widetilde{p}_{\Lambda }^{B}\left( \beta
,\mu ;c^{\#}\right) \right] \leq p^{B}\left( \beta ,\mu \right) .
\]
Then definitions (\ref{eq for not equality with P0}), (\ref{definition de
D}) imply $D_{0}\subseteq D$, whereas (\ref{eq 3.42}) implies that $%
D_{0}=D$, which proves (\ref{eq 3.2}). Hence, we have
\begin{equation}
p^{B}\left( \beta ,\mu \right) = \ \stackunder{c\in \Bbb{C}}{\sup }\widetilde{p}%
^{B}\left( \beta ,\mu ;c^{\#}\right) \text{ for }\left( \theta ,\mu \right)
\in Q\backslash \partial D.  \label{eq 3.60}
\end{equation}
\end{corollary}
%%%%%%%%%%%%%%%%%%%%%%%%%%%%%%%%%%%%%%%%%%%%%%%%%%%%%%%%%%%%%%%%%%%%%%%%%%%%%%%%%%%%%%%%%%%%%%%%%%%
\begin{remark}
Since (\ref{eq 2.28}) implies that
\begin{equation}
\widetilde{p}_{\Lambda }^{B}\left( \beta ,\mu ;c^{\#}=0\right) = p_{\Lambda
}^{PBG}\left( \beta ,\mu \right) ,  \label{eq 3.61}
\end{equation}
by (\ref{eq for not equality with P0}), (\ref{definition de Dbis}) and (\ref
{eq 3.2}) we get
\begin{equation}
D_{0}=\left\{ \left( \theta ,\mu \right) :\text{ }\left| \widehat{c}\left(
\beta ,\mu ;\nu \right) \right| >0\right\} = \left\{ \left( \theta ,\mu
\right) :\text{ }\rho _{0}^{B}\left( \beta ,\mu \right) >0\right\} =D.
\label{eq 3.62}
\end{equation}
Therefore, (see Remark \ref{remark 2.18}) the condition (C) is sufficient
and necessary for $D\neq \left\{ \emptyset \right\} $.
\end{remark}

%%%%%%%%%%%%%%%%%%%%%%%%%%%%%%%%%%%%%%%%%%%%%%%%%%%%%%%%%%%%%%%%%%%%%%%%%%%%%%%%%%%%%%%%%%%%%%%%%%%
%\setcounter{proposition}{0}
\setcounter{equation}{0}
\section{Non-conventional condensate in Weakly Imperfect Bose-Gas}\label{section 4}
%%%%%%%%%%%%%%%%%%%%%%%%%%%%%%%%%%%%%%%%%%%%%%%%%%%%%%%%%%%%%%%%%%%%%%%%%%%%%%%%%%%%%%%%%%%%%%%%%%%
Since the pressure $\widetilde{p}_{\Lambda }^{B}$ (\ref{eq 2.28}) and $%
\stackunder{\Lambda }{\lim }\widetilde{p}_{\Lambda }^{B}=\widetilde{p}^{B}$
are known explicitly:
\begin{eqnarray}
&&\widetilde{p}^{B}\left( \beta ,\mu ;c^{\#},\nu ^{\#}\right) =\frac{1}{%
\beta \left( 2\pi \right) ^{3}}\stackunder{\Bbb{R}^{3}}{\int }d^{3}k\ln
\left( 1-e^{-\beta E_{k}\left( \left| c\right| ^{2}\right) }\right) ^{-1} -
\label{eq 4.1} \\
&&-\frac{1}{\beta \left( 2\pi \right) ^{3}}\stackunder{\Bbb{R}^{3}}{\int }%
d^{3}k\left[ E_{k}\left( \left| c\right| ^{2}\right) -f_{k}\left( \left|
c\right| ^{2}\right) \right]
+\mu \left| c\right| ^{2}-\frac{1}{2}v\left( 0\right) \left| c\right|
^{4}+\left( \nu \overline{c}+\overline{\nu }c\right) , \nonumber
\end{eqnarray}
Theorem \ref{theorem 3.16} and Corollaries \ref{corollary 3.17}, \ref
{corollary 3.18} give exact solution of the model (\ref{hamiltonien de
Bogoliubov}) on the level of thermodynamics. Therefore, (\ref{eq 3.60})
gives access to thermodynamic properties of the Weakly Imperfect Bose-Gas
for all $\left( \theta,\mu \right) \in Q$ except the line of transitions $%
\partial D$, Figure 4.

\hspace{0.4cm}
The aim of this section is to discuss
thermodynamic properties of the model (\ref{hamiltonien de Bogoliubov}) and
in particular the Bose-condensate which appears in domain $D$. The first
statement concerns the gauge-symmetry breaking in domain $D$.
%%%%%%%%%%%%%%%%%%%%%%%%%%%%%%%%%%%%%%%%%%%%%%%%%%%%%%%%%%%%%%%%%%%%%%%%%%%%%%%%%%%%%%%%%%%%
\begin{figure}[p]

\vspace{-24cm}

\centering
\includegraphics[width=1.9 \linewidth]{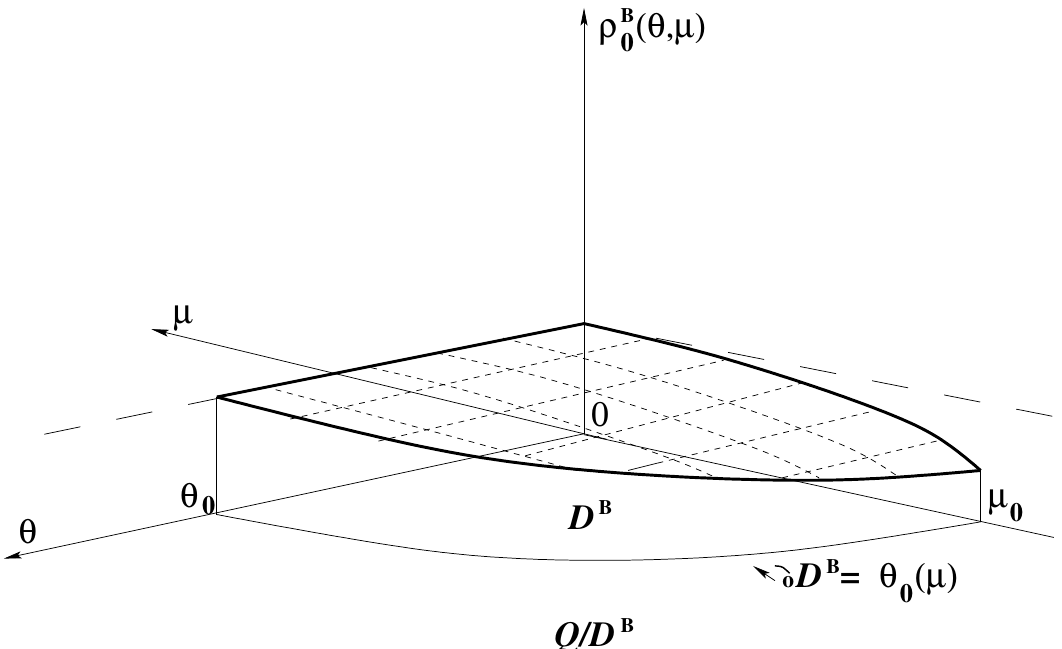}

%\end{figure}

\vspace{1.0cm}

{\bf Figure 4.}
Non-conventional dynamical condensate density $\rho_{0}^{B}\left(\theta ,\mu \right)$ as a
function of the chemical potential $\mu $ and the temperature $\theta $ for WIBG.
	
\end{figure}
%%%%%%%%%%%%%%%%%%%%%%%%%%%%%%%%%%%%%%%%%%%%%%%%%%%%%%%%%%%%%%%%%%%%%%%%%%%%%%%%%%%%%%%%%%%%%
%%%%%%%%%%%%%%%%%%%%%%%%%%%%%%%%%%%%%%%%%%%%%%%%%%%%%%%%%%%%%%%%%%%%%%%%%%%%%%%%%%%%%%%%%%%%
\begin{figure}[p]

\vspace{-2cm}

\centering
\includegraphics[width=1.2 \linewidth]{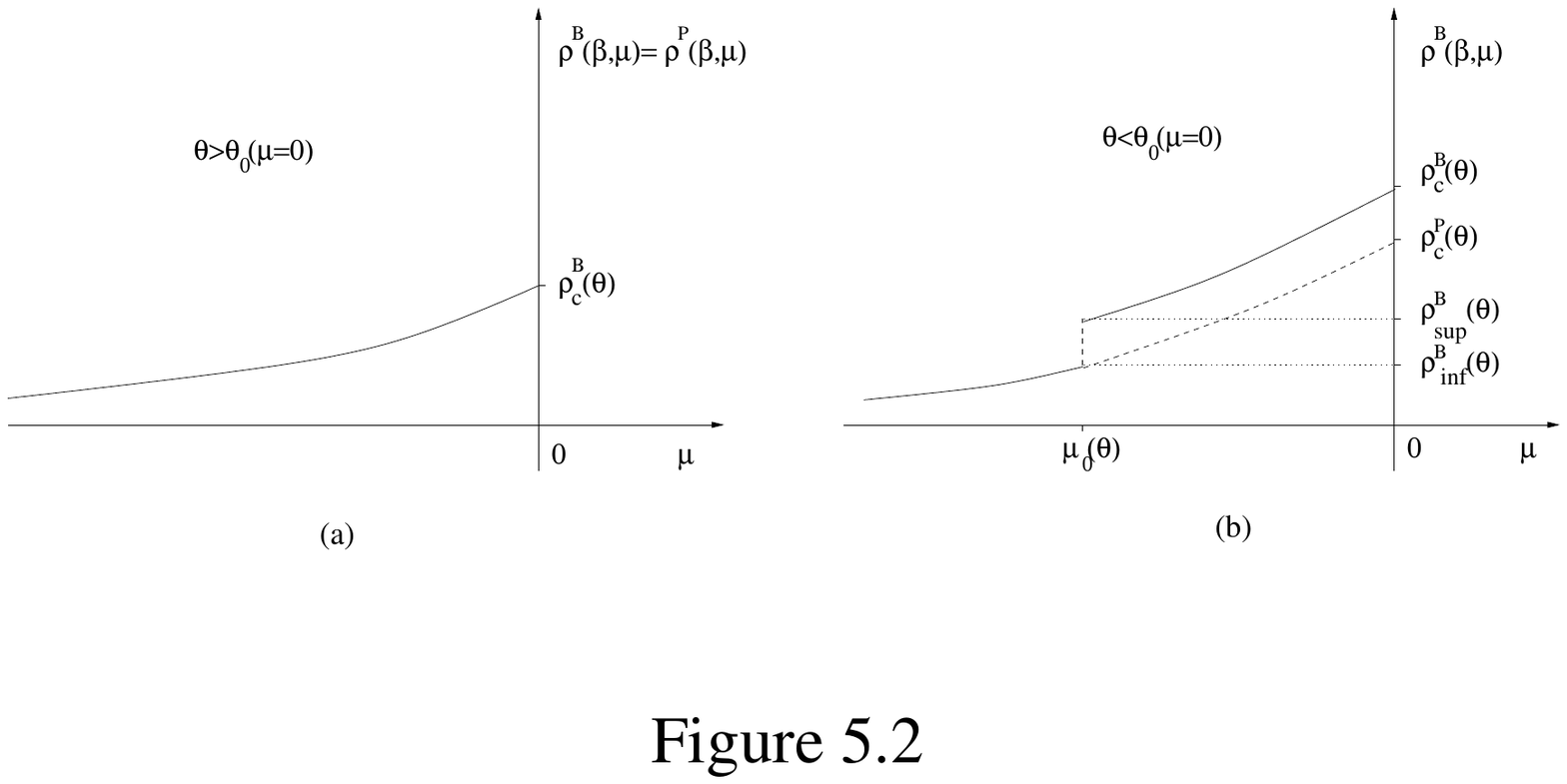}

%\end{figure}

\vspace{-5.0cm}

{\bf Figure 5.}
Illustration of the grand-canonical total particle
density for the model $H_{\Lambda }^{B}$ as a function of the chemical
potential $\mu $ at fixed temperature $\theta =\beta ^{-1}:\left( a\right) $
if $\theta >\theta _{0}\left( 0\right) :$ the graph of $\rho ^{B}\left(
\beta ,\mu \right) =\rho ^{PBG}\left( \beta ,\mu \right) ,$ where $\rho
_{c}^{PBG}\left( \theta \right) $ is the critical density for the PBG (here
$\rho _{c}^{B}\left( \theta \right) =\rho _{c}^{PBG}\left( \theta \right)
:= \rho ^{PBG}\left( \beta ,0\right) $); $\left( b\right) $ if $\theta
<\theta _{0}\left( 0\right) :$ the graph of $\rho ^{B}\left( \beta ,\mu
\right) \geq \rho ^{PBG}\left( \beta ,\mu \right) ,$ note that in this case
$\rho _{c}^{B}\left( \theta \right) >\rho _{c}^{PBG}\left( \theta \right)$.
\end{figure}
%%%%%%%%%%%%%%%%%%%%%%%%%%%%%%%%%%%%%%%%%%%%%%%%%%%%%%%%%%%%%%%%%%%%%%%%%%%%%%%%%%%%%%%%%%%%%
\begin{theorem}
\label{theorem 4.1}Let $D\neq \left\{ \emptyset \right\} $. Then
quasi-averages
\begin{equation}
\stackunder{\left\{ \nu \rightarrow 0:\arg \nu =\varphi \right\} }{\lim }%
\stackunder{\Lambda }{\lim }\left\langle a_{0}^{\#}/\sqrt{V}\right\rangle
_{H_{\Lambda }^{B}\left( \nu ^{\#}\right) }=e^{\pm i\varphi }\left| \widehat{%
c}\left( \beta ,\mu \right) \right| =\left\{
\begin{array}{c}
\neq 0,\left( \theta ,\mu \right) \in D \\
=0,\left( \theta ,\mu \right) \notin D
\end{array}
\right\}   \label{eq 4.2}
\end{equation}
\end{theorem}
%%%%%%%%%%%%%%%%%%%%%%%%%%%%%%%%%%%%%%%%%%%%%%%%%%%%%%%%%%%%%%%%%%%%%%%%%%%%%%%%%%%%%%%%%%%%
\emph{Proof.}
As in the proof of Theorem \ref{theorem 3.16} by the gauge transformation
\[
\mathcal{U}_{\varphi }a_{0}\mathcal{U}_{\varphi }^{*}=a_{0}e^{-i\varphi }=%
\widetilde{a}_{0},\text{ }\varphi =\arg \nu
\]
we get
\begin{eqnarray}
\widetilde{H}_{\Lambda }^{B}\left( \mu ,r\right) &=&\mathcal{U}_{\varphi
}H_{\Lambda }^{B}\left( \mu ,\nu ^{\#}\right) \mathcal{U}_{\varphi }^{*}=%
\widetilde{H}_{\Lambda }^{B}-\mu \widetilde{N}_{\Lambda }-\sqrt{V}r\left(
\widetilde{a}_{0}+\widetilde{a}_{0}^{*}\right) ,  \nonumber \\
p_{\Lambda }\left[ H_{\Lambda }^{B}\left( \nu ^{\#}\right) \right]
&=&p_{\Lambda }\left[ \mathcal{U}_{\varphi }H_{\Lambda }^{B}\left( \mu ,\nu
^{\#}\right) \mathcal{U}_{\varphi }^{*}\right] =p_{\Lambda }^{B}\left( \beta
,\mu ;r=\left| \nu \right| \right) .  \label{eq 4.3}
\end{eqnarray}
By virtue of
\[
0=\left\langle \left[ \widetilde{H}_{\Lambda }^{B}\left( \mu ,r\right) ,%
\widetilde{N}_{\Lambda }\right] \right\rangle _{\widetilde{H}_{\Lambda
}^{B}\left( r\right) }=r\sqrt{V}\left\langle \widetilde{a}_{0}-\widetilde{a}%
_{0}^{*}\right\rangle _{\widetilde{H}_{\Lambda }^{B}\left( r\right) },
\]
and (cf. (\ref{eq 3.10}))
\[
0\leq \left\langle \left[ \widetilde{N}_{\Lambda },\left[ \widetilde{H}%
_{\Lambda }^{B}\left( \mu ,r\right) ,\widetilde{N}_{\Lambda }\right] \right]
\right\rangle _{\widetilde{H}_{\Lambda }^{B}\left( r\right) }=r\sqrt{V}%
\left\langle \widetilde{a}_{0}+\widetilde{a}_{0}^{*}\right\rangle _{%
\widetilde{H}_{\Lambda }^{B}\left( r\right) },
\]
we get that
\begin{equation}
\left\langle \widetilde{a}_{0}\right\rangle _{\widetilde{H}_{\Lambda
}^{B}\left( r\right) }=\left\langle \widetilde{a}_{0}^{*}\right\rangle _{%
\widetilde{H}_{\Lambda }^{B}\left( r\right) }\geq 0.  \label{eq 4.4}
\end{equation}
Since (cf. (\ref{eq 3.9}))
\begin{eqnarray}\label{eq 4.5}
&&\partial _{r}^{2}p_{\Lambda }^{B}\left( \beta ,\mu ;r\right) =  \\
&&\beta \left(
\{ \left( \widetilde{a}_{0}+\widetilde{a}_{0}^{*}\right) -\left\langle
\widetilde{a}_{0}+\widetilde{a}_{0}^{*}\right\rangle_{\widetilde{H}%
_{\Lambda }^{B}( r) }\} , \{ \left( \widetilde{a}_{0}+%
\widetilde{a}_{0}^{*}\right) - \left\langle \widetilde{a}_{0}+\widetilde{a}%
_{0}^{*}\right\rangle_{\widetilde{H}_{\Lambda }^{B}(r)}\}
\right)_{\widetilde{H}_{\Lambda }^{B}\left( r\right) }
\geq 0 \nonumber \ ,
\end{eqnarray}
by the Theorem \ref{theorem 3.16} and Corollary \ref{corollary 3.17} the
sequence of the convex (for $r\geq 0$) functions $\left\{ p_{\Lambda
}^{B}\left( \beta ,\mu ;r\right) \right\} _{\Lambda }$ converges to the
(convex function)
\begin{eqnarray}
\widehat{p}^{B}\left( \beta ,\mu ;r\right) &:= &\stackunder{c\in \Bbb{C}%
}{\sup }\widetilde{p}^{B}\left( \beta ,\mu ;c^{\#},\nu ^{\#}\right) =%
\stackunder{\psi =\arg c}{\stackunder{\left| c\right| \geq 0}{\sup }}%
\widetilde{p}^{B}\left( \beta ,\mu ;\left| c\right| e^{\pm i\psi },\left|
\nu \right| e^{\pm i\varphi }\right)  \nonumber \\
&=&\widetilde{p}^{B}\left( \beta ,\mu ;\left| \widehat{c}\left( \beta ,\mu
;r\right) \right| e^{\pm i\varphi },\left| \nu \right| e^{\pm i\varphi
}\right) ,  \label{eq 4.6}
\end{eqnarray}
see (\ref{eq 3.49}) and (\ref{eq 4.1}), uniformly on $D\times \left[
0,r_{0}\right] $. By explicit calculations one gets that derivatives
\begin{eqnarray}
0 &\leq &\partial _{r}\widehat{p}^{B}\left( \beta ,\mu ;r\right) =2\left|
\widehat{c}\left( \beta ,\mu ;r\right) \right| \leq C_{1},  \nonumber \\
0 &\leq &\partial _{r}^{2}\widehat{p}^{B}\left( \beta ,\mu ;r\right)
=2\partial _{r}\left| \widehat{c}\left( \beta ,\mu ;r\right) \right| \leq
C_{2}  \label{eq 4.7}
\end{eqnarray}
are continuous and bounded in $D\times \left[ 0,r_{0}\right] $. Therefore,
by the Griffiths lemma
\[
\stackunder{\Lambda }{\lim }\partial _{r}p_{\Lambda }\left[ \widetilde{H}%
_{\Lambda }^{B}\left( r\right) \right] = \ \stackunder{\Lambda }{\lim }%
\left\langle \frac{\widetilde{a}_{0}+\widetilde{a}_{0}^{*}}{\sqrt{V}}%
\right\rangle _{\widetilde{H}_{\Lambda }^{B}\left( r\right) }=2\left|
\widehat{c}\left( \beta ,\mu ;r\right) \right| ,
\]
or by (\ref{eq 4.4}),
\begin{equation}
\begin{array}{l}
\stackunder{\Lambda }{\lim }\left\langle \widetilde{a}_{0}/\sqrt{V}%
\right\rangle _{\widetilde{H}_{\Lambda }^{B}\left( r\right) }=\left|
\widehat{c}\left( \beta ,\mu ;r\right) \right| , \\
\stackunder{\Lambda }{\lim }\left\langle \widetilde{a}_{0}^{*}/\sqrt{V}%
\right\rangle _{\widetilde{H}_{\Lambda }^{B}\left( r\right) }=\left|
\widehat{c}\left( \beta ,\mu ;r\right) \right| .
\end{array}
\label{eq 4.8}
\end{equation}
Returning in (\ref{eq 4.8}) back to original creation/annihilation
operators, one gets
\begin{equation}
\begin{array}{l}
\stackunder{\Lambda }{\lim }\left\langle a_{0}/\sqrt{V}\right\rangle
_{H_{\Lambda }^{B}\left( \nu ^{\#}\right) }=e^{+i\varphi }\left| \widehat{c}%
\left( \beta ,\mu ;r\right) \right| , \\
\stackunder{\Lambda }{\lim }\left\langle a_{0}^{*}/\sqrt{V}\right\rangle
_{H_{\Lambda }^{B}\left( \nu ^{\#}\right) }=e^{-i\varphi }\left| \widehat{c}%
\left( \beta ,\mu ;r\right) \right| .
\end{array}
\label{eq 4.9}
\end{equation}
Then the first part of the statement (\ref{eq 4.2}) follow from (\ref{eq 4.9}%
) and the continuity of the solution $\widehat{c}\left( \beta ,\mu ;r\right)
$ at $r=0$. Whereas the second part follows from (\ref{eq 3.62}). \hfill $\square$
%%%%%%%%%%%%%%%%%%%%%%%%%%%%%%%%%%%%%%%%%%%%%%%%%%%%%%%%%%%%%%%%%%%%%%%%%%%%%%%%%%%%%%%%%%%%%%%
\begin{corollary}
Note that by the gauge invariance
\begin{equation}
\left\langle \frac{a_{0}^{\#}}{\sqrt{V}}\right\rangle _{H_{\Lambda
}^{B}\left( \nu ^{\#}=0\right) }=0.  \label{eq 4.10}
\end{equation}
Therefore, we have the gauge-symmetry breaking :
\begin{equation}
\stackunder{\nu \rightarrow 0}{\lim }\stackunder{\Lambda }{\text{ }\lim }%
\left\langle \frac{a_{0}^{\#}}{\sqrt{V}}\right\rangle _{H_{\Lambda
}^{B}\left( \nu ^{\#}\right) }\neq \stackunder{\Lambda }{\lim }\text{ }%
\stackunder{\nu \rightarrow 0}{\lim }\left\langle \frac{a_{0}^{\#}}{\sqrt{V}}%
\right\rangle _{H_{\Lambda }^{B}\left( \nu ^{\#}\right) },  \label{eq 4.11}
\end{equation}
as soon as the Bose-condensation $\rho _{0}^{B}\left( \beta ,\mu \right)
\neq 0$.
\end{corollary}
%%%%%%%%%%%%%%%%%%%%%%%%%%%%%%%%%%%%%%%%%%%%%%%%%%%%%%%%%%%%%%%%%%%%%%%%%%%%%%%%%%%%%%%%%%%%%%%%%%
\begin{corollary}
Since by (\ref{eq 4.5}), (\ref{eq 4.7})
\[
\partial _{r}^{2}\left( \stackunder{\left| c\right| }{\inf }\widetilde{%
\Delta }_{\Lambda }\left( r\right) \right) =\partial _{r}^{2}\left(
p_{\Lambda }^{B}\left( \beta ,\mu ;r\right) -\widehat{p}^{B}\left( \beta
,\mu ;r\right) \right) \geq -C_{2},
\]
the Kolmogorov lemma {\rm{\cite{Kolm39}}} implies that
\begin{equation}
\left| \left\langle \frac{\widetilde{a}_{0}}{\sqrt{V}}\right\rangle _{%
\widetilde{H}_{\Lambda }^{B}\left( r\right) }-\left| \widehat{c}_{\Lambda
}\left( \beta ,\mu ;r\right) \right| \right| \leq 2\sqrt{\delta _{\Lambda
}C_{2}}  \label{eq 4.12}
\end{equation}
for $r\in \left[ l_{\Lambda },r_{0}-l_{\Lambda }\right] $, $l_{\Lambda }=2%
\sqrt{\delta _{\Lambda }/C_{2}}$, see (\ref{eq 3.54}) and (\ref{eq 4.8}).
\end{corollary}

\hspace{0.4cm}
Note that the Cauchy-Shwartz inequality gives
\[
\left\langle \frac{a_{0}^{*}}{\sqrt{V}}\right\rangle _{H_{\Lambda
}^{B}\left( \nu ^{\#}\right) }\left\langle \frac{a_{0}}{\sqrt{V}}%
\right\rangle _{H_{\Lambda }^{B}\left( \nu ^{\#}\right) }\leq \left\langle
\frac{a_{0}^{*}a_{0}}{V}\right\rangle _{H_{\Lambda }^{B}\left( \nu
^{\#}\right) }.
\]
Hence, by (\ref{definition de densite de condensant}) and (\ref{eq 4.2}) one
gets
\begin{equation}
\left| \widehat{c}_{\Lambda }\left( \beta ,\mu \right) \right| ^{2}\leq
\stackunder{\nu \rightarrow 0}{\lim }\text{ }\stackunder{\Lambda }{\lim }%
\left\langle \frac{a_{0}^{*}a_{0}}{V}\right\rangle _{H_{\Lambda }^{B}\left(
\nu ^{\#}\right) }=\rho _{0}^{B}\left( \beta ,\mu \right) ,  \label{eq 4.13}
\end{equation}
which is in coherence with definitions of domains $D_{0}$ and $D$, cf.
Theorem \ref{th for not equality with P0} and Corollary \ref{corollary 2.17}.
To prove equality in (\ref{eq 4.13}) we proceed as follows.
%%%%%%%%%%%%%%%%%%%%%%%%%%%%%%%%%%%%%%%%%%%%%%%%%%%%%%%%%%%%%%%%%%%%%%%%%%%%%%%%%%%%%%%%%%%%%%%%%%%
\begin{theorem}
\label{theorem 4.4}Let
\begin{equation}
\begin{array}{l}
H_{\Lambda ,\alpha }^{B}=H_{\Lambda }^{B}+\alpha a_{0}^{*}a_{0} \\
H_{\Lambda ,\alpha }^{B}\left( \nu ^{\#}\right) =H_{\Lambda ,\alpha }^{B}-%
\sqrt{V}\left( \nu a_{0}^{*}+\overline{\nu }a_{0}\right)
\end{array}
\label{eq 4.14}
\end{equation}
for $\alpha \in \Bbb{R}^{1}$. Then
\begin{equation}
p_{\alpha }^{B}\left( \beta ,\mu ;\nu ^{\#}\right) = \ \stackunder{\Lambda }{%
\lim }p_{\Lambda }\left[ H_{\Lambda ,\alpha }^{B}\left( \nu ^{\#}\right)
\right] = \ \stackunder{\Lambda }{\lim }\left[ \stackunder{c\in \Bbb{C}}{\sup }%
\widetilde{p}_{\Lambda ,\alpha }^{B}\left( \beta ,\mu ;c^{\#},\nu
^{\#}\right) \right] ,  \label{eq 4.15}
\end{equation}
for $\left| \nu \right| \leq r_{0},r_{0}>0$ and $\left( \theta ,\mu \right)
\in Q\backslash \partial D_{\alpha }$ where domain
\begin{equation}
D_{\alpha }:= \left\{ \left( \theta ,\mu \right) :\text{ }p_{\alpha
}^{B}\left( \beta ,\mu ;\nu ^{\#}=0\right) >p^{PBG}\left( \beta ,\mu \right)
\right\} .  \label{eq 4.16}
\end{equation}
\end{theorem}
%%%%%%%%%%%%%%%%%%%%%%%%%%%%%%%%%%%%%%%%%%%%%%%%%%%%%%%%%%%%%%%%%%%%%%%%%%%%%%%%%%%%%%%%%%%%%%%%%%
\begin{remark}
Since $H_{\Lambda ,\alpha = \varphi\left( 0\right)/2  }^{B}=\widehat{%
H}_{\Lambda }^{B}$, see (\ref{eq 2.48}), by the Theorem \ref{theorem 2.20}
we get that $D_{\alpha = \varphi \left( 0\right)/2 }=\left\{
\emptyset \right\} $.
\end{remark}

\hspace{0.4cm}
Our reasoning below is a translation of some results of Sections \ref
{section 2} and \ref{section 3} to perturbed Hamiltonian $H_{\Lambda,\alpha }^{B}$
for small $\alpha $.
%%%%%%%%%%%%%%%%%%%%%%%%%%%%%%%%%%%%%%%%%%%%%%%%%%%%%%%%%%%%%%%%%%%%%%%%%%%%%%%%%%%%%%%%%%%%%%%%%%%
\begin{lemma}
\label{lemma 4.6}If potential $v\left( k\right) $ satisfies (A), (B) and
(C), then
\begin{equation}
D_{0\alpha }:= \left\{ \left( \theta ,\mu \right) :\text{ }\stackunder{%
c\in \Bbb{C}}{\sup }\widetilde{p}_{\alpha }^{B}\left( \beta ,\mu
;c^{\#}\right) > p^{PBG}\left( \beta ,\mu \right) \right\} \neq \left\{
\emptyset \right\} .  \label{eq 4.17}
\end{equation}
for $\alpha <-\mu _{0}$, where $\mu _{0}$ is defined in Lemma \ref{lemma
pour etha}.
\end{lemma}
\emph{Proof.}
Since the $\eta _{\Lambda ,\alpha }\left( \mu ;x\right) $ for the
Hamiltonian (\ref{eq 4.14}) (cf. (\ref{eq 2.28})) has the form
\begin{equation}
\eta _{\Lambda ,}\left( \mu ;x\right) =-\frac{1}{2V}\stackunder{k\in \Lambda
^{*},k\neq 0}{\sum }\left( E_{k}-f_{k}\right) +\left( \mu -\alpha \right) x-%
\frac{1}{2}v\left( 0\right) x^{2},  \label{eq 4.18}
\end{equation}
one can follow the line reasoning of Lemma \ref{lemma pour etha} and Theorem
\ref{th for not equality with P0} to get (\ref{eq 4.17}) for $\mu \leq 0$
such that $\left( \mu -\alpha \right) >\mu _{0}$. Therefore, the value of $%
\mu _{0}+\alpha $ must be negative. \hfill $\square$

\hspace{0.4cm}
By continuity of (\ref{eq 4.18}) which respect to $\alpha $ it is clear that
$\stackunder{\alpha \rightarrow 0}{\lim }D_{0\alpha }=D_{0}$. Now we return to
the \\
\emph{Proof of Theorem \ref{theorem 4.4} }:

$\left( 1\right) $ Since the Bogoliubov approximation (\ref{definition de
approx de bogoliubov}) gives the estimate of the pressure $p_{\Lambda
}\left[ H_{\Lambda ,\alpha }^{B}\left( \nu ^{\#}\right) \right] $ from below
(see Proposition \ref{pression inf}):
\[
\stackunder{c\in \Bbb{C}}{\sup }\widetilde{p}_{\Lambda ,\alpha }^{B}\left(
\beta ,\mu ;c^{\#},\nu ^{\#}\right) \leq p_{\Lambda }\left[ H_{\Lambda
,\alpha }^{B}\left( \nu ^{\#}\right) \right] ,
\]
by the Bogoliubov inequality we get (cf. (\ref{eq 3.4}))
\begin{eqnarray}
0 &\leq &\Delta _{\Lambda ,\alpha }\left( \beta ,\mu ;c^{\#},\nu
^{\#}\right) := p_{\Lambda }\left[ H_{\Lambda ,\alpha }^{B}\left( \nu
^{\#}\right) \right] -\widetilde{p}_{\Lambda ,\alpha }^{B}\left( \beta ,\mu
;c^{\#},\nu ^{\#}\right)  \nonumber \\
&\leq &\frac{1}{V}\left\langle H_{\Lambda ,\alpha }^{B}\left( \widehat{c}%
^{\#},\mu ,\nu ^{\#}\right) -H_{\Lambda ,\alpha }^{B}\left( \mu ,\nu
^{\#}\right) \right\rangle _{H_{\Lambda ,\alpha }^{B}\left( \nu ^{\#}\right)
}.  \label{eq 4.19}
\end{eqnarray}

$\left( 2\right) $ For operators $A^{\#}:= a_{0}^{\#}-\sqrt{V}c^{\#}$
and for a Taylor expansion of $H_{\Lambda ,\alpha }^{B}\left( \widehat{c}%
^{\#},\mu ,\nu ^{\#}\right) $ around $a_{0}^{\#}$ one gets the estimate
\begin{eqnarray}
0 &\leq &\stackunder{c\in \Bbb{C}}{\inf }\Delta _{\Lambda ,\alpha }\left(
\beta ,\mu ;c^{\#},\nu ^{\#}\right) =\Delta _{\Lambda ,\alpha }\left( \beta
,\mu ;\widehat{c}_{\Lambda ,\alpha }^{\#}\left( \beta ,\mu ;\nu ^{\#}\right)
,\nu ^{\#}\right) \leq u_{\alpha }+  \nonumber \\
&&+\frac{w_{\alpha }}{2}\left\langle \left\{ \left( a_{0}^{*}-\sqrt{V}%
\overline{c}\right) ,\left( a_{0}-\sqrt{V}c\right) \right\} \right\rangle
_{H_{\Lambda }^{B}\left( \nu ^{\#}\right) }.  \label{eq 4.20}
\end{eqnarray}
by carrying through verbatim the arguments developed starting on Remark \ref
{remark 3.2} and finishing by Remark \ref{remark 3.15}. The only difference
with the case $\alpha =0$ comes from
\[
\left[ A,\left[ H_{\Lambda ,\alpha }^{B}\left( \mu ,\nu ^{\#}\right)
,A^{*}\right] \right] =\left[ A,\left[ H_{\Lambda }^{B}\left( \mu ,\nu
^{\#}\right) ,A^{*}\right] \right] +\alpha ,
\]
cf. (\ref{eq 3.34}) and the note that $\stackunder{\alpha \rightarrow 0}{%
\lim }u_{\alpha }=u$ and $\stackunder{\alpha \rightarrow 0}{\lim }w_{\alpha
}=w$.

$\left( 3\right) $ Put $c^{\#}=\left\langle a_{0}^{\#}/\sqrt{V}\right\rangle
_{H_{\Lambda ,\alpha }^{B}\left( \nu ^{\#}\right) }$ in the left-hand side
of (\ref{eq 4.20}). The same line of reasoning as in Theorem \ref{theorem
3.16} gives asymptotic estimate
\begin{equation}
0\leq \stackunder{c\in \Bbb{C}}{\inf }\Delta _{\Lambda ,\alpha }\left( \beta
,\mu ;c^{\#},\nu ^{\#}\right) \leq \delta _{\Lambda ,\alpha }=const.\frac{1}{%
\sqrt{V}}  \label{eq 4.21}
\end{equation}
valid for $\left( \theta ,\mu \right) \in Q\backslash \partial D_{\alpha }$,
$\left| \alpha \right| <-\mu _{0}$, and $\left| \nu \right| \leq r_{0}$
which ensures the proof of (\ref{eq 4.15}) for $D_{\alpha }\neq \left\{
\emptyset \right\} $. \hfill $\square$
%%%%%%%%%%%%%%%%%%%%%%%%%%%%%%%%%%%%%%%%%%%%%%%%%%%%%%%%%%%%%%%%%%%%%%%%%%%%%%%%%%%%%%%%%%%%%%%%%
\begin{corollary}
Since
\begin{eqnarray*}
&& \partial _{\alpha }^{2}p_{\Lambda }\left[ H_{\Lambda ,\alpha }^{B}\left( \nu
^{\#}\right) \right]  = \\
&&\frac{\beta }{V}\left( (a_{0}^{*}a_{0}-
\left\langle a_{0}^{*}a_{0}\right\rangle_{H_{\Lambda ,\alpha}^{B}
( \nu ^{\#}) }) , ( a_{0}^{*}a_{0}-
\left\langle a_{0}^{*}a_{0}\right\rangle_{H_{\Lambda ,\alpha }^{B}( \nu^{\#}) })
\right)_{H_{\Lambda ,\alpha }^{B}( \nu^{\#}) }
\geq 0 \, ,
\end{eqnarray*}
functions $\left\{ p_{\Lambda }\left[ H_{\Lambda ,\alpha }^{B}\left( \nu
^{\#}=0\right) \right] \right\} _{\Lambda }$ are convex for $\alpha \in \Bbb{%
R}^{1}$. The same is obviously true for (cf. (\ref{eq 4.1}), (\ref{eq 4.14})
and (\ref{eq 4.15}))
\begin{eqnarray}
&&\stackunder{\Lambda }{\lim }p_{\Lambda }\left[ H_{\Lambda ,\alpha
}^{B}\left( \nu ^{\#}=0\right) \right] =  \stackunder{c\in \Bbb{C}}{\sup }%
\widetilde{p}_{\alpha }^{B}\left( \beta ,\mu ;c^{\#},\nu ^{\#}=0\right) =
\label{eq 4.22} \\
&&\widetilde{p}_{\alpha }^{B}\left( \beta ,\mu ;\widehat{c}_{\alpha
}^{\#}\left( \beta ,\mu \right) ,0\right)
=\widetilde{p}^{B}\left( \beta ,\mu ;\widehat{c}_{\alpha }^{\#}\left(
\beta ,\mu \right) ,0\right) -\alpha \left| \widehat{c}_{\alpha }\left(
\beta ,\mu \right) \right| ^{2}.  \nonumber
\end{eqnarray}
By explicit calculations one gets that
\begin{equation}
\partial _{\alpha }\widetilde{p}_{\alpha }^{B}\left( \beta ,\mu ;\widehat{c}%
_{\alpha }^{\#}\left( \beta ,\mu \right) ,0\right) =-\left| \widehat{c}%
_{\alpha }\left( \beta ,\mu \right) \right| ^{2}<const \, ,  \label{eq 4.23}
\end{equation}
for $\left( \theta ,\mu \right) \in Q$ and $\left| \alpha \right| \leq -\mu
_{0}$. Therefore, by the Griffiths lemma (Section \ref{Griffiths lemma}) we obtain :
\begin{eqnarray}
\stackunder{\Lambda }{\lim }\partial _{\alpha }p_{\Lambda }\left[ H_{\Lambda
,\alpha }^{B}\left( \nu ^{\#}=0\right) \right]  &=&\stackunder{\Lambda }{%
\lim }\left( -\left\langle \frac{a_{0}^{*}a_{0}}{V}\right\rangle
_{H_{\Lambda ,\alpha }^{B}( \nu ^{\#}=0) }\right)   \nonumber \\
&=&-\left| \widehat{c}_{\alpha }\left( \beta ,\mu \right) \right|^{2}.
\label{eq 4.24}
\end{eqnarray}
\end{corollary}
%%%%%%%%%%%%%%%%%%%%%%%%%%%%%%%%%%%%%%%%%%%%%%%%%%%%%%%%%%%%%%%%%%%%%%%%%%%%%%%%%%%%%%%%%%%%%%%%%%
\begin{corollary}
By the continuity in $\alpha \rightarrow 0$, equations (\ref{eq 4.2}) and (%
\ref{eq 4.24}) imply that
\begin{equation}
\rho _{0}^{B}\left( \beta ,\mu \right) = \ \stackunder{\Lambda }{\lim }%
\left\langle \frac{a_{0}^{*}a_{0}}{V}\right\rangle _{H_{\Lambda }^{B}}=%
\stackunder{\Lambda }{\lim }\left\langle \frac{a_{0}^{*}}{\sqrt{V}}%
\right\rangle _{H_{\Lambda }^{B}}\stackunder{\Lambda }{\lim }\left\langle
\frac{a_{0}}{\sqrt{V}}\right\rangle _{H_{\Lambda }^{B}}=\left| \widehat{c}%
\left( \beta ,\mu \right) \right| ^{2}.  \label{eq 4.25}
\end{equation}
\end{corollary}

\hspace{0.4cm}
We conclude this section by analysis of the non-conventional Bose-condensate $\rho
_{0}^{B}\left( \beta ,\mu \right) $ behaviour. In virtue of (\ref{eq 4.25})
it reduces to the analysis of the behaviour of $\left| \widehat{c}\left(
\beta ,\mu \right) \right| $ which corresponds to the $\stackunder{c\in \Bbb{%
C}}{\sup }$ of the trial pressure (\ref{eq 4.1}) :
\begin{equation}
\widetilde{p}^{B}\left( \beta ,\mu ;c^{\#},\nu ^{\#}=0\right) =\xi \left(
\beta ,\mu ;x:= \left| c\right| ^{2}\right) +\eta \left( \mu ;x:=
\left| c\right| ^{2}\right) := \widetilde{p}^{B}\left( \beta ,\mu
;c^{\#}\right) ,  \label{eq 4.26}
\end{equation}
where (cf. (\ref{eq 2.28}),(\ref{eq 2.29}))
\begin{eqnarray}
&&\xi \left( \beta ,\mu ;x\right) =\frac{1}{\left( 2\pi \right) ^{3}\beta }%
\stackunder{\Bbb{R}^{3}}{\int }d^{3}k\ln \left( 1-e^{-\beta E_{k}}\right)
^{-1}  \nonumber \\
&&\eta \left( \mu ;x\right) =\frac{1}{2\left( 2\pi \right) ^{3}}\stackunder{%
\Bbb{R}^{3}}{\int }d^{3}k\left( f_{k}-E_{k}\right) +\mu x-\frac{1}{2}v\left(
0\right) x^{2},  \label{eq 4.27} \\
&&f_{k} =\varepsilon _{k}-\mu +x\left[ v\left( 0\right) +v\left( k\right)
\right], \ h_{k}=xv\left( k\right), \ E_{k}=\sqrt{f_{k}^{2}-h_{k}^{2}}\ .
\nonumber
\end{eqnarray}
Below we collect some properties of the trial pressure (\ref{eq 4.26}) :

$\left( 1\right) $ For $\mu \leq 0$ the function (\ref{eq 4.26}) is
differentiable with respect to $x=\left| c\right| ^{2}\geq 0$ and
\begin{equation}
\stackunder{\left| c\right| ^{2}\rightarrow \infty }{\lim }\widetilde{p}%
^{B}\left( \beta ,\mu ;c^{\#}\right) =-\infty ,  \label{eq 4.28}
\end{equation}
Hence, $\stackunder{x\geq 0}{\sup }\left( \xi +\eta \right) \left( \beta
,\mu ;x\right) $ is attained either at $x=0$, or at a positive solution of
the equation
\begin{eqnarray}
0 &=&\partial _{x}\left( \xi +\eta \right) \left( \beta ,\mu ;x\right) =%
\frac{1}{\left( 2\pi \right) ^{3}}\stackunder{\Bbb{R}^{3}}{\int }%
d^{3}k\left( 1-e^{\beta E_{k}}\right) ^{-1}\partial _{x}E_{k}  \nonumber \\
&&-\frac{1}{2\left( 2\pi \right) ^{3}}\stackunder{\Bbb{R}^{3}}{\int }%
d^{3}k\left( \partial _{x}E_{k}-\partial _{x}f_{k}\right) +\mu -xv\left(
0\right) .  \label{eq 4.29}
\end{eqnarray}

$\left( 2\right) $ By definitions (\ref{eq 4.27}) and the properties (A) and
(B) of the potential $v\left( k\right) $ one gets that
\[
\partial _{x}f_{k}=v\left( 0\right) +v\left( k\right) ,\partial
_{x}E_{k}=E_{k}^{-1}\left( f_{k}v\left( 0\right) +\left( f_{k}-h_{k}\right)
v\left( k\right) \right) \geq 0,
\]
for $\mu \leq 0$, $x\geq 0$ and any $k\in \Bbb{R}^{3}$. Therefore, by (\ref
{eq 4.29}) we have
\begin{equation}
\partial _{x}\widetilde{p}^{B}\left( \beta ,\mu ;c^{\#}=0\right) \leq
\partial _{x}\eta \left( \mu ;x=0\right) := \partial _{x}\widetilde{p}%
^{B}\left( \beta =\infty ,\mu ;c^{\#}=0\right) =\mu \text{ .}
\label{eq 4.30}
\end{equation}

$\left( 3\right) $ By explicit calculation one finds that $\partial _{\mu
}\partial _{x}\eta \left( \mu ;x\right) \geq 0$ for $\mu \leq 0$ and $x\geq
0 $. Hence
\begin{equation}
\partial _{x}\eta \left( \mu ;x\right) \leq \partial _{x}\eta \left( \mu
=0;x\right) ,  \label{eq 4.31}
\end{equation}
and $\partial _{x}\eta \left( \mu =0;x\right) $ is a concave function of $%
\left( 0,\infty \right) .$

$\left( 4\right) $ Now let potential $v\left( k\right) $ satisfy the
condition (C). Then
\begin{equation}
\partial _{x}^{2}\eta _{\Lambda }\left( \mu =0;x\right) =-v\left( 0\right) +%
\frac{1}{2\left( 2\pi \right) ^{3}}\int_{\Bbb{R}^{3}}\frac{\left[
v(k)\right] ^{2}}{\varepsilon _{k}}d^{3}k\geq 0,  \label{eq 4.32}
\end{equation}
Since $\eta _{\Lambda }\left( \mu =0;x=0\right) =0$, (\ref{eq 4.32}) means
that the trivial pressure
\[
\widetilde{p}^{B}\left( \beta =\infty ,\mu ;c^{\#}\right) =\eta _{\Lambda
}\left( \mu =0;x\right)
\]
attains $\stackunder{x\geq 0}{\sup }$ for $\widehat{x}\left( \theta =0,\mu
=0\right) >0$, and, by continuity for $\left( \theta \geq 0,\mu \geq
0\right) $, the domain
\[
D_{0}=\left\{ \left( \theta ,\mu \right) :\text{ }\widehat{x}\left( \theta
,\mu \right) >0\right\} \neq \left\{ \emptyset \right\} ,
\]
see Lemma \ref{lemma pour etha}, Theorem \ref{th for not equality with P0}.

$\left( 5\right) $ Fix $\mu \in D_{0}$ and $\theta =0$. Then, according to (%
\ref{eq 4.30}),
\[
\partial _{x}\eta _{\Lambda }\left( \mu ;x=0\right) =\partial _{x}\widetilde{%
p}^{B}\left( \beta =\infty ,\mu ;c^{\#}=0\right) =\eta _{\Lambda }\left( \mu
=0;x\right) <0,
\]
but $\partial _{x}^{2}\widetilde{p}^{B}\left( \beta =\infty ,\mu ;c^{\#},\nu
^{\#}=0\right) >0$ ensures $\left| \widehat{c}\left( \beta =\infty ,\mu
\right) \right| ^{2}=\widehat{x}\left( \theta =0,\mu \right) :=
\overline{x}\left( \mu \right) >0$ (see Fig. 2), i.e.
\begin{equation}
\widetilde{p}^{B}\left( \beta =\infty ,\mu ;c^{\#}=0\right) <\widetilde{p}%
^{B}\left( \beta =\infty ,\mu ;\left| \widehat{c}\left( \beta =\infty ,\mu
\right) \right| ^{2}\right) .  \label{eq 4.33}
\end{equation}

$\left( 6\right) $ Since $\partial _{x}\xi \left( \beta ,\mu ;x\right) <0$
(see (\ref{eq 4.29})) and
\begin{equation}
\partial _{\theta }\partial _{x}\xi \left( \beta ,\mu ;x\right) =\frac{%
\left( -1\right) }{\left( 2\pi \right) ^{3}}\stackunder{\Bbb{R}^{3}}{\int }%
d^{3}k\frac{\beta ^{2}E_{k}e^{\beta E_{k}}}{\left( 1-e^{\beta E_{k}}\right)
^{2}}\partial _{x}E_{k}<0,  \label{eq 4.34}
\end{equation}
there is $\theta _{0}\left( \mu \right) $ (cf. Theorem \ref{th for not
equality with P0}) such that for $\mu \in D_{0}$, one gets :
\begin{eqnarray}
\stackunder{x\geq 0}{\sup }\left[ \xi \left( \beta _{0}\left( \mu \right)
,\mu ;x\right) +\eta \left( \mu ;x\right) \right] &=&\xi \left( \beta
_{0}\left( \mu \right) ,\mu ;0\right) +\eta \left( \mu ;0\right)  \nonumber
\\
&=&\xi \left( \beta _{0}\left( \mu \right) ,\mu ;\widehat{x}\left( \theta
_{0}\left( \mu \right) ,\mu \right) >0\right)  \nonumber \\
&&+\eta \left( \mu ;\widehat{x}\left( \theta _{0}\left( \mu \right) ,\mu
\right) >0\right) ,  \label{eq 4.35}
\end{eqnarray}
whereas for $\theta <\theta _{0}\left( \mu \right) $ the supremum is
attained at $x=\widehat{x}\left( \theta ,\mu \right) >0$ and for $\theta
>\theta _{0}\left( \mu \right) $ it "jumps" to $\widehat{x}\left( \theta
,\mu \right) =0$.

\hspace{0.4cm}
Note that by definition of $\widehat{x}\left( \theta,\mu \right)$ and by (\ref{eq 4.25}) one gets
\[
\widehat{x}\left( \theta,\mu \right) =
\left| \widehat{c}\left( \beta,\mu \right) \right| ^{2} =
\rho _{0}^{B}\left(\beta, \mu \right) \, .
\]
Therefore, we have just proved the following assertion.
%%%%%%%%%%%%%%%%%%%%%%%%%%%%%%%%%%%%%%%%%%%%%%%%%%%%%%%%%%%%%%%%%%%%%%%%%%%%%%%%%%%%%%%%%%%%
\begin{theorem}
\label{theorem 4.7}If interaction potential $v\left( k\right) $ satisfies
conditions {\rm{(A), (B)}}, and {\rm{(C)}}, then domain $D\neq \left\{ \emptyset \right\} $
and dynamical condensate undergo a jump on the boundary $\partial D$ :
\begin{equation}
\rho_{0}^{B}\left(\beta = \theta^{-1} , \mu \right) =\left\{
\begin{array}{c}
\hspace{-0.5cm}>0\text{, } \left( \theta ,\mu \right) \in D \\
=0 \text{, }\left( \theta ,\mu \right) \in Q\backslash \overline{D}
\end{array}
\right\} .  \label{eq 4.36}
\end{equation}
\end{theorem}

\hspace{0.4cm}
Behaviour of the non-conventional dynamical condensate (\ref{eq 4.36}) is illustrated by Figure 4,
where its density  is denoted  by $\rho_{0}^{B}(\theta, \mu)$.

%%%%%%%%%%%%%%%%%%%%%%%%%%%%%%%%%%%%%%%%%%%%%%%%%%%%%%%%%%%%%%%%%%%%%%%%%%%%%%%%%%%%%%%%%%%%%%%%%%
\setcounter{equation}{0}
\section{Conventional BEC in Weakly Imperfect Bose-Gas}\label{section 5}
%%%%%%%%%%%%%%%%%%%%%%%%%%%%%%%%%%%%%%%%%%%%%%%%%%%%%%%%%%%%%%%%%%%%%%%%%%%%%%%%%%%%%%%%%%%%%%%%%%
First we establish that (similar to the PBG) the total particle density $\rho
^{B}\left( \theta ,\mu \right) $ of the WIBG is saturated when $\mu \rightarrow - \,0$
(or $\mu \uparrow 0$), i.e.
there exists a critical particle density $\rho _{c}^{B}\left(
\theta \right) = \ \stackunder{\mu \, \uparrow \, 0}{\lim }\rho ^{B}\left(
\theta ,\mu \right) $. Indeed, using the Griffiths Lemma (Section \ref{Griffiths lemma})
and Theorem \ref{theorem 3.16}, Theorem \ref {theorem 4.7}, we obtain for the
grand-canonical total particle density in the WIBG:
\begin{eqnarray}
\rho ^{B}\left( \theta ,\mu \right) &\equiv &\stackunder{\Lambda }{\lim }%
\omega _{\Lambda }^{B}\left( \frac{N_{\Lambda }}{V}\right) = \ \stackunder{%
\Lambda }{\lim }\frac{1}{V}\stackunder{k\in \Lambda ^{*}}{\sum }\omega
_{\Lambda }^{B}\left( N_{k}\right) =  \nonumber \\
&=&\stackunder{\Lambda }{\lim }\partial _{\mu }p_{\Lambda }^{B}\left( \beta
,\mu \right) =\partial _{\mu }\widetilde{p}^{B}\left( \beta ,\mu ;0\right) =
\nonumber \\
&=&\frac{1}{\left( 2\pi \right) ^{3}}\stackunder{\Bbb{R}^{3}}{\int }\left(
e^{\beta \left( \varepsilon _{k}-\mu \right) }-1\right) ^{-1}d^{3}k,
\label{eq cond15}
\end{eqnarray}
for $\left( \theta ,\mu <0\right) \in Q\backslash \overline{D}$, whereas for
$\left( \theta ,\mu <0\right) \in D$ one has:
\begin{eqnarray}
\rho ^{B}\left( \theta ,\mu \right) &=&\partial _{\mu }\widetilde{p}%
^{B}\left( \beta ,\mu ;\widehat{c}^{\#}\left( \theta ,\mu \right) \neq
0\right)  \nonumber \\
&=&\frac{1}{\left( 2\pi \right) ^{3}}\stackunder{\Bbb{R}^{3}}{\int }\left[
\frac{f_{k}}{E_{k}}\left( e^{\beta E_{k}}-1\right) ^{-1}+\frac{h_{k}^{2}}{%
2E_{k}\left( f_{k}+E_{k}\right) }\right] _{c=\widehat{c}\left( \theta ,\mu
\right) }d^{3}k  \nonumber \\
&&+\left| \widehat{c}\left( \theta ,\mu \right) \right| ^{2}.
\label{eq cond16}
\end{eqnarray}
Then, from (\ref{eq cond15}) and (\ref{eq cond16}), we see that the total
particle density $\rho ^{B}\left( \theta ,\mu \right) $ reaches its maximal
(critical) value $\rho _{c}^{B}\left( \theta \right) \equiv \rho ^{B}\left(
\theta ,\mu =0\right) $ at $\mu =0$ :

$\left( i\right) $ for $\theta >\theta _{0}\left( \mu =0\right) $ one gets
\begin{equation}
\rho _{c}^{B}\left( \theta \right) =\frac{1}{\left( 2\pi \right) ^{3}}%
\stackunder{\Bbb{R}^{3}}{\int }\left( e^{\beta \varepsilon _{k}}-1\right)
^{-1}d^{3}k=\rho _{c}^{P}\left( \theta \right) <+\infty ,  \label{eq cond17}
\end{equation}

$\left( ii\right) $ for $\theta <\theta _{0}\left( \mu =0\right) $ one has
\begin{eqnarray}
\rho _{c}^{B}\left( \theta \right) &=&\frac{1}{\left( 2\pi \right) ^{3}}%
\stackunder{\Bbb{R}^{3}}{\int }\left[ \frac{f_{k}}{E_{k}}\left( e^{\beta
E_{k}}-1\right) ^{-1}+\frac{h_{k}^{2}}{2E_{k}\left( f_{k}+E_{k}\right) }%
\right] \Sb c=\widehat{c}\left( \theta ,0\right)  \\ \mu =0  \endSb d^{3}k
\nonumber \\
&&+\left| \widehat{c}\left( \theta ,\mu =0\right) \right| ^{2}<+\infty ,
\label{eq cond18}
\end{eqnarray}
since the non-conventional condensation density $\rho_{0}^{B}(\theta, \mu)$ is saturated
for $\mu =0$ by $\left| \widehat{c}\left( \theta ,\mu =0\right)\right| ^{2}$, see
Theorem \ref{theorem 4.7} and Figure 5.

\hspace{0.4cm}
Note that convexity of $p^{B}\left( \beta,\mu \right) $ with respect to $\mu $ yields
that the function $\rho^{B}\left( \theta ,\mu \right) $ is monotonous and
\begin{equation}
\stackunder{\mu \, \uparrow \, \mu _{0}\left( \theta \right)}{\lim }\rho
^{B}\left( \theta ,\mu \right) =: \rho _{\inf }^{B}\left( \theta \right)
<\stackunder{\mu \, \downarrow \,  \mu _{0}\left( \theta \right)}{\lim }\rho
^{B}\left( \theta ,\mu \right) =: \rho _{\sup }^{B}\left( \theta \right)
,  \label{eq cond18bis}
\end{equation}
where $\mu _{0}\left( \theta \right) $ is the inverse function of $\theta
_{0}\left( \mu \right) $ and
\begin{equation}
\stackunder{\theta \, \downarrow \, \theta _{0}\left( 0\right)}{\lim }\rho_{c}^{B}
\left( \theta \right) <\stackunder{\theta \, \uparrow \,\theta_{0}\left( 0\right)}
{\lim }\rho_{c}^{B}\left( \theta \right) .
\label{eq cond19}
\end{equation}
The $\mu \mapsto \rho ^{B}\left( \theta ,\mu \right)$ for the \textit{total} particle density of
the Weakly Imperfect Bose-Gas is illustrated by Figure 5.

\hspace{0.4cm}
Now we consider the WIBG model for temperatures and total particle
densities as given parameters. Note that for any finite domain $\Lambda$ there exists
\[
\varepsilon _{\Lambda ,1}\in \left[ \stackunder{k\neq 0}{\inf }\left(
\varepsilon _{k}-\frac{v\left( k\right) }{2V}\right) , \widehat{\varepsilon }%
_{\Lambda ,1} = \stackunder{k\neq 0}{\inf }\varepsilon _{k} = \varepsilon_{\left\| k\right\| =
\frac{2\pi }{L}}\right] ,
\]
(see \textbf{7.1}) such that for $\mu <\varepsilon _{\Lambda ,1}<\varepsilon _{\left\| k\right\| =
\frac{2\pi }{L}}, $ one gets
\[
\omega _{\Lambda }^{B}\left( \frac{N_{\Lambda }}{V}\right) < + \infty ,
\]
although the limit
\begin{equation}
\stackunder{\mu \, \uparrow \, \varepsilon _{\Lambda ,1}}{\lim }\omega
_{\Lambda }^{B}\left( \frac{N_{\Lambda }}{V}\right) = + \infty ,
\label{eq condsup5}
\end{equation}
see Lemma D.1 in \cite{BZ00}.
Therefore, for any $\rho >0,$ there is a unique value of the chemical
potential $\mu _{\Lambda }^{B}\left( \theta ,\rho \right) <\varepsilon
_{\Lambda ,1}$, which satisfies the equation:
\begin{equation}
\left\langle \frac{N_{\Lambda }}{V}\right\rangle _{H_{\Lambda }^{B}}\left(
\beta ,\mu _{\Lambda }^{B}\left( \theta ,\rho \right) \right) =\omega
_{\Lambda }^{B}\left( \frac{N_{\Lambda }}{V}\right) =\rho .
\label{eq cond42}
\end{equation}

\hspace{0.4cm}
Note that for $\rho <\rho _{c}^{B}\left( \theta \right) $ the
monotonicity of $\rho ^{B}\left( \theta ,\mu \right) $ for $\mu \leq 0$
implies that (\ref{eq cond42}) has a unique solution
\[
\mu ^{B}\left( \theta ,\rho \right) =\text{ }\stackunder{\Lambda }{\lim }\mu
_{\Lambda }^{B}\left( \theta ,\rho \right) <0 \, ,
\]
\textit{independent} of the presence of the \textit{non-conventional} condensation,
see (\ref{eq cond15}) and (\ref{eq cond16}). Therefore, below the saturation
limit $\rho _{c}^{B}\left( \theta \right) $ only the {non-conventional} condensate
(\ref{eq 4.36}) is possible.

\hspace{0.4cm}
In the rest of this section we consider the case $\rho \geq \rho _{c}^{B}\left(
\theta \right) $. In general, for any $\rho \geq \rho _{c}^{B}\left( \theta
\right) $, one gets by (\ref{eq condsup5}) and (\ref{eq cond42}) that
$\mu_{\Lambda }^{B}\left( \theta ,\rho \right) \gtrless 0$ and
\begin{equation}
\stackunder{\Lambda }{\lim }\mu _{\Lambda }^{B}\left( \theta ,\rho \geq \rho
_{c}^{B}\left( \theta \right) \right) =0.  \label{eq cond43bis}
\end{equation}
From now on we set for the Gibbs state at a fixed density $\rho$:
\begin{equation}
\omega _{\Lambda ,\rho }^{B}\left( -\right) := \omega _{\Lambda
}^{B}\left( -\right) \mid _{\mu =\mu _{\Lambda }^{B}\left( \theta ,\rho
\right) }.  \label{eq cond43}
\end{equation}

\hspace{0.4cm}
According to Section \ref{section 4} the WIBG non-conventional condensate
in the mode $k=0$ is saturated for $\mu \, \uparrow \, 0$ either by
$\left|\widehat{c}\left( \theta ,0\right) \right| ^{2}>0$ (for $\theta <\theta
_{0}\left( 0\right) $), or by $\left| \widehat{c}\left( \theta ,0\right)
\right| ^{2}=0$ (for $\theta >\theta _{0}\left( 0\right) $), see (\ref{eq 4.36}) and Figure 4.
Therefore, (\ref{eq cond15})-(\ref{eq cond18}) and saturation of the total particle
density $\rho^{B}\left( \theta ,\mu \right) $ for $\mu \, \uparrow \, 0$ yield
a \textit{conventional} Bose-condensation in modes \textit{next} to $k=0$.

\hspace{0.4cm}
For discussion of coexistence of these \textit{two} kind of condensations in
the framework of our toy model see Section \ref{sec:3}.

\hspace{0.4cm}
To control the conventional condensation for $k\neq 0$ we introduce an auxiliary Hamiltonian
\[
H_{\Lambda ,\alpha }^{B}=H_{\Lambda }^{B}-\alpha \stackunder{k\in \Lambda
^{*},\ a<\left\| k\right\| <b}{\sum }a_{k}^{*}a_{k},
\]
for $0<a<b$. We set
\begin{equation}
p_{\Lambda }^{B}\left( \beta ,\mu ,\alpha \right) \equiv \frac{1}{\beta V}%
\ln Tr_{\mathcal{F}_{\Lambda }}e^{-\beta H_{\Lambda ,\alpha }^{B}\left( \mu
\right) },  \label{eq cond44}
\end{equation}
and
\[
\omega _{\Lambda }^{B,\alpha }\left( -\right) := \left\langle
-\right\rangle _{H_{\Lambda ,\alpha }^{B}}\left( \beta ,\mu \right)
\]
for grand-canonical Gibbs state corresponding to $H_{\Lambda ,\alpha}^{B}\left( \mu \right) $.

Recall that $\mu _{0}\left( \theta \right)$ is the function (inverse to
$\theta _{0}\left( \mu \right) $), which defines
the borderline of domain $D$, see Figure 4.
%%%%%%%%%%%%%%%%%%%%%%%%%%%%%%%%%%%%%%%%%%%%%%%%%%%%%%%%%%%%%%%%%%%%%%%%%%%%%%%%%%%%%%%%%%%%%%%%
\begin{proposition}\label{theorem cond4}
Let $\alpha \in \left[ -\delta ,\delta \right] $ where $%
0\leq \delta \leq \varepsilon _{a}/2$ and $\varepsilon _{a}= \ \stackunder{%
\left\| k\right\| \geq a}{\inf }\varepsilon _{k}$. Then there exists a
domain $D_{\delta }\subset D$:
\begin{equation}
D_{\delta }\equiv \left\{ \left( \theta ,\mu \right) :\mu _{0}<\mu
_{0}\left( \delta \right) \leq \mu \leq 0,\text{ }0\leq \theta \leq \theta
_{0}\left( \mu ,\delta \right) <\theta _{0}\left( \mu \right) \right\}
\label{eq condbis53}
\end{equation}
such that
\begin{equation}
\left| p_{\Lambda }^{B}\left( \beta ,\mu ,\alpha \right) -\stackunder{c\in
\Bbb{C}}{\sup }\widetilde{p}_{\Lambda }^{B}\left( \beta ,\mu ,\alpha
;c^{\#}\right) \right| \leq \frac{K\left( \delta \right) }{\sqrt{V}}
\label{eq cond53}
\end{equation}
for $V$ sufficiently large, uniformly in $\alpha \in \left[ -\delta ,\delta
\right] $ and for:
\begin{equation}
\begin{array}{l}
\left( i\right) \text{ }\left( \theta ,\mu \right) \in D_{\delta }\text{, if
}\mu _{\Lambda }^{B}\left( \theta ,\rho \geq \rho _{c}^{B}\left( \theta
\right) \right) \leq 0; \\
\text{ \qquad or } \\
\left( ii\right) \text{ }\left( \theta ,\mu \right) \in D_{\delta }\cup
\left\{ \left( \theta ,\mu \right) :0\leq \mu \leq \mu _{\Lambda }^{B}\left(
\theta ,\rho \geq \rho _{c}^{B}\left( \theta \right) \right) ,\text{ }0\leq
\theta \leq \theta _{0}\left( \mu =0,\delta \right) \right\} \text{, } \\
\text{if }\mu _{\Lambda }^{B}\left( \theta ,\rho \geq \rho _{c}^{B}\left(
\theta \right) \right) \geq 0.
\end{array}
\label{eq cond53bis}
\end{equation}
\end{proposition}
%%%%%%%%%%%%%%%%%%%%%%%%%%%%%%%%%%%%%%%%%%%%%%%%%%%%%%%%%%%%%%%%%%%%%%%%%%%%%%%%%%%%%%%%%%%%%
\textit{Proof}. The existence of the domain $D_{\delta }$ follows from the
proof of Theorem \ref{theorem 3.16}. This means that the estimate (\ref{eq cond53}%
) is stable with respect to local perturbations of the free-particle
spectrum: $\varepsilon _{k}\rightarrow \varepsilon _{k}-\alpha \chi _{\left(
a,b\right) }\left( \left\| k\right\| \right) $ for $\left| \alpha \right|
\leq \delta \leq \varepsilon _{a}/2$ in a reduced domain $D_{\delta }\subset
D$. Here $\chi _{\left( a,b\right) }\left( \left\| k\right\| \right) $ is
the characteristic function of interval $\left( a,b\right) \subset \Bbb{R}$.
Extension in (\ref{eq cond53bis}) is due to continuity of the pressure $%
p_{\Lambda }^{B}\left( \beta ,\mu ,\alpha \right) $ and the trial pressure $%
\widetilde{p}_{\Lambda }^{B}\left( \beta ,\mu ,\alpha ;c^{\#}\right) $ in
parameters $\alpha \in \left[ -\delta ,\delta \right] $ and $\mu \leq \mu
_{\Lambda }^{B}\left( \theta ,\rho \geq \rho _{c}^{B}\left( \theta \right)
\right) $, see (\ref{eq cond42}), (\ref{eq cond43bis}). \hfill $\square$

\begin{corollary}
\label{corollary cond6}Let $\rho \geq \rho _{c}^{B}\left( \theta \right) $,
see (\ref{eq cond17}), (\ref{eq cond18}). Then for $\theta <\theta
_{0}\left( 0\right) $ one has
\begin{eqnarray}
&&\stackunder{\Lambda }{\lim }\frac{1}{V}\stackunder{k\in \Lambda
^{*}, \ a<\left\| k\right\| <b}{\sum }\omega _{\Lambda ,\rho }^{B}\left(
N_{k}\right) = \label{eq cond50.1} \\
&&\frac{1}{\left( 2\pi \right) ^{3}}\stackunder{a<\left\|
k\right\| <b}{\int } d^{3}k \left[ \frac{f_{k}}{E_{k}}\left( e^{\beta
E_{k}}-1\right) ^{-1}+\frac{h_{k}^{2}}{2E_{k}\left( f_{k}+E_{k}\right) }%
\right] \Sb c=\widehat{c}\left( \theta ,0\right)  \\ \mu =0 \endSb  \ ,
\nonumber
\end{eqnarray}
whereas for $\theta >\theta _{0}\left( 0\right) $ one gets
\begin{equation}
\stackunder{\Lambda }{\lim }\frac{1}{V}\stackunder{k\in \Lambda
^{*},\ a<\left\| k\right\| <b}{\sum }\omega _{\Lambda ,\rho }^{B}\left(
N_{k}\right) =\frac{1}{\left( 2\pi \right) ^{3}}\stackunder{a<\left\|
k\right\| <b}{\int } d^{3}k \left( e^{\beta \varepsilon _{k}}-1\right) ^{-1} \ .
\label{eq cond50.2}
\end{equation}
\end{corollary}
%%%%%%%%%%%%%%%%%%%%%%%%%%%%%%%%%%%%%%%%%%%%%%%%%%%%%%%%%%%%%%%%%%%%%%%%%%%%%%%%%%%%%%%%%%
\textit{Proof.} Consider the sequence of functions $\left\{ p_{\Lambda
}^{B}\left( \beta ,\mu _{\Lambda }^{B}\left( \theta ,\rho \right) ,\alpha
\right) \right\} _{\Lambda }$ defined by (\ref{eq cond44}), where chemical
potential is a solution of (\ref{eq cond42}), for the corresponding
Hamiltonian and $\alpha \in \left[ -\delta ,\delta \right] $. Since by (\ref
{eq cond44})
\begin{equation}
\partial _{\alpha }p_{\Lambda }^{B}\left( \beta ,\mu _{\Lambda }^{B}\left(
\theta ,\rho \right) ,\alpha \right) =\frac{1}{V}\stackunder{k\in \Lambda
^{*},a<\left\| k\right\| <b}{\sum }\omega _{\Lambda ,\rho }^{B,\alpha
}\left( N_{k}\right) , \label{eq cond50.3}
\end{equation}
and $\left\{ p_{\Lambda }^{B}\left( \beta ,\mu _{\Lambda }^{B}\left( \theta
,\rho \right) ,\alpha \right) \right\} _{\Lambda }$ are convex functions of $%
\alpha \in \left[ -\delta ,\delta \right] $, Proposition \ref{theorem cond4}
and the Griffiths Lemma imply
\begin{eqnarray}
&&\stackunder{\Lambda }{\lim }\partial _{\alpha }p_{\Lambda }^{B}\left( \beta
,\mu _{\Lambda }^{B}\left( \theta ,\rho \right) ,\alpha \right) = \
\stackunder{\Lambda }{\lim }\frac{1}{V}\stackunder{k\in \Lambda
^{*},a<\left\| k\right\| <b}{\sum }\omega _{\Lambda ,\rho }^{B,\alpha
}\left( N_{k}\right) = \nonumber \\
&&\partial _{\alpha }\stackunder{\Lambda }{\lim }\text{ }\stackunder{c\in
\Bbb{C}}{\sup }\widetilde{p}_{\Lambda }^{B}\left( \beta ,\mu _{\Lambda
}^{B}\left( \theta ,\rho \right) ,\alpha ;c^{\#}\right) ,
\label{eq cond50.5}
\end{eqnarray}
for $\alpha \in \left[ -\delta ,\delta \right] $. By explicit calculations
in the right-hand side of (\ref{eq cond50.5}) one obtains for $\alpha =0$
equalities (\ref{eq cond50.1}) and (\ref{eq cond50.2}). \hfill $\square$

\begin{remark}
\label{remark cond8}Note that mean particle values $\omega _{\Lambda
}^{B}\left( N_{k}\right) =\left\langle N_{k}\right\rangle _{H_{\Lambda
}^{B}}\left( \beta ,\mu \right) $ (and similar $\omega _{\Lambda ,\rho
}^{B}\left( N_{k}\right) =\left\langle N_{k}\right\rangle _{H_{\Lambda
}^{B}}\left( \beta ,\mu _{\Lambda }^{B}\left( \theta ,\rho \right) \right) $%
) are defined on the discrete set $\Lambda ^{*}$ (\ref{dual-Lambda}). Below
we denote by $\left\{ \omega _{\Lambda }^{B}\left( N_{k}\right) \right\}
_{k\in \Bbb{R}^{3}}$ a \textit{continuous interpolation }of these values
from the set $\Lambda ^{*}$ to $\Bbb{R}^{3}$.
\end{remark}

\hspace{0.4cm}
Now we are in position to prove the main statement of this
section about non-conventional and conventional condensations showing up in the
WIBG for densities $\rho >\rho _{c}^{B}\left( \theta \right) $.
%%%%%%%%%%%%%%%%%%%%%%%%%%%%%%%%%%%%%%%%%%%%%%%%%%%%%%%%%%%%%%%%%%%%%%%%%%%%%%%%%%%%%%%%%%%%%%%%%%%%
\begin{theorem}
\label{theorem cond5}Let $\rho >\rho _{c}^{B}\left( \theta \right) $. Then
we have that :\newline
$\left( i\right) $%
\begin{equation}
\rho _{0}^{B}\left( \theta ,0\right) =\text{ }\stackunder{\Lambda }{\lim }%
\omega _{\Lambda ,\rho }^{B}\left( \frac{a_{0}^{*}a_{0}}{V}\right) =\left\{
\begin{array}{c}
\left| \widehat{c}\left( \theta ,0\right) \right| ^{2}\text{, }\theta
<\theta _{0}\left( 0\right)  \\
0\text{, }\theta >\theta _{0}\left( 0\right)
\end{array}
\right\} ;  \label{eq cond68.1}
\end{equation}
\newline
$\left( ii\right) $ for any $k\in \Lambda ^{*},$ such that $\left\|
k\right\| >\frac{2\pi }{L},$%
\begin{equation}
\stackunder{\Lambda }{\lim }\omega _{\Lambda ,\rho }^{B}\left( \frac{N_{k}}{V%
}\right) =0;  \label{eq cond68.2}
\end{equation}
\newline
$\left( iii\right) $ for $\theta <\theta _{0}\left( 0\right) $ and for all $%
k\in \Lambda ^{*},$ such that $\left\| k\right\| >\delta >0$%
\begin{equation}
\stackunder{\Lambda }{\lim }\omega _{\Lambda ,\rho }^{B}\left( N_{k}\right)
=\left[ \frac{f_{k}}{E_{k}}\left( e^{\beta E_{k}}-1\right) ^{-1}+\frac{%
h_{k}^{2}}{2E_{k}\left( f_{k}+E_{k}\right) }\right] \Sb c=\widehat{c}\left(
\theta ,0\right)  \\ \mu =0 \endSb   \label{eq cond68.2.1}
\end{equation}
whereas for $\theta >\theta _{0}\left( 0\right) $%
\begin{equation}
\stackunder{\Lambda }{\lim }\omega _{\Lambda ,\rho }^{B}\left( N_{k}\right) =%
\frac{1}{e^{\beta \varepsilon _{k}}-1};  \label{eq cond68.2.2}
\end{equation}
\newline
$\left( iv\right) $ the double limit
\begin{equation}
\widetilde{\rho }_{0}^{B}\left( \theta \right) := \text{ }\stackunder{%
\delta \rightarrow 0^{+}}{\lim }\stackunder{\Lambda }{\lim }\frac{1}{V}%
\stackunder{\left\{ k\in \Lambda ^{*},0<\left\| k\right\| \leq \delta
\right\} }{\sum }\omega _{\Lambda ,\rho }^{B}\left( N_{k}\right) =\rho -\rho
_{c}^{B}\left( \theta \right) ,  \label{eq cond68.3}
\end{equation}
which means that the WIBG manifests a \textit{conventional} \textit{%
(generalised)} Bose condensation $\widetilde{\rho }_{0}^{B}\left( \theta
\right) >0$ in modes next to the zero-mode due to particle density
saturation.
\end{theorem}
%%%%%%%%%%%%%%%%%%%%%%%%%%%%%%%%%%%%%%%%%%%%%%%%%%%%%%%%%%%%%%%%%%%%%%%%%%%%%%%%%%%%%%%%%%%%%%%%%%
\textit{Proof. }$\left( i\right)$ Since by (\ref{eq cond43bis}) we have
\begin{equation}
\stackunder{\Lambda }{\lim }\mu _{\Lambda }^{B}\left( \theta ,\rho \right)
=0,  \label{eq cond66.1}
\end{equation}
the thermodynamic limit (\ref{eq cond68.1}) results from Theorem 4.4 and
Corollary 4.8 of \cite{BZ98JP}, see (\ref{eq 4.36}) for $\mu =0$.

$\left( ii\right) $ Since $\left\| k\right\| > {2\pi }/{L}$ and $\Lambda
=L\times L\times L$ is a cube, which excludes generalized Bose-Einstein
condensation due to anisotropy (see Section \ref{classification des
condensations}), the thermodynamic limit (\ref{eq cond68.2}) follows from $%
\mu _{\Lambda }^{B}\left( \theta ,\rho \right) <\varepsilon _{\left\|
k\right\| = {2\pi }/{L}}$ and estimate (D.10) in Lemma D.2 of \cite{BZ00}.

$\left( iii\right) $ Let us consider $g_{\theta }\left( k\right) $ defined
for $k\in \Bbb{R}^{3}$, $\left\| k\right\| >\delta >0$ by
\begin{equation}
g_{\theta }\left( k\right) := \text{ }\stackunder{\Lambda }{\lim }\omega
_{\Lambda ,\rho }^{B}\left( N_{k}\right) ,  \label{eq cond66.2}
\end{equation}
where the state $\omega _{\Lambda ,\rho }^{B}\left( -\right) $ stands for $%
\omega _{\Lambda }^{B}\left( -\right) $ with $\mu =\mu _{\Lambda }^{B}\left(
\theta ,\rho \right) $, cf. (\ref{eq cond43}). Note that by Lemma D.2 of \cite{BZ00}
and the fact that
\[
\mu _{\Lambda }^{B}\left( \theta ,\rho \right) <\varepsilon _{\Lambda ,1}<%
\text{ }\stackunder{k\neq 0}{\inf }\varepsilon _{k}=\varepsilon _{\left\|
k\right\| = {2\pi }/{L}},
\]
the thermodynamic limit (\ref{eq cond66.2}) exists and it is informly
bounded for $\left\| k\right\| >\delta >0$. Moreover, for any interval $%
\left( a>\delta ,b\right) $ we have
\[
\stackunder{\Lambda }{\lim }\frac{1}{V}\stackunder{k\in \Lambda ^{*},\left\|
k\right\| \in \left( a,b\right) }{\sum }\omega _{\Lambda ,\rho }^{B}\left(
N_{k}\right) =\frac{1}{\left( 2\pi \right) ^{3}}\stackunder{\left\|
k\right\| >\delta }{\int } d^{3}k \ g_{\theta }\left( k\right) \chi _{\left(
a,b\right) }\left( \left\| k\right\| \right) \ ,
\]
where again $\chi _{\left( a,b\right) }\left( \left\| k\right\| \right) $ is
the characteristic function of $\left( a,b\right) $. Then Corollary \ref
{corollary cond6} implies that
\begin{equation}
\frac{1}{\left( 2\pi \right) ^{3}}\stackunder{\left\| k\right\| >\delta }{
\int }  d^{3}k g_{\theta }\left( k\right) \chi _{\left( a,b\right) }\left( \left\|
k\right\| \right) = \frac{1}{\left( 2\pi \right) ^{3}}\stackunder{%
\left\| k\right\| >\delta }{\int } d^{3}k \ f_{\theta }\left( k\right) \chi _{\left(
a,b\right) }\left( \left\| k\right\| \right) \ ,  \label{eq cond66.3}
\end{equation}
where $f_{\theta }\left( k\right) $ is a continuous function on $k\in \Bbb{R}%
^{3}$ defined by (\ref{eq cond50.1}), (\ref{eq cond50.2}), i.e.,
\begin{equation}
f_{\theta }\left( k\right):= \frac{1}{\left( 2\pi \right) ^{3}}\left[
\frac{f_{k}}{E_{k}}\left( e^{\beta E_{k}}-1\right) ^{-1}+\frac{h_{k}^{2}}{%
2E_{k}\left( f_{k}+E_{k}\right) }\right] \Sb c=\widehat{c}\left( \theta
,0\right)  \\ \mu =0  \endSb \ ,  \label{eq cond66.4}
\end{equation}
for $\theta <\theta _{0}\left( 0\right) $ and
\begin{equation}
f_{\theta }\left( k\right) := \frac{1}{\left( 2\pi \right) ^{3}}\left(
e^{\beta \varepsilon _{k}}-1\right) ^{-1},  \label{eq cond66.5}
\end{equation}
for $\theta >\theta _{0}\left( 0\right) $. Since the relation (\ref{eq
cond66.3}) is valid for any interval $\left( a>\delta ,b\right) \subset \Bbb{%
R}$ one gets
\[
g_{\theta }\left( k\right) =f_{\theta }\left( k\right) ,\text{ }k\in \Bbb{R}%
^{3},\text{ }\left\| k\right\| >\delta >0.
\]
By this and (\ref{eq cond66.2})-(\ref{eq cond66.5}) we deduce (\ref{eq
cond68.2.1}) and (\ref{eq cond68.2.2}).

$\left( iv\right) $ Since the total density $\rho $ is fixed, by definition (%
\ref{eq cond43}) we have
\begin{equation}
\frac{1}{V}\stackunder{\left\{ k\in \Lambda ^{*},0<\left\| k\right\| \leq
\delta \right\} }{\sum }\omega _{\Lambda ,\rho }^{B}\left( N_{k}\right)
=\rho -\omega _{\Lambda ,\rho }^{B}\left( \frac{a_{0}^{*}a_{0}}{V}\right) -%
\frac{1}{V}\stackunder{\left\{ k\in \Lambda ^{*}:\left\| k\right\| >\delta
\right\} }{\sum }\omega _{\Lambda ,\rho }^{B}\left( N_{k}\right) .
\label{eq cond68.9}
\end{equation}
By Corollary \ref{corollary cond6} for $a=\delta $ and $b\rightarrow +\infty
$ we obtain for $\theta <\theta _{0}\left( 0\right) $
\begin{equation}
\stackunder{\Lambda }{\lim }\frac{1}{V}\stackunder{\left\{ k\in \Lambda
^{*}:\left\| k\right\| >\delta \right\} }{\sum }\omega _{\Lambda ,\rho
}^{B}\left( N_{k}\right) =\frac{1}{\left( 2\pi \right) ^{3}}\stackunder{%
\left\| k\right\| >\delta }{\int } d^{3}k \ \left[ \frac{f_{k}}{E_{k}}\left( e^{\beta
E_{k}}-1\right) ^{-1}+\frac{h_{k}^{2}}{2E_{k}\left( f_{k}+E_{k}\right) }%
\right] \Sb c=\widehat{c}\left( \theta ,0\right)  \\ \mu =0  \endSb  \ ,
\label{eq cond68.11}
\end{equation}
and for $\theta >\theta _{0}\left( 0\right) $%
\begin{equation}
\stackunder{\Lambda }{\lim }\frac{1}{V}\stackunder{k\in \Lambda ^{*},\left\|
k\right\| >\delta }{\sum }\omega _{\Lambda ,\rho }^{B}\left( N_{k}\right) =%
\frac{1}{\left( 2\pi \right) ^{3}}\stackunder{\left\| k\right\| >\delta }{%
\int }d^{3}k \, \left( e^{\beta \varepsilon _{k}}-1\right) ^{-1} \ .
\label{eq cond68.11.1}
\end{equation}
Now, from (\ref{eq cond17}), (\ref{eq cond18}), (\ref{eq cond68.1}), (\ref
{eq cond68.9})-(\ref{eq cond68.11.1}) we deduce (\ref{eq cond68.3}) by
taking the limit $\delta \, \downarrow\, 0$. \hfill $\square $

\hspace{0.4cm}
Therefore, according to (\ref{eq cond68.3})
%(and in a close similarity to \cite{11})
for $\theta >\theta _{0}\left( 0\right) $ and $\rho
>\rho _{c}^{B}\left( \theta \right) $ the WIBG manifests only \textit{one
kind} of condensation, namely the \textit{conventional} Bose-Einstein
condensation which occurs in modes $k\neq 0$, whereas for $\theta <\theta
_{0}\left( 0\right) $ it manifests for $\rho >\rho _{c}^{B}\left( \theta
\right) $ this kind of condensation at the \textit{second} stage after the
\textit{non-conventional} Bose condensation $\left| \widehat{c}\left( \theta
,0\right) \right| ^{2}$, see (\ref{eq cond68.1}). For classification of
different types of condensations see Section \ref{classification des condensations}.
%%%%%%%%%%%%%%%%%%%%%%%%%%%%%%%%%%%%%%%%%%%%%%%%%%%%%%%%%%%%%%%%%%%%%%%%%%%%%%%%%%%%%%%%%%%%%%%%%%
\begin{remark}
\label{remark cond7}In domain: $\theta <\theta _{0}\left( 0\right) $, $\rho
>\rho _{c}^{B}\left( \theta \right) $, we have coexistence of these two
kinds of condensations, namely:\newline
- the non-conventional one which starts when $\rho $ becomes larger than $%
\rho _{\sup }^{B}\left( \theta \right) $, see (\ref{eq cond18bis}) and
Figure 2 (b), and which reaches its maximal value $\rho _{0}^{B}\left( \theta
,0\right) $ for $\rho \geq \rho _{c}^{B}\left( \theta \right) >\rho _{\sup
}^{B}\left( \theta \right) ;$\newline
- and the conventional Bose condensation $\widetilde{\rho }_{0}^{B}\left(
\theta \right) $ which appears when $\rho >\rho _{c}^{B}\left( \theta
\right) $, see (\ref{eq cond68.3}).
\end{remark}

\hspace{0.4cm}
Since the Bose-Einstein condensation (\ref{eq cond68.3}) occurs
in modes $k\neq 0$, it should be classified as a \textit{generalised }%
condensation. According to the van den Berg-Lewis-Pul\`{e} classification
(see Section \ref{classification des condensations}), from (\ref
{eq cond68.2}) and (\ref{eq cond68.3}) we can deduce only that the \textit{%
generalised conventional} condensation in the WIBG can be either a
condensation of type I in modes $\left\| k\right\| =2\pi /L$, or a
condensation of type III if modes $\left\| k\right\| =2\pi /L$ are not
macroscopically occupied (non-extensive condensation), or finally it can be
a combination of the two.

\begin{corollary}
\label{corollary cond5}For $\rho >\rho _{c}^{B}\left( \theta \right) $ and
periodic boundary conditions on $\partial \Lambda$ the (generalised)
conventional condensation (\ref{eq cond68.3}) is of type I in the first $%
2d(=6)$ modes next to the zero-mode $k=0$, i.e.
\begin{equation}
\widetilde{\rho }_{0}^{B}\left( \theta \right) =\text{ }\stackunder{\Lambda
}{\lim }\frac{1}{V}\stackunder{\left\{ k\in \Lambda ^{*},\left\| k\right\| =%
\frac{2\pi }{L}\right\} }{\sum }\omega _{\Lambda ,\rho }^{B}\left(
a_{k}^{*}a_{k}\right) =\rho -\rho _{c}^{B}\left( \theta \right) .
\label{eq condnew8}
\end{equation}
\end{corollary}
%%%%%%%%%%%%%%%%%%%%%%%%%%%%%%%%%%%%%%%%%%%%%%%%%%%%%%%%%%%%%%%%%%%%%%%%%%%%%%%%%%%%%%%%%%%%
\textit{\textit{Proof. }}Since for $\delta >0$%
\begin{eqnarray*}
\frac{1}{V}\stackunder{\left\{ k\in \Lambda ^{*},\left\| k\right\| =\frac{%
2\pi }{L}\right\} }{\sum }\omega _{\Lambda ,\rho }^{B}\left( N_{k}\right)
&=&\rho -\omega _{\Lambda ,\rho }^{B}\left( \frac{a_{0}^{*}a_{0}}{V}\right) -%
\frac{1}{V}\stackunder{\left\{ k\in \Lambda ^{*},\frac{2\pi }{L}<\left\|
k\right\| <\delta \right\} }{\sum }\omega _{\Lambda ,\rho }^{B}\left(
N_{k}\right) \\
&&-\frac{1}{V}\stackunder{\left\{ k\in \Lambda ^{*}:\left\| k\right\| \geq
\delta \right\} }{\sum }\omega _{\Lambda ,\rho }^{B}\left( N_{k}\right) ,
\end{eqnarray*}
by Lemma D.2 \cite{BZ00} we obtain
\begin{eqnarray}
\frac{1}{V}\stackunder{\left\{ k\in \Lambda ^{*},\left\| k\right\| =\frac{%
2\pi }{L}\right\} }{\sum }\omega _{\Lambda ,\rho }^{B}\left( N_{k}\right)
&\geq &\rho -\frac{1}{V}\stackunder{\left\{ k\in \Lambda ^{*},\frac{2\pi }{L}%
<\left\| k\right\| <\delta \right\} }{\sum }\frac{1}{e^{B_{k}\left( \mu
_{\Lambda }^{B}\left( \theta ,\rho \right) \right) }-1}  \nonumber \\
&&-\omega _{\Lambda ,\rho }^{B}\left( \frac{a_{0}^{*}a_{0}}{V}\right) \left[
1+\frac{\beta }{2V}\stackunder{\left\{ k\in \Lambda ^{*},\frac{2\pi }{L}%
<\left\| k\right\| <\delta \right\} }{\sum }\frac{v\left( k\right) }{%
1-e^{-B_{k}\left( \mu _{\Lambda }^{B}\left( \theta ,\rho \right) \right) }}%
\right]  \nonumber \\
&&-\frac{1}{V}\stackunder{\left\{ k\in \Lambda ^{*}:\left\| k\right\| \geq
\delta \right\} }{\sum }\omega _{\Lambda ,\rho }^{B}\left( N_{k}\right) ,
\label{eq condnew10}
\end{eqnarray}
where
\[
B_{k}\left( \mu =\mu _{\Lambda }^{B}\left( \theta ,\rho \right) \right)
= \beta \left[ \varepsilon _{k}-\mu _{\Lambda }^{B}\left( \theta ,\rho
\right) -\frac{v\left( k\right) }{2V}\right] .
\]

\hspace{0.4cm}
Since by Lemma D.1 \cite{BZ00} one has
\[
\mu _{\Lambda }^{B}\left( \theta ,\rho \right) <\varepsilon _{\Lambda ,1}<%
\text{ }\stackunder{k\neq 0}{\inf }\varepsilon _{k}=\varepsilon _{\left\|
k\right\| =\frac{2\pi }{L}},
\]
from (\ref{eq cond17}), (\ref{eq cond18}) and (\ref{eq cond68.11}) we deduce
that
\begin{equation}
\stackunder{\Lambda }{\lim }\frac{1}{V}\stackunder{\left\{ k\in \Lambda
^{*},\left\| k\right\| =\frac{2\pi }{L}\right\} }{\sum }\omega _{\Lambda
,\rho }^{B}\left( a_{k}^{*}a_{k}\right) \geq \rho -\rho _{c}^{B}\left(
\theta \right)  \label{eq condnew11}
\end{equation}
by taking the limit $\delta \, \downarrow \, 0$ in the right-hand side of (%
\ref{eq condnew10}) \textit{after} the thermodynamic limit. Hence combining
the inequality
\[
\stackunder{\Lambda }{\lim }\frac{1}{V}\stackunder{\left\{ k\in \Lambda
^{*},\left\| k\right\| =\frac{2\pi }{L}\right\} }{\sum }\omega _{\Lambda
,\rho }^{B}\left( N_{k}\right) \leq \text{ }\stackunder{\Lambda }{\lim }%
\frac{1}{V}\stackunder{\left\{ k\in \Lambda ^{*},0<\left\| k\right\| <\delta
\right\} }{\sum }\omega _{\Lambda ,\rho }^{B}\left( N_{k}\right)
\]
with (\ref{eq cond68.3}) and (\ref{eq condnew11}), we obtain (\ref{eq
condnew8}). \hfill $\square $

\hspace{0.4cm}
Therefore, for temperature $\theta $ and total particle density $%
\rho $ as parameters, we obtain \textit{three regimes} in thermodynamic
behaviour of the WIBG when $\theta <\theta _{0}\left( 0\right) $\textit{\ }%
(see Figures 4 and 5):

$\left( i\right) $ for $\rho \leq \rho _{\inf }^{B}\left( \theta \right) $,
there is no condensation;

$\left( ii\right) $ for $\rho _{\sup }^{B}\left( \theta \right) \leq \rho
\leq \rho _{c}^{B}\left( \theta \right) $, there is a \textit{%
non-conventional} condensation (\ref{eq 4.36})\textit{\ }in the mode $k=0$
due to non-diagonal interaction in the Bogoliubov Hamiltonian, see Figure 4;

$\left( iii\right) $ for $\rho _{c}^{B}\left( \theta \right) \leq \rho $,
there is a \textit{second} kind of condensation: the \textit{conventional}
type I Bose-Einstein condensation which occurs \textit{after }the\textit{\
non-conventional} one; it appears due to the standard mechanism of the total
particle density saturation (Corollary \ref{corollary cond5}).

\hspace{0.4cm}
When $\theta >\theta _{0}\left( 0\right) $, there are only \textit{two }%
types of thermodynamic behaviour: they correspond to $\rho \leq \rho
_{c}^{B}\left( \theta \right) $ with no condensation and to $\rho
_{c}^{B}\left( \theta \right) <\rho $ with a \textit{conventional}
condensation as in $\left( iii\right) $. Hence, for $\theta >\theta
_{0}\left( 0\right) $ the condensation in the WIBG coincides with the type I
generalised Bose-Einstein condensation in the PBG with \textit{excluded }%
mode $k=0$, see Theorem \ref{theorem cond5} $\left( iii\right) $.
%%%%%%%%%%%%%%%%%%%%%%%%%%%%%%%%%%%%%%%%%%%%%%%%%%%%%%%%%%%%%%%%%%%%%%%%%%%%%%%%%%%%%%%%%%%%%%%%%%
\setcounter{equation}{0}
\section{Conclusion}\label{section 6}

The paper presents a review of results regarding the non-conventional dynamical condensation versus
conventional Bose-Einstein condensation, including the generalised BEC \`{a} la
van den Berg-Lewis-Pul\'{e}.
It is based on discussion of two models: a simple toy model and the
Bogoliubov Weakly Imperfect Bose-Gas model, which was invented for explanation of superfluity
of ${\rm{^4He}}$, but which is also instructive for analysis of non-conventional dynamical
condensation versus recent reinterpretations of experimental data, see \cite{VYIC}, \cite{Blag20}.

\hspace{0.4cm}
We forewarn the reader about another usage of expression "non-conventional BEC", e.g.,
in the preprint arXiv:200101315v1, \textit{New scenario for the emergence of
non-conventional Bose-Einstein Condensation. Beyond the notion of energy gap},
by Marco Corgini.

%%%%%%%%%%%%%%%%%%%%%%%%%%%%%%%%%%%%%%%%%%%%%%%%%%%%%%%%%%%%%%%%%%%%%%%%%%%%%%%%%%%%%%%%%%%%%%%%%%%%%%%
\bigskip

\noindent
%%%%%%%%%%%%%%%%%%%%%%%%%%%%%%%%%%%%%%%%%%%%%%%%%%%%%%%%%%%%%%%%%%%%%%%%%%%%%%%%%%%%%%%%%%%%
\textbf{Acknowledgements}

\smallskip

This review was motivated by my lecture at the Conference (Bogoliubov Laboratory of
Theoretical, JINR-Dubna, held on 10 September 2019) dedicated to Viatcheslav Borisovich Priezzhev.

\hspace{0.4cm}
I am very thankful to organisers: Viatcheslav Spiridonov and Alexander Povolotsky,
for invitation and hospitality.

%%%%%%%%%%%%%%%%%%%%%%%%%%%%%%%%%%%%%%%%%%%%%%%%%%%%%%%%%%%%%%%%%%%%%%%%%%%%%%%%%%%%%%%%%%%%%%%%%%%%

%%%%%%%%%%%%%%%%%%%%%%%%%%%%%%%%%%%%%%%%%%%%%%%%%%%%%%%%%%%%%%%%%%%%%%%%%%%%%%%%%%%%%%%%%%%%%%%%%%%%
\end{document}